\documentclass[12pt]{article}
\pdfoutput=1
\usepackage{makeidx}
\makeindex
\usepackage[a4paper]{geometry}
\usepackage{jheppub,amsmath,amssymb,amsfonts,amsxtra, mathrsfs,graphics,graphicx,amsthm,epsfig,ytableau,bm,longtable,float,color,tikz,mathtools,xfrac,footnote,rotating,lscape}
\usepackage{dsfont}
\usepackage{textgreek}
\usetikzlibrary{decorations.pathmorphing}
\usetikzlibrary{decorations.markings}
\usetikzlibrary{arrows, decorations.markings, calc, fadings, decorations.pathreplacing, patterns, decorations.pathmorphing, positioning}
\usepackage{tikz-cd}

\usetikzlibrary{positioning,shapes}
\usetikzlibrary{chains}
\usetikzlibrary{arrows,fit,decorations.pathreplacing}
\tikzstyle{every picture}+=[remember picture]
\tikzstyle{na} = [baseline=-.5ex]

\usepackage{empheq}
\usepackage{multirow}
\usepackage{booktabs}
\usepackage[american]{babel}

\usepackage[latin1]{inputenc}

\makeatletter
\newcommand{\vast}{\bBigg@{1}}
\newcommand{\Vast}{\bBigg@{5}}
\makeatother

\usepackage{hyperref}

\numberwithin{equation}{section}

\numberwithin{equation}{section}

\newcommand{\be}{\begin{equation}} \newcommand{\ee}{\end{equation}}
\newcommand{\bea}{\begin{equation} \begin{aligned}} \newcommand{\eea}{\end{aligned} \end{equation}}

\def\U{\mathrm{U}}

\newcommand{\dvol}{d\mathrm{vol}}

\newcommand{\Vol}{\mathrm{Vol}}

\newcommand{\parfrac}[2]{\frac{\partial #1}{\partial #2}}

\newcommand{\wb}{\overline}
\newcommand{\wt}{\widetilde}

\newcommand{\pdag}{{\phantom{\dag}}}

\DeclareMathOperator{\Tr}{Tr}

\DeclareMathOperator{\sign}{sign}

\DeclareMathOperator{\re}{\mathbb{R}e}
\DeclareMathOperator{\im}{\mathbb{I}m}

\DeclareMathOperator{\Li}{Li}

\newcommand{\cC}{\mathcal{C}}

\newcommand{\cF}{\mathcal{F}}

\newcommand{\cI}{\mathcal{I}}

\newcommand{\cK}{\mathcal{K}}
\newcommand{\cL}{\mathcal{L}}
\newcommand{\cM}{\mathcal{M}}
\newcommand{\cN}{\mathcal{N}}

\newcommand{\cQ}{\mathcal{Q}}
\newcommand{\cR}{\mathcal{R}}

\newcommand{\cW}{\mathcal{W}}

\newcommand{\cZ}{\mathcal{Z}}

\newcommand{\bC}{\mathbb{C}}

\newcommand{\bR}{\mathbb{R}}
\newcommand{\bZ}{\mathbb{Z}}

\newcommand{\fg}{\mathfrak{g}}

\newcommand{\fh}{\mathfrak{h}}

\newcommand{\fm}{\mathfrak{m}}

\newcommand{\fn}{\mathfrak{n}}

\newcommand{\fp}{\mathfrak{p}}
\newcommand{\fq}{\mathfrak{q}}

\newcommand{\fR}{\mathfrak{R}}

\newcommand{\ft}{\mathfrak{t}}

\usepackage[Symbol]{upgreek}
\usepackage{bm}

\usepackage{calligra}
\DeclareMathAlphabet{\mathcalligra}{T1}{calligra}{m}{n}

\makeatletter
\def\@fpheader{\relax}
\makeatother

\renewcommand{\citep}{\cite}
\renewcommand{\citealt}{\cite}
\title{Lectures on AdS Black Holes, Holography  and Localization }

\author[a,b]{Alberto Zaffaroni}
\affiliation[a]{Dipartimento di Fisica, Universit\`a di Milano - Bicocca, I-20126 Milano, Italy}
\affiliation[b]{INFN, sezione di Milano-Bicocca, I-20126 Milano, Italy}
\emailAdd{alberto.zaffaroni@mib.infn.it}

\abstract{In these lectures I review some recent progresses in counting the number of microstates of AdS supersymmetric black holes 
       in dimension equal or greater than four using holography. The counting  is obtained by applying localization and matrix model techniques to the dual field theory.
       I cover in details the case of dyonic AdS$_4$ black holes, corresponding to a twisted compactification of the dual field theory, and I discuss
       the state of the art for rotating AdS$_5$ black holes.}

\begin{document}

\setcounter{tocdepth}{2}
\maketitle

%
%

\date{Dated: \today}

\section{Introduction}\label{sec:intro}

We know since the seventies that black holes in general relativity are thermodynamic objects, in particular they have a temperature and an entropy \citep{Bekenstein:1972tm,Bardeen:1973gs,Hawking:1974sw}. One of the most intriguing and celebrated relation in theoretical physics is the \emph{Bekenstein--Hawking formula} expressing the entropy of a black hole in terms of the area $A$ of the event horizon and the natural constants $c,\hbar, k_{{\rm B}}, G_{{\rm N}}$
\bea   S_{{\rm BH}} = k_{{\rm B}} \frac{  c^3 A}{4 \hbar  G_{{\rm N}}} \, ,\eea
merging in a single expression gravity, relativity, statistical mechanics and quantum theory.
Using a standard statistical mechanics interpretation,   we are led to write the entropy as
\bea S_{{\rm BH}} = k_{{\rm B}} \log n \, ,\eea
in terms of the number $n$ of microscopic degrees of freedom of the system,  the  microstates of the black hole.   It is a main challenge for all theories of quantum gravity to give an explanation of the Bekenstein--Hawking formula and to identify the corresponding microstates. 

The Bekenstein--Hawking formula suggests that the microstates are localized on the event horizon. This is an instance of the \emph{holographic principle} \citep{tHooft:1993dmi,Susskind:1994vu}, which states that a volume of space in quantum gravity can be described  just in terms of   boundary degrees of freedom. A concrete incarnation of this general principle is the AdS/CFT correspondence \citep{Maldacena:1997re}, a cornerstone of modern theoretical physics. The correspondence explicitly identifies a theory of quantum gravity in Anti-de-Sitter  space-time (AdS) with a dual conformal quantum field theory (CFT), which we may naively think of as living on  the boundary of AdS. String theory provides many explicit examples of AdS backgrounds and dual CFTs. The most famous is the original example of dual pairs,  type IIB string theory on AdS$_5\times S^5$ and the maximally supersymmetric gauge theory in four dimensions, ${\cal N} = 4$ super-Yang--Mills (SYM). Impressive checks of the correctness of this duality have been made over the last twenty years.

In this context, one of the great successes of string theory is the microscopic explanation of the entropy of certain  asymptotically flat black holes.  The first result was obtained in \cite{Strominger:1996sh}, more than twenty years ago, and has been followed by an immense literature, which would be too long to refer to. However, quite curiously, no similar results exist for asymptotically AdS black holes in dimension four or greater until very recently. Since holography suggests that the microstates of the black hole correspond to states in a dual conformal field theory, the AdS/CFT correspondence is the natural setting where to explain the black hole entropy in terms of a microscopical theory. In the past, various attempts have been made to derive the entropy of a class of rotating black holes in AdS$_5\times S^5$ in terms of states of the dual ${\cal N} = 4$ SYM theory in the large $N$ limit,  but none was completely successful. The more recent advent of localization techniques  for supersymmetric quantum field theories, in the spirit of \cite{Nekrasov:2002qd,Pestun:2007rz}, opens a new perspective on this problem. In these lectures we discuss how to use localization to derive the entropy for a class of supersymmetric black holes in AdS$_4$ and AdS$_5$ and discuss the current status for other black holes appearing in holography. We will work in dimension equal or greater than four. AdS$_3$ is somehow special, and well-studied in the literature,  and it will not be discussed in these notes. 

\subsection{Content of the lectures}
The first microscopic counting for AdS  black holes in dimension equal or greater than four was performed in \cite{Benini:2015eyy}, considering asymptotically AdS$_4$ static supersymmetric black holes. One of the main characteristics of this class of black holes is the presence of magnetic charges that correspond to a topological twist in the dual field theory. The black holes considered in \cite{Benini:2015eyy} can be embedded in M theory and are asymptotic to AdS$_4\times S^7$. They are  dual to a topologically twisted compactification of the ABJM theory in three dimensions \citep{Aharony:2008ug}, and their entropy scales as $N^{3/2}$ at large $N$, as familiar from three-dimensional holography. In these lectures, we will focus on this example as a prototype for many similar computations.  The entropy can be extracted from the (regularized) Witten index of the quantum mechanics obtained by compactifying ABJM on a Riemann surface $\Sigma_\fg$. Holographically,  the quantum mechanics describes the physics of the near horizon geometry AdS$_2\times \Sigma_\fg$ of the black holes. The index can be computed, using localization,  as the three-dimensional supersymmetric partition function of ABJM on $\Sigma_\fg\times S^1$, topologically twisted along the Riemann surface $\Sigma_\fg$.  From this perspective, this computation can be generalized to more general  domain wall solutions interpolating between AdS$_{d+n}$ and $AdS_d\times \cM_n$, with a topological twist along the $n$-dimensional compact manifold  $\cM_n$, thus providing general tests of holography.

In the second part of these lectures, we also  discuss the analogous problem for supersymmetric rotating  electrically charged black holes in AdS$_d$. The main difference  with the previous case is the absence of a topological twist. The entropy for such black holes should be obtained by counting states with given electric charge and spin in the dual field theory and the natural observable to consider is the superconformal index, which receives contributions precisely from the BPS states of the theory. 
Recent results in this direction have been obtained starting with the work   \citep{Cabo-Bizet:2018ehj,Choi:2018hmj,
Benini:2018ywd} and we will discuss these recent progresses obtained in various  overlapping limits but all pointing towards a unified picture.

Many of the relevant field theory computations are performed using localization. This allows to reduce exact path integrals in quantum field theory to matrix models, which  can be solved in the large $N$ limit combining standard and more recent techniques. One successful approach for the physics of black holes, that works both in four \citep{Benini:2015eyy,Hosseini:2016cyf} and five dimensions \citep{Benini:2018ywd}, involves writing the matrix model partition function as a sum of Bethe vacua \citep{Nekrasov:2009uh} of an auxiliary theory. Some technical aspects of this approach are discussed in Sect.~\ref{sec:lecture2}.
 
Let us stress that many derivations of the entropy for asymptotically flat black holes involve the use of the Cardy formula for the asymptotic growing of states of a two-dimensional CFT.
In the localization approach for AdS black holes, we directly count the number of microstates using an index.\footnote{Although some results about rotating  electrically charged black holes have been obtained in a Cardy limit, which provides a generalization of the Cardy formula to higher dimensions.} We will briefly make contact with the original Cardy approach based on a two-dimensional CFT in Sect.~\ref{sec:lecture4} where we discuss black strings.


These lectures assume some familiarity with supersymmetry and the main examples of holographic dualities in various dimensions. We assume that the reader knows that ${\cal N}=4$ SYM in four dimensions is  dual to AdS$_5\times S^5$, the ABJM theory to  AdS$_4\times S^7$ and the so-called $(2,0)$ theory in six dimensions   to AdS$_7\times S^4$. Some preliminary exposure to localization computation\footnote{We refer to \cite{Marino:2011nm} for a nice introduction and \cite{Pestun:2016jze} for a more comprehensive review.}  would be also useful although not necessary. We review instead in Sect.~\ref{subsec:BH} the
elements of gauged supergravity that are needed for these lectures. 

The lectures are organized as follows. In Sect.~\ref{sec:lecture1} we give a general overview of the various classes of supersymmetric black holes that are relevant for holography, stressing that they fall into two main classes, distinguished by the presence or absence of magnetic charges (or more precisely of a twist). We also discuss how we should compute their entropy using field theory methods and we introduce the concepts of entropy functional and attractor mechanism that prove useful in the comparison between gravity and field theory. 
In Sect.~\ref{sec:lecture2} and  \ref{sec:lecture3} we discuss in details the example of dyonic black holes asymptotic to AdS$_4\times S^7$ and dual to a twisted compactification of ABJM.  In Sect.~\ref{sec:lecture2} we discuss the field theory aspects of the story, introducing the topologically twisted index and showing how to evaluate it using localization. 
 In Sect.~\ref{sec:lecture3} we perform the large $N$ limit of the resulting matrix model and we compare with gravity.  In Sect.~\ref{sec:lecture4} we discuss black string solutions interpolating between AdS$_5$ and AdS$_3\times \Sigma_\fg$, as a prototype of more general domain walls interpolating between AdS spaces that can be studied with these techniques. In Sect.~\ref{sec:lectureSCI} and  \ref{sec:lecture5} we discuss the case of  rotating electrically charged black holes in AdS$_d$. In Sect.~\ref{sec:lectureSCI} we discuss the field theory aspects of the story, introducing the superconformal index.  Finally, in  Sect.~\ref{sec:lecture5} we discuss the state of the art of the comparison between field theory and gravity for such black holes.

\section{AdS black holes in $d\ge4$}
\label{sec:lecture1}

In these lectures we are interested in supersymmetric black holes that can be embedded in string theory or M-theory and are asymptotic to AdS$_d$ vacua with a known field theory dual. There are many such black holes that can be embedded in maximally supersymmetric backgrounds.  For example, we can find supersymmetric black holes  in AdS$_5\times S^5$, AdS$_4\times S^7$ and AdS$_7\times S^4$, whose dual field theories are well known. Supersymmetric black holes are extremal and have zero temperature. They also satisfy a BPS condition that relates their mass to the other conserved charges. In the limit where gravity is weakly coupled, the entropy of a black hole can be computed with the Bekenstein--Hawking formula
\be\label{BH} S= \frac{A}{4 G_{\text{N}}}\, ,\ee
where $A$ is the area of the horizon and  $G_{\text{N}}$ is the Newton constant.  We set $c=\hbar=k_{{\rm B}}=1$.

We now discuss some general features of these black holes and their holographic interpretation. %

\subsection{AdS black holes and holography} \label{subsec:AdSBH}

For the purposes of holography, we can divide the known supersymmetric black holes in dimension $d\ge4$ into two main classes, distinguished by how supersymmetry is realized and the  holographic interpretation.  In particular, they are distinguished by existence (or absence) of certain magnetic charges corresponding to a topological twist. 

\subsubsection{Kerr--Newman black holes and generalizations} \label{subsubsec:electric}

The first class of black holes consists of supersymmetric electrically charged  rotating black holes (Kerr--Newman). The most famous examples are the type IIB supergravity  black holes asymptotic to AdS$_5\times S^5$ found in \cite{Gutowski:2004yv,Gutowski:2004ez,Chong:2005da,Chong:2005hr,Kunduri:2006ek}. They depend on two angular momenta corresponding to  two Cartan isometries of AdS$_5$
\be  (j_1\, ,  j_2)  \qquad \qquad U(1)^2 \subset SO(4) \subset SO(2,4) \, ,\ee
 and three electric charges  under the Cartan isometries of $S^5$
\be   (q_1\, ,  q_2\, , q_3) \qquad \qquad U(1)^3 \subset SO(6) \, ,\ee
parameterizing rotations in the internal space $S^5$. Supersymmetry actually imposes a  relation among the conserved charges, $f(j_1,j_2,q_1,q_2,q_3)=0$, so that there are only four independent parameters.\footnote{For many different families of BPS black holes,   supersymmetry  imposes constraints among the charges.  The reason why this happens is still unclear. In the case of AdS$_5\times S^5$,  BPS hairy black holes depending on all the charges have been  found  in \cite{Markeviciute:2018yal,Markeviciute:2018cqs}.}   These black holes preserve two real supercharges out of the original thirty-two of type IIB supergravity on AdS$_5\times S^5$.
The five-dimensional part of the metric is asymptotic to AdS$_5$ with $\mathbb{R}\times S^3$ as conformal boundary. As well known, type IIB string theory on AdS$_5\times S^5$  is dual to ${\cal N}=4$ SYM in four dimensions.  It is then a natural expectation that the black holes correspond holographically to an ensemble of states of ${\cal N}=4$ SYM on  $\mathbb{R}\times S^3$ that preserve the same supersymmetries and have the same electric charges and the same angular momenta.  It is also natural to expect that, by counting all the $1/16$ BPS states of ${\cal N}=4$ SYM on  $\mathbb{R}\times S^3$ with electric charges $(q_1 ,  q_2 , q_3)$ and spin $(j_1 ,  j_2)$, we should be able to reproduce the entropy of these black holes. We will work under these assumptions.
We are interested in macroscopic black holes whose entropy, when expressed in terms of field theory data,
 scales as $O(N^2)$, where $N$ is the number of colors of the dual field theory.    

The situation is analogous in other dimensions \citep{Chong:2004dy,Cvetic:2005zi,Chow:2007ts,Chow:2008ip,Hristov:2019mqp}. Consider the maximally supersymmetric backgrounds AdS$_4\times S^7$ and AdS$_7\times S^4$ in M-theory. The isometry of AdS$_4\times S^7$ is $SO(2,3)\times SO(8)$ and we can find 
 electrically charged  rotating black holes depending on one angular momentum $j$  and four electric charges $(q_1,q_2,q_3,q_4)$ 
 with a constraint. They preserve two real supercharges.   We expect to reproduce the entropy of such black holes by counting all $1/16$ BPS states of the dual field theory on  $\mathbb{R}\times S^2$ with the same quantum numbers.  As well-known, the dual of M-theory on AdS$_4\times S^7$  is the three-dimensional ABJM theory at Chern--Simons level $k=1$ \citep{Aharony:2008ug}. The entropy of these black holes scales as $O(N^{3/2})$. Similarly,  since the isometry of AdS$_7\times S^4$ is  $SO(2,6)\times SO(5)$, there are black holes depending on three angular momenta $(j_1,j_2,j_3)$  and two electric charges $(q_1,q_2)$ with a constraint, again preserving two real supercharges.\footnote{Actually,  only black holes with equal charges or equal momenta  have been studied.  However, we expect  a family with at least four independent parameters to exist.}  In this case, the entropy, which scales as $O(N^{3})$, should be reproduced by counting states in  the  ${\cal N}=(2,0)$ theory in six dimensions \citep{Witten:1995zh}  on  $\mathbb{R}\times S^5$. 
 
 Notice that  all these supersymmetric black holes   rotate. If we turn off the angular momenta $j_i$,  we find singularities.
  
In principle, although there are not so many examples  in the literature, we expect the existence of similar supersymmetric black holes in
more general type II or M-theory backgrounds with an AdS$_d$ vacuum, rotating in AdS$_d$ and charged under the isometries of the compactification manifold. The holographic interpretation is similar. For example, for type IIB black holes in  AdS$_5\times SE_5$ \citep{Klebanov:1998hh}, where $SE_5$ is a five-dimensional Sasaki--Einstein manifold, we should try to match the entropy by counting $1/4$ BPS states of the dual ${\cal N}=1$ superconformal field theory on     $\mathbb{R}\times S^3$.

\subsubsection{Magnetically charged black holes with a twist} \label{subsubsec:magnetic}

The second class of black holes are characterized by the existence of certain magnetic charges. We should more properly refer to such black holes as solutions where supersymmetry is realized with a topological twist, as we will see.   Although there are examples in higher dimensions,\footnote{There are exotic $d$-dimensional solutions with horizon AdS$_2\times \cM_{d-2}$, where $\cM_{d-2}$ is a compact manifold, with non-zero fluxes of the gauge fields on $\cM_{d-2}$.}  we will focus on four dimensions, where these black holes arise naturally.  Indeed we can have both magnetic and electric charges in $d=4$ and it is natural to consider dyonic black holes.  There exists a family of  BPS black holes in  AdS$_4\times S^7$ depending on one angular momentum $j_1$ in AdS$_4$ and on electric and  magnetic charges 
\be (q_1\, ,  q_2\, , q_3, q_4) \qquad \qquad (p^1,\, p^2,\, p^3,\, p^4)\, ,\ee
under the abelian $U(1)^4\subset SO(8)$ isometries of $S^7$ \citep{Cacciatori:2009iz,DallAgata:2010ejj,Hristov:2010ri,Katmadas:2014faa,Halmagyi:2014qza,Hristov:2018spe}. Supersymmetry requires a linear constraint among the magnetic charges $p^a$ and non-linear ones among the conserved charges,  so that we have a six-dimensional family of rotating, dyonic black holes.  For this  class of solutions, we can also turn off rotation and have static supersymmetric black holes.\footnote{These BPS black holes have been found in ${\cal N}=2$ gauged supergravity in $d=4$ with vector multiplets and later uplifted to M-theory. The first static example, with a hyperbolic horizon,  was found in \cite{Sabra:1999ux} in minimal gauged supergravity. The first static spherically symmetric example was found in \cite{Cacciatori:2009iz}, further discussed in \cite{DallAgata:2010ejj,Hristov:2010ri} and generalized to the dyonic case in \cite{Katmadas:2014faa,Halmagyi:2014qza}. The rotating case has been discussed in \cite{Hristov:2018spe} (for other examples, see also \citep{Daniele:2019rpr}).}
 
In general, AdS$_4$ black holes with magnetic charges are qualitatively different from  those with zero magnetic charge, as first noticed in \cite{Romans:1991nq} and elaborated in \cite{Hristov:2011ye,Hristov:2011qr}.  The difference is well explained using holography.    
Consider the black holes as solutions of an effective four-dimensional theory with a AdS$_4$ vacuum dual to a boundary conformal field theory. For most of these lectures, the CFT will be ABJM, but the following arguments apply to black holes in general compactifications and more general  CFTs with at least ${\cal N}=2$ supersymmetry.   
For our purposes, we need the terms of the effective  theory describing the dynamics of the metric and of the vector fields $A_\mu^a$ corresponding to the abelian isometries of the internal manifold.  Their dynamics is described by an Einstein-Maxwell theory 
\be {\cal L}= \sqrt{g} \left ( R + g_{ab}(\phi_i) F_{\mu\nu}^a F^{\mu\nu b} + \ldots \, \right )\ee
where, in general,  the matrix of coupling constants   depends on the scalar fields $\phi_i$ of the theory. According to the rules of holography, gauge fields in the bulk correspond to global symmetries in the boundary CFT. For example, in the case of AdS$_4\times S^7$
we are interested in the four fields $A_\mu^a$ corresponding to the abelian isometries $U(1)^4\subset SO(8)$ and they couple to the field theory conserved currents  $J^{\mu a}$
\be \int d^4 x A_\mu^a J^{\mu a} \, \ee
associated to the Cartan generators of the $SO(8)$ R-symmetry of ABJM at Chern--Simons level $k=1$. Focusing for simplicity on the static case, one finds that, near the boundary, the black hole solutions behave as
\bea\label{asympt}
ds^2 \hskip 0.3truecm &= \frac{dr^2}{r^2} + r^2 ds^2_{\cM_3} +\ldots \, , \\
A_\mu^a(x,r) &= \hat A_\mu^a(x) +\ldots \, ,
\eea
where $r$ is some large radial coordinate, the ellipsis refers to terms suppressed by inverse powers of $r$, and $x$ are coordinates on the boundary manifold. For spherically symmetric black holes the boundary manifold is $\cM_3= \mathbb{R}\times S^2$. However we can have more exotic solutions with horizon AdS$_2\times \Sigma_\fg$, where $\Sigma_\fg$ is a Riemann surface of genus $g$, and, in this case,  $\cM_3= \mathbb{R}\times \Sigma_\fg$.  Holography tells us that we should interpret \eqref{asympt} as the dual of our CFT defined on the
curved manifold $\cM_3$. What can we say  about $A_\mu^a$? Recall again the basic rules of the AdS/CFT correspondence \citep{Klebanov:1999tb}.  Any field $\phi(x,r)$ in AdS is associated with an operator $O(x)$ in the dual CFT.  If we have an expansion
\be \phi(x,r) = r^{\alpha_1} \phi_0(x) + r^{\alpha_2} \phi_1(x) \, ,
\ee
of the solution of the second order equations of motion for $\phi$,\footnote{The values of $\alpha_i$ are related to the conformal dimension of $O$.}  we  interpret the non-normalizable piece, $\phi_0(x)$,  as a deformation of the original CFT with the corresponding operator $O$,
\be {\cal L}_{CFT}(x) \rightarrow {\cal L}_{CFT}(x) + \phi_0(x) O(x) \, ,\ee
while we interpret the normalizable one, $\phi_1(x)$, as a vacuum expectation value (vev) for $O$, $\langle O\rangle \ne 0$. More precisely, 
if $\phi_0(x)\ne 0$, we are deforming the CFT with $O$; if $\phi_0(x)=0$ and $\phi_1(x)\ne 0$ we have a state of the CFT with non zero vev for $O$. There are situations where both modes $\phi_0(x)$ and $\phi_1(x)$ are normalizable (or better have finite energy). In this case there are different possible quantizations of the same theory and we have to choose who plays the role of $\phi_0$. Massless vector fields in AdS$_4$ allow for different types of quantizations, related to electric/magnetic duality in the bulk, and this leads to interesting applications, but this is not strictly the most important point. What is important is that, in the expansion  \eqref{asympt} for $A_\mu^a$ both leading and sub-leading terms are turned on. The field $A_\mu^a$ has a leading contribution for $r\gg 1$ that approaches a constant value on the boundary $\cM_3= \mathbb{R}\times \Sigma_\fg$, corresponding to  the magnetic charge of the black hole
\be\label{magnetic} \frac{1}{2\pi} \int_{\Sigma_\fg} F^a = p^a \, ,\ee
and sub-leading terms (the ellipsis in  \eqref{asympt}) that encode information about the electric charges. This means that a dyonic black hole is holographically dual to a deformation of the dual CFT. In the natural quantization of the theory, the non-zero value of $A_\mu^a$ at the boundary  corresponds to the deformation
\be\label{def}  {\cal L}_{CFT}(x) \rightarrow {\cal L}_{CFT}(x) + \hat A_\mu^a (x) J^{\mu a}(x)  \, .\ee
This deformation in field theory is equivalent to turning on a background gauge field for a global symmetry. 
For example, on $S^2$ we would turn on a background  which is just the familiar  Dirac monopole $\hat A_\mu^a = -\frac12 p ^a \cos\theta d\phi$. On a torus $T^2$ we would just turn on a background constant magnetic field. Fields satisfying \eqref{magnetic} can be  also written explicitly for all $\Sigma_\fg$ but their expression  is not particular illuminating.  We will see in Sect.~\ref{sec:lecture2} that the deformation   \eqref{def} is compatible with  supersymmetry.

To understand better what is going on, it is useful to have a look at how supersymmetry is preserved in the presence of a generic assignment of magnetic charges.
We will be very schematic here just to convey the general idea. Consider the case where our effective theory is  a certain ${\cal N}=2$ gauged supergravity in four dimensions, corresponding to a ${\cal N}=2$ three-dimensional CFT.  The effective theory contains   the graviphoton field, $A_\mu^R$, holographically dual  to the $U(1)$ R-symmetry of the theory.  In general, $A_\mu^R$ is a linear combination of the vector fields  $A_\mu^a$ corresponding to the  isometries of the internal manifold. 
For the solution to be supersymmetric,  all fermion variations in the black hole background must be zero.  The gravitino variation in ${\cal N}=2$ gauged supergravity is schematically given by
\be\label{var} \delta \psi_\mu = \partial_\mu \epsilon + \frac 14 \omega_\mu^{ab} \Gamma_{ab} \epsilon - i A_\mu^R \epsilon  +\ldots \, .\ee
The magnetically charged static black holes of interest in this section   satisfy the BPS condition $\delta \psi_\mu=0$ by \emph{cancelling the spin connection with a background field for the R-symmetry}.  More precisely, we can regard the spin connection $\omega_\mu$ along  $\Sigma_\fg$ as a $U(1)$ gauge field. An explicit computation shows that $\omega_\mu$ is just a monopole of charge $2-2 \fg$, as the familiar relation $\frac{1}{2\pi}\int_{\Sigma_\fg} R = 2 -2 \fg$, with $R=d\omega$, clearly  shows.  Since $A_\mu^R$ is a linear combination of the $A_\mu^a$, it is  also a monopole, with charge given by a linear combination of the $p^a$. By appropriately choosing this linear combination\footnote{This is the linear constraint on the magnetic charges of the black hole that we mentioned before.}  and the spinor $\epsilon$, we can cancel the second and the  third term on the right hand side in \eqref{var}. We will come back to more precise expressions in Sect.~\ref{sec:lecture2}.  For the dyonic static black holes, restricting the index $\mu$ to lie along $\Sigma_g$,  one discovers that  the ellipsis cancels independently and we are left with the equation
\be \delta\psi_\mu =\partial_\mu \epsilon =0 \, .\ee
This equation is solved by taking $\epsilon$ constant along $\Sigma_\fg$. One also finds that the other components of the supersymmetry variations imply that $\epsilon$ is  time-independent but, in general, has a non-trivial profile in $r$. 

This discussion can be also applied to the dual CFT. By restricting the variations to the boundary, we see that the field theory  on $\mathbb{R}\times \Sigma_\fg$  is invariant under supersymmetry transformations with  a constant spinor.  The very same mechanism   is at work on the boundary:  we are turning on a  magnetic background for the R-symmetry that compensates the spin connection. In quantum field theory, this construction is well-known \citep{Witten:1988ze}. It is called \emph{topological twist}, as we will discuss in details in Sect.~\ref{sec:lecture2}. The conclusion is that  the dual CFT is  deformed by the presence of magnetic fluxes for all the global and R-symmetries, and, in particular, it is  topologically twisted by the magnetic flux for the R-symmetry. Notice that our argument was based on ${\cal N}=2$ supersymmetry with a $U(1)$ R-symmetry.  Theories can have a larger R-symmetry group, like ABJM, or many flavor symmetries. In these cases,  the choice of a $U(1)$ R-symmetry is not unique.  Each choice corresponds  to a different twist. We can indeed think of the magnetic charges $p^a$ as parameterizing a family of inequivalent twists. It is important to remember, however, that a linear combination of the $p^a$ is fixed   by the condition that the  background for the selected $U(1)$ R-symmetry  cancels the spin connection. There are only $n_V-1$ independent magnetic charges, where $n_V$ is the number of massless vectors. This number is  $n_V=4$ for ABJM.

The interpretation of the general rotating dyonic black hole is more complicated but similar in spirit. We can have rotation only in the spherically symmetric case, where $j$ is the spin along $S^2$. Rotations in the bulk correspond to turning on an Omega-background  \citep{Nekrasov:2003rj} in the boundary theory on $S^2$. The theory is still topologically twisted.

All this should be contrasted  with the black holes discussed in Sects.~\ref{subsubsec:electric}  where there is  no cancellation between the spin connection and the R-symmetry.  It can be expressed more formally in a difference between the supersymmetry algebra, as discussed in \cite{Hristov:2011ye,Hristov:2011qr}.  The real discriminant among the two class of black holes is the topological twist, or equivalently    a magnetic charge associated with the R-symmetry. There exist black holes with non-zero magnetic charges for the flavor symmetries only.\footnote{See \cite{Hristov:2019mqp} for example of rotating dyonic black holes with flavor magnetic charges.} They correspond to a CFT in a magnetic background but with no topological twist. From the point of view of micro-state counting  using holography, they are more similar in spirit to the black holes discussed in Sect.~\ref{subsubsec:electric}.

Now it is clear what we should do in order to compute the entropy of magnetically charged black holes (with a twist) using field theory: enumerate all the states 
with electric charges $q_i$ and angular momentum $j$ and the right amount of supersymmetry  in the twisted CFT on $\mathbb{R}\times \Sigma_\fg$. The theory is topologically twisted\footnote{And also Omega-deformed if there is rotation.} by the magnetic background for a $U(1)$ R-symmetry and possibly deformed by magnetic fluxes for all other global symmetries.

\subsection{Computing the entropy}\label{subsubsec:entropy}
It is reasonable to expect that we can recover the entropy of the two classes of AdS$_d$ black holes by enumerating supersymmetric states in the dual field theory on $\mathbb{R}\times \cM_{d-2}$. For all our examples, the preserved supersymmetry $\cQ$ satisfies an algebra  of the form
 \bea \{ {\cal Q}^\dagger \, , {\cal Q}\}= H - \mu_a Q_a -\nu_i J_i \, ,\eea
where $H$ is the Hamiltonian and $Q_a$ and $J_i$ are the charge operators associated with the global symmetries and the angular momenta, respectively,  with certain constants $\mu_a$ and $\nu_i$ that depend on the model. The explicit form of the algebra is different for different types of black holes and it will be discussed in  details in the rest of these lectures. For the moment, we notice that the R-symmetry charge  enters in the supersymmetry algebra for Kerr-Newmann black holes but not for topologically twisted ones. Supersymmetric states are annihilated by $\cQ$ and their energy is determined by  the  BPS condition\footnote{It follows from the algebra that all states in the theory satisfy the  BPS bound $E\ge \mu_a q_a +\nu_i j_i$.} 
\bea \label{BPSbound} E = \mu_a q_a +\nu_i j_i \, .\eea

\subsubsection{The grand canonical partition function}\label{grancan}
 Enumerating BPS states is equivalent to knowing the grand canonical partition function
\be\label{BPSpf}  {\cal Z}(\Delta_a, \omega_i) = \Tr \Big |_{{\cal Q}=0} e^{i (\Delta_a Q_a +\omega_i J_i)} = \sum_{q_a,j_i} c(q_a,j_i) e^{i (\Delta_a q_a +\omega_i j_i)} \, ,\ee
where the trace is taken over the Hilbert space of states on $\cM_{d-2}$ that preserves the same amount of supersymmetry of the black hole,  and $\Delta_a$ and $\omega_i$  chemical potentials conjugated to $Q_a$ and $J_i$, respectively. In practical applications, ${\cal Z}$ is a function of complex chemical potentials and converges in an appropriate domain of the complex plane for the  fugacities $y_a=e^{i\Delta_a}$, $\zeta_i=e^{i \omega_i}$.  In the previous formula, $c(q_a,j_i)$ is the number of supersymmetric states of electric charge $q_a$ and angular momentum $j_i$.   Electric and magnetic charges enter in an asymmetric way in this construction. The magnetic charges $p^a$ enter explicitly as a set of couplings in the Lagrangian of the deformed CFT, while the electric charges $q_a$ are introduced through chemical potentials. In particular, for magnetically charged black holes with a twist, the trace must be taken in the topologically twisted theory. 

The grand canonical partition function \eqref{BPSpf} should also enumerate the BPS states in the dual gravity theory. Our working assumption is that, for large charges (scaling with appropriate powers of $N$) the supersymmetric density of states is dominated by the macroscopic black holes  discussed before.  Under this assumption, by the very definition of entropy, the entropy of the black hole is given by  
\be S(q_a,j_i) = \log c(q_a,j_i)\, ,\ee
where the dependence on the magnetic charges $p^a$, if present, is hidden in the form of the function $c$.
If $\cZ(\Delta_a, \omega_i)$ is known,  the entropy can be extracted as a Fourier  coefficient
\be e^{S(q_a,j_i)} =  c(q_a,j_i) = \int \frac{d\Delta_a}{2\pi} \frac{d\omega_i}{2\pi}  \cZ(\Delta_a, \omega_i) e^{-i (\Delta_a q_a +\omega_i j_i)} \, ,\ee
with an appropriate integration contour. In the limit of large charges,  this can be evaluated by a saddle point approximation
\be\label{entropyind} S(q_a,j_i) = \log \cZ(\Delta_a, \omega_i) - i (\Delta_a q_a +\omega_i j_i) \Big |_{\bar \Delta_a,\bar \omega_i} \, ,\ee
where $\bar \Delta_a$ and $\bar \omega_i$ are obtained by extremizing the functional
\be {\cal I}(\Delta_a, \omega_i)=  \log \cZ(\Delta_a, \omega_i) - i (\Delta_a q_a +\omega_i j_i)  \, ,\ee
with respect to $ \Delta_a$ and $ \omega_i$,
\be \partial_{\Delta_a}  {\cal I}(\Delta_a, \omega_i) = \partial_{\omega_i}  {\cal I}(\Delta_a, \omega_i) = 0  \Big |_{\bar \Delta_a,\bar \omega_i} \, .\ee
We see that the entropy is just the Legendre transform of the logarithm of the partition function.

To understand better this point, recall that we are interested in extremal supersymmetric black holes. In particular, they have zero temperature.
The standard thermodynamics relation for the grand canonical partition function of a system with  temperature $T$,  
\bea \log \cZ = -(E- T S - i \tilde \Delta_a q_a -i \tilde\omega_i j_i)/T \, , \eea
 looks singular in  the zero-temperature limit. However,  supersymmetric states  satisfy the BPS condition \eqref{BPSbound}, namely $E = \mu_a q_a +\nu_i j_i$. When we take the  zero temperature limit, we need also to scale $\tilde \Delta_a(T)=- i \mu_a +  \Delta_a T$ and $\omega_i(T)=- i \nu_i +  \omega_i T$. In this way we obtain  the Legendre transform $S= \log \cZ  -  i\Delta_a q_a -i \omega_i j_i$.\footnote{\label{limitT}  For explicit examples of this zero-temperature limit  from the gravitational point of view  see \cite{Silva:2006xv}, and, in particular, for  AdS$_5$ black holes, see \cite{Cabo-Bizet:2018ehj,Choi:2018hmj}.}

 The entropy is also obtained via a Legendre transform in many other approaches,  as  the OSV conjecture \citep{Ooguri:2004zv} and the Sen's quantum entropy functional  \citep{Sen:2005wa,Sen:2008vm,Sen:2009vz} for  asymptotically flat black holes. 

\subsubsection{The supersymmetric index}\label{index}

So far everything was simple. The problem is that $\cZ(\Delta_a, \omega_i)$ is too hard to compute, in general. For electrically charged rotating black holes in AdS$_5\times S^5$,  computing $\cZ(\Delta_a, \omega_i)$ would correspond to enumerate all the 1/16 BPS states of ${\cal N}=4$ SYM.
For comparison, in a four-dimensional theory with ${\cal N}=1$ supersymmetry, this would correspond to count all the 1/4 BPS states. While almost everything is known about the counting of 1/2 BPS states in an ${\cal N}=1$ theory \citep{Kinney:2005ej,Benvenuti:2006qr}, the analogous problem for 1/4 BPS states is still open. 

What we can instead compute  is a supersymmetric index
\bea\label{BPSindex}  \cZ_{\rm index}(\Delta_a, \omega_i)  = \Tr \Big |_{{\cal Q}=0} (-1)^F e^{i (\Delta_a Q_a +\omega_i J_i)} \, , \eea
with the insertion of the fermionic number $(-1)^F$. 
Standard arguments tell us  that we can also write
\bea\label{indexsusy}
\cZ_{\rm index}(\Delta_a, \omega_i) = \Tr  (-1)^F e^{-\beta \{ {\cal Q}^\dagger \, , {\cal Q}\}} e^{i (\Delta_a Q_a +\omega_i J_i)} =
Z^{susy}_{S^1\times \cM_{d-1}}(\Delta_a, \omega_i)  \, ,\eea
if $Q_a$ and $J_i$ commute with ${\cal Q}$. In the first step of the previous identification we used the fact that  states with ${\cal Q}\ne 0$ do not contribute to the trace, since bosonic and fermionic states are paired and contribute with opposite sign.\footnote{This is the logic of the Witten index \citep{Witten:1982df}.  For every  state $|\Omega \rangle$ not annihilated by $\cQ$  there is a  state ${\cal Q} |\Omega \rangle$ of opposite statistics and the same value of $ \{ {\cal Q}^\dagger \, , {\cal Q}\}$. Since these states have opposite value of $(-1)^F$, their contribution cancels in the trace in  \eqref{indexsusy}.  Therefore, only supersymmetric states, annihilated by ${\cal Q}$, contribute to the Witten index.  See also the discussion in Sect.~\ref{subsec:QM}.}   In particular, the index is independent of $\beta$. In the second step, we identified the  trace at temperature $1/\beta$ with the Euclidean path integral $Z^{susy}_{S^1\times \cM_{d-1}}$ of the theory
compactified on a circle of radius $\beta$.\footnote{\label{foot1} It is a standard textbook result that the finite temperature partition function $\Tr e^{-\beta H}$ can be expressed as an Euclidean partition function with time  compactified on a circle of radius $\beta$ and periodic boundary conditions for bosons and anti-periodic boundary conditions for fermions. In a supersymmetric partition functions, bosons and fermions should have the same boundary conditions and this is enforced by the fermionic number $(-1)^F$. The extra insertion of $ e^{\beta(\mu_a {\cal R}_a -\nu_i {\cal J}_i)}$  just introduces twisted boundary conditions along $S^1$ for the symmetries associated with  ${\cal R}_a$  and ${\cal J}_i$.} 
 The latter partition function can be computed using  localization techniques in quantum field theory as we discuss in Sect.~\ref{sec:lecture2}.

In general, $\cZ_{\rm index}(\Delta_a, \omega_i)\ne \cZ(\Delta_a, \omega_i)$. First of all, the index can accommodate fugacities only for the conserved charges that commute with  ${\cal Q}$ and, in general, contains less parameters than the BPS partition function. We will see that this is not a major problem for the black holes considered in this paper since they  also have a constraint among charges.  More importantly, the entropy should count all the BPS ground states of the theory, while the index counts bosonic ground states with a plus and fermionic ground states with a minus. However, it may happen for particular theories that   the majority of ground states   are of definite statistic. In this case there is no cancellation between bosonic and fermionic ground states and the index correctly reproduces the entropy in a suitable limit. This typically happens for  asymptotically flat black holes in the limit of large charges, and we may hope that the same is true for asymptotically AdS black holes in the large $N$ limit. In the case of certain  asymptotically flat spherically symmetric black holes, there is an extra symmetry that implies $(-1)^F=1$ on the relevant set of states \citep{Sen:2009vz} and one can prove that $\cZ_{\rm index}(\Delta_a, \omega_i)= \cZ(\Delta_a, \omega_i)$. A similar argument for asymptotically  AdS black holes is more subtle and it is discussed in Sect.~\ref{subsec:Iextremization}.  More generally, we can always rely on an explicit computation, and, as we will see in the rest of these lectures:

\begin{itemize}
\item for magnetically charged  black holes in AdS$_4$, the dual field theory is topologically twisted. $\cZ_{\rm index}(\Delta_a, \omega_i)$ is the so-called  topologically twisted index that we define and study in Sect.~\ref{sec:lecture2}. There is no cancellation between bosonic and fermionic ground states and the index correctly reproduces the entropy at large $N$, as we will see in Sect.~\ref{sec:lecture3};
\item for electrically charged rotating black holes, we are just counting states of the CFT on $\mathbb{R}\times S^{d-1}$. $\cZ_{\rm index}(\Delta_a, \omega_i)$ is the superconformal  index, whose properties we discuss in Sect.~\ref{sec:lectureSCI}.   The superconformal index is known to have large cancellations between bosons and fermions for real values of the fugacities
\citep{Kinney:2005ej}, and this would suggest that  we really need to compute 
the original BPS partition function $\cZ(\Delta_a, \omega_i)$. 
However, it has been recently realized that introducing phases for the fugacities can obstruct the cancellation between bosonic and fermionic states and that the entropy of KN black holes is indeed correctly captured by the index for complex values of the chemical potentials. We will discuss  these results  in Sect.~\ref{sec:lecture5}.
%
\end{itemize} 

Notice also that the entropy and the index  of asymptotically flat black holes coincide at leading order in the charges but are in general different when corrections are included \citep{Dabholkar:2010rm}. We might expect the same for asymptotically AdS black holes.

\subsection{Entropy functionals and the attractor mechanism}\label{subsec:attractor}

In the limit where gravity is weakly coupled, we can compute the entropy of a black hole from the area of the horizon using the Bekenstein--Hawking formula \eqref{BH}.
The area can be extracted from the explicit  solution of the relevant gauged supergravity, which typically contains many scalars $X^I(r)$ varying with the radial distance. In principle, the area may depend on many parameters including  asymptotic moduli. However, the microscopic entropy
of our black holes is just a function of the conserved charges $q_a$ and $j_i$. This is explicitly realized through   an \emph{attractor mechanism}: independently of the asymptotic moduli, the scalar fields approach  a value at the horizon, $X^I_*(q_a,j_i)$, that is a function of the conserved charges only. Moreover, this mechanism often allows to express the area, and therefore  the entropy, in terms of the values of the scalar fields at the horizon with simple algebraic equations.  This is the case of the  attractor mechanism for ${\cal N}=2$  supergravity discovered in \cite{Ferrara:1996dd,Ferrara:1995ih}. It is also the idea behind  Sen's quantum entropy function \citep{Sen:2005wa} that allows to find the entropy
of black holes with AdS$_2$ horizon including higher derivative corrections. In all these cases, one can define some sort of entropy functional $S(X^a, \Omega^i, q_a,j_i)$, which is a function of the conserved charges and the \emph{horizon value} $X^a,\, \Omega^a$ of the scalar fields and possibly other modes, and whose extremization with respect to  $X^a$ and $\Omega^i$  reproduces the entropy. This extremization is expected to be the gravity
analog of \eqref{entropyind}. 

In order to make the  comparison between gravity and field theory more manifest it is convenient to write the entropy functional  as a Legendre transform
\bea S(X^a, \Omega^i, q_a,j_i) ={\cal E} (X^a, \Omega^i) -  i (X^a q_a + \Omega^i j_i) \, , \eea
where $X^a$ and $\Omega^i$ are now interpreted as the black hole chemical potentials and ${\cal E} (X^a, \Omega^i) $ as the black hole gran-canonical partition function.
According to standard arguments \citep{Gibbons:1976ue}, ${\cal E} (X^a, \Omega^i)$ can be identified with the on-shell action 
of the Euclidean continuation of the black hole. 

This description fully agrees with the field theory picture if we identify $X^a$ and $\Omega^i$  with $\Delta^a$ and $\omega_i$ and ${\cal E} (X^a, \Omega^i)$ with  $\log \cZ(\Delta_a, \omega_i)$. The latter is indeed the field theory gran-canonical partition function and, according to the rule of holography, should be also identified with the on-shell action of the gravity solution corresponding to the chemical potentials $\Delta^a$ and $\omega_i$.

The attractor mechanism has played an important role in the interpretation of the field theory result
leading to the  black hole entropy.   For example, the attractor mechanism in N = 2 gauged supergravity \citep{Cacciatori:2009iz,DallAgata:2010ejj} for static dyonic black holes in AdS$_4\times S^7$ predicts 
\be
\label{Z large N0}
S(X^a)  = - \frac{2\sqrt2 N^\frac32}3   \sum_{a=1}^4 \fp_a \parfrac{}{X^a} \sqrt{X^1 X^2 X^3 X^4} - i \sum_{a=1} X^a q_a \;,
\ee
with the constraint $\sum_{a=1}^a X^a=2 \pi$ and this result perfectly matches with the field theory computation based on the topologically twisted index  \citep{Benini:2015eyy}  as we will discuss in Sect.~\ref{sec:lecture3} (see formula \eqref{Z large N}).  For other  black holes the attractor mechanism is not known in supergravity, and the entropy functional  has been  written using combined field theory and gravity intuition. This approach was successfully used in \cite{Hosseini:2017mds} to write an entropy functional for KN black holes in  AdS$_5\times S^5$,
\bea\label{entropyind00} S(X^a,\Omega^i) =  -  i \frac{ N^2}{2}  \frac{X^1 X^2 X^3}{\Omega^1\Omega^2} -i \left(\sum_{a=1}^3 X^a q_a +\sum_{i=1}^2\Omega^i j_i\right )  \, ,\eea
with the constraint $\sum_{a=1}^3 X^a -\sum_{i=1}^2 \Omega^i = 2\pi$. This result has been instrumental in the later developments and has been matched with field theory computations based on the superconformal  index \citep{Cabo-Bizet:2018ehj,Choi:2018hmj,Benini:2018ywd}, as we will discuss in Sect.~\ref{sec:lecture5}. Entropy functional for other electrically charged and rotating black holes in diverse dimensions has been later found in \cite{Hosseini:2018dob,Choi:2018fdc} and, in some cases, successfully compared to quantum field theory expectations. 
These entropy functionals can be also obtained by computing the zero-temperature limit of the on-shell action of a class of supersymmetric but non-extremal Euclidean black holes \citep{Cabo-Bizet:2018ehj,Cassani:2019mms}. 

A general field theory inspired formula for an entropy functional that generalizes \eqref{Z large N0} and \eqref{entropyind00} and covers all existing black hole solutions in four and five dimensions has been discussed in \cite{Hosseini:2019iad}.

\subsection{$\cI$-extremization}\label{subsec:Iextremization}

We can provide a perhaps more speculative but intriguing explanation of the extremization principle \eqref{entropyind} based on the renormalization group flow. All the black holes we will consider have a (possibly warped) AdS$_2$ factor in the near-horizon region. This suggests the existence of a superconformal quantum mechanics describing the near-horizon
degrees of freedom and whose ground states correspond to the black holes microstates. The superconformal algebra associated with AdS$_2$ is su$(1,1|1)$ and contains the bosonic factor $SL(2,\mathbb{R})\times U(1)_R$, where $U(1)_R$ is the R-symmetry.  From the field theory point of view, the quantum mechanics arises from the reduction of the dual CFT on $\cM_{d-2}$.  We can see the black hole as a solution interpolating between AdS$_{d}$ and AdS$_2$ with the dual interpretation of a renormalization group flow across dimensions where a $d-1$-dimensional CFT compatified on $\cM_{d-2}$ flows to a infrared superconformal quantum mechanics. The index $\cZ_{\rm index}(\Delta_a, \omega_i)$ is invariant under renormalization group flow and can  be interpreted as the Witten index of the infrared quantum mechanics. 

We can write the index as
\bea\label{trace000} \cZ_{\rm index}(\Delta_a, \omega_i) =  \Tr \Big |_{{\cal Q}=0} (-1)^F e^{i (\Delta_a Q_a +\omega_i J_i)}=  \Tr \Big |_{{\cal Q}=0} e^{\pi i R(\Delta,\omega)}  e^{- \im (\Delta_a Q_a +\omega_i J_i)} \, , \eea
where
\bea R(\Delta,\omega) = F + \frac{\re \Delta_a}{\pi} Q_a + \frac{\re \omega_i}{\pi} J_i \, .\eea 
In the examples that we will discuss the fermion number $F$ acts precisely as a particular R-symmetry, $F\equiv R_0$.  In a general supersymmetric theory with global symmetries, the R-symmetry is not unique. If $R_0$ is a R-symmetry and $Q$ is a global symmetry, also $R_0+Q$ is a R-symmetry. We see that $R(\Delta, \omega)$ is a R-symmetry of our quantum mechanics, and given the symmetries of the problem, we  expect it to be the most general R-symmetry we can write. However, when a supersymmetric theory is also conformal there exists the notion of the \emph{exact} R-symmetry, which is the one singled out by the superconformal algebra. The problem of finding the exact R-symmetry is central in supersymmetric quantum field theory and it is usually associated with an extremization problem. We have indeed $c$-extremization in two dimensions  \citep{Benini:2012cz}, $F$-maximization in three dimensions \citep{Jafferis:2010un} and $a$-maximization in four dimensions \citep{Intriligator:2003jj}.  It is tempting to propose that  the extremization \eqref{entropyind}  selects among all $R(\Delta, \omega)$  precisely the exact R-symmetry of our quantum mechanics. This principle has been called $\cI$-\emph{extremization} in \cite{Benini:2015eyy,Benini:2016rke}.  It is  suggested by the fact that, in odd dimensions, the exact R-symmetry is obtained by extremizing the supersymmetric sphere partition function, which, in the case of one dimension, is just the Witten index. It would be interesting to prove or disprove this principle for generic holographic theories.  

This interpretation would explain why there is no cancellation in the index between vacua of different statistic for large charges. Adapting an argument in \cite{Sen:2009vz}, we might expect that the microstates are invariant under the superconformal algebra su$(1,1|1)$ and this implies that they have zero exact R-charge. This is certainly true if the quantum mechanics consists of a set of degenerate ground states with an energy gap to the first excited state.  Under this assumption and using also large $N$ factorization of the correlation functions,  the trace \eqref{trace000} becomes
\bea\label{trace02} \cZ_{\rm index}(\Delta_a, \omega_i) = e^{- \im (\Delta_a \langle Q_a \rangle +\omega_i \langle J_i\rangle )}  \Tr \Big |_{{\cal Q}=0} 1 =  e^{- \im (\Delta_a q_a +\omega_i j_i )} e^{S(q_a,j_i)} \, , \eea
where we used that, for the values of $\Delta_a$ and $\omega_i$ that select the exact R-charge, all states have $R(\Delta,\omega)=0$.
We then see that the cancelation due to $(-1)^F$ is balanced by the non-zero phases of the fugacities at the saddle point and all the ground states contribute with the same sign. 
The previous equation is consistent with  \eqref{entropyind}. Indeed, by taking the logarithm of \eqref{trace02}, we find
\bea S(q_a,j_i) = \re \left( \log \cZ_{\rm index}(\Delta_a, \omega_i) - i (\Delta_a q_a +\omega_i j_i ) \right) \, .\eea
In all our example, the extremum of \eqref{entropyind} is actually real and the  real part in the previous formula is superfluous.\footnote{That the extremum of \eqref{entropyind} is real is usually part of the attractor mechanism. As we will see, the fact that the extremization of \eqref{entropyind} leads to a real number is equivalent to the non-linear constraint among charges imposed by supersymmetry.} 

Let us also notice that $c$-extremization and $F$- and $a$-maximization have a well-known gravitational dual \citep{Martelli:2005tp,Couzens:2018wnk,Gauntlett:2018dpc} which allows to determine the exact  R-symmetry in terms of geometrical data of the supergravity background. A gravitational dual for $I$-extremization has been also proposed in  \cite{Couzens:2018wnk} and further discussed in \cite{Gauntlett:2019roi,Hosseini:2019ddy,Kim:2019umc}.

\section{The topologically twisted index}\label{sec:lecture2}

As we discussed in Sect.~\ref{sec:lecture1}, magnetically charged black holes in AdS$_4$ are dual to  topologically twisted CFT$_3$. In this section, we discuss the \emph{topologically twisted index} in three dimensions,  $\cZ_{\rm index}(\Delta_a, \omega_i)$, defined as the supersymmetric partition function $Z^{susy}_{\Sigma_\fg\times S^1}(\Delta_a, \omega_i)$  with a topological A-twist along $\Sigma_\fg$. We will discuss the case of a generic ${\cal N}=2$ Yang--Mills--Chern--Simons theory in three dimensions with an R-symmetry and we will specialize to the ABJM theory in Sect.~\ref{sec:lecture3}.   

The index  can be computed in many  different ways.  We will discuss the localization approach here, following \cite{Benini:2015noa,Benini:2016hjo}.  The index  has been first derived  by  topological field theory arguments in various examples in \cite{Okuda:2012nx,Okuda:2013fea} and discussed in general in \cite{Nekrasov:2014xaa}. In this second approach, further discussed and generalized in \cite{,Gukov:2015sna,Okuda:2015yea,Closset:2016arn,Gukov:2016gkn,Closset:2017zgf,Closset:2017bse,Closset:2018ghr}, the index is written as a sum of contributions coming from the Bethe vacua, the critical points of the twisted superpotential of the two-dimensional theory obtained by compactifying on $S^1$ \citep{Nekrasov:2009uh,Nekrasov:2009rc}. We will discuss the connections between the two approaches in Sect.~\ref{subsec:BE}. The reader that is not interested in quantum field theory properties of the index can just have a quick look at Sect.~\ref{subsec:loc}, the first part of Sect.~\ref{subsec:localization} and Sect.~\ref{subsec:BE}.


\subsection{The topological twist}\label{subsec:loc}
Consider an ${\cal N}=2$ quantum field theory in three dimensions.  The  ${\cal N}=2$ supersymmetry multiplets are:

\begin{itemize}
\item the vector multiplet, $(A_\mu,\lambda,\sigma,D)$, where $\lambda$ is a Dirac spinor and $\sigma$ and $D$ are real scalars. $D$ is an auxiliary field;
\item the chiral multiplet, $(\phi,\psi,F)$, where $\psi$ is a Dirac spinor and $\phi$ and $F$ are complex scalars. $F$ is an auxiliary field.
\end{itemize} 
These multiplets can be obtained by dimensional reduction from the corresponding ${\cal N}=1$ multiplets in four dimensions. We assume that the theory has an R-symmetry\footnote{The sign of the charges is somehow unconventional (for example, $\lambda$ and $\epsilon$ have charge $-1$) but we keep it for consistency with \cite{Benini:2015noa,Benini:2016hjo}.}
\be \lambda \rightarrow e^{-i\alpha} \lambda\, ,\qquad (\phi,\psi,F) \rightarrow (e^{i r_\phi \alpha} \phi \, ,  e^{i (r_\phi-1) \alpha} \psi \, ,  e^{i (r_\phi-2) \alpha} F) \, ,  \ee 
with  charges $r_\phi$ for the chiral fields. The charges should be  integrally quantized, as we will discuss in the following. We want to define a supersymmetric theory  on $\Sigma_\fg\times S^1$ using a non-trivial  background for the R-symmetry. Let us start, for simplicity, with the case of $S^2\times S^1$ with metric
\bea\label{background}  ds^2 &= R^2 (d\theta^2+\sin\theta^2 d \phi^2) + \beta^2 d t^2 \, ,\eea
and a non-trivial R-symmetry background field $A_\mu^R$. 

To see if supersymmetry is preserved we can use the approach of \cite{Festuccia:2011ws}. We promote the metric $g_{\mu\nu}$  and the R-symmetry background field $A_\mu^R$ to dynamical fields by  coupling the theory to  supergravity, using an appropriate off-shell formulation. The theory coupled to gravity is invariant under  supersymmetry transformations with a local spinorial parameter $\epsilon(x)$. We can recover the rigid theory by  freezing  the supergravity multiplet to a background value. This can be done,  for example,  by sending the Planck mass  to infinity. In this process we set all supergravity fermionic fields to zero, while keeping a non-trivial background for the metric and $A_\mu^R$, and possibly  some auxiliary fields. The rigid theory so obtained is no more invariant under  local supersymmetries. However, it is still invariant under those transformations 
that preserve the background fields.   The supersymmetry variation of the bosonic supergravity fields is automatically zero since it is proportional to the supergravity fermions that vanish in the background. On the other hand, the vanishing of the fermionic  variations gives a differential equation for $\epsilon(x)$. The solutions to this equation determine the   rigid supersymmetries that are preserved by the curved background.\footnote{These conditions impose constraints on the manifold and the choice of background fields. For examples related to our context see \cite{Festuccia:2011ws,Klare:2012gn,Dumitrescu:2012ha,Closset:2012ru,Hristov:2013spa}.}

In any supergravity with a R-symmetry gauge field, when all the other supergravity fields are set to zero,  the fermionic variations have the universal form
\be \delta \psi_\mu = D_\mu \epsilon = \partial_\mu \epsilon +\frac14 \omega_\mu ^{ab} \gamma_{ab} \epsilon + i A_\mu^R \epsilon =0 \, . \ee
In three dimensions, we can choose $\gamma_a=\sigma_a$, where $\sigma_a$ are the Pauli matrices.  The non-trivial components of the spin connection are easily computed\footnote{We use the frame $e^1=R d\theta, e^2=R \sin\theta d\phi$ and $e^3=\beta dt$.} to be
$\omega^{12} = -\cos \theta d \phi$. If we take $\gamma_3\epsilon=\epsilon$, so that $\gamma_{12}\epsilon=i\epsilon$, we see that 
the background field
\be\label{mon}  A^R= \frac12 \cos \theta d \phi\ee precisely cancels the spin connection.  The equation reduces to
\be \delta \psi_\mu  = \partial_\mu \epsilon  =0 \, , \ee
which is  solved by a constant spinor $\epsilon$. We thus see that the background \eqref{mon} allows to define a supersymmetric theory on $S^2\times S^1$. 
The generalization of the above discussion to $\Sigma_\fg \times S^1$ is straightforward: we just turn on a background for the R-symmetry with $A^R=-\omega/2$ and everything else works in the same way. 

This way of preserving supersymmetry corresponds to a \emph{topological twist} along $\Sigma_\fg$ in the sense of \cite{Witten:1988ze,Witten:1991zz}.  To make contact with the language used in \cite{Witten:1988ze,Witten:1991zz},  we can interpret the background field for $A^R$ as effectively changing the spin of the fields in the theory, thus transforming $\epsilon$ into a scalar.   

A supersymmetric Lagrangian for a Yang--Mills--Chern--Simons theory with gauge group $G$ and chiral matter  transforming in a representation ${\cal R}$ on $\Sigma_\fg \times S^1$ 
can be written as  $\cL=\cL_\text{YM} + \cL_{CS} + \cL_\text{mat}+ \cL_\text{W}$ with\footnote{This can be obtained for example by taking the rigid limit of supergravity in the background \eqref{background}, as suggested in \cite{Festuccia:2011ws}.} 
\bea\label{Lagr}
\cL_\text{YM} &= \Tr\bigg[ \frac14 F_{\mu\nu} F^{\mu\nu} + \frac12 D_\mu\sigma D^\mu\sigma + \frac12 D^2 - \frac i2 \lambda^\dag \gamma^\mu D_\mu\lambda - \frac i2 \lambda^\dag [\sigma,\lambda] \bigg]  \\
\cL_{\text{CS}} &= - \frac{ik}{4\pi} \Tr \bigg[ \epsilon^{\mu\nu\rho} \Big( A_\mu \partial_\nu A_\rho - \frac{2i}3 A_\mu A_\nu A_\rho \Big) + \lambda^\dag \lambda + 2D\sigma \bigg]  \\
\cL_\text{mat} &= D_\mu \phi_i^\dag D^\mu\phi_i + \phi_i^\dag \big( \sigma^2 + iD - r_\phi F^R_{12} \big) \phi_i + F_i^\dag F_i  \\ &+ i \psi_i^\dag ( \gamma^\mu D_\mu -\sigma ) \psi_i - i \psi_i^\dag \lambda\phi_i + i \phi_i^\dag \lambda^\dag \psi_i  \\
\cL_{W} &= i\left (\parfrac{W}{\Phi_i} F_i - \frac12\, \parfrac{^2W}{\Phi_i \partial\Phi_j} \psi_j^{c\dag} \psi_i\pdag + \parfrac{\wb W}{\Phi_i^\dag} F_i^\dag - \frac12\, \parfrac{^2 \wb W}{\Phi_i^\dag \partial\Phi_j^\dag} \psi_j^\dag \psi_i^c \right )\, ,
\eea
where the superpotential $W(\phi_i)$ is a holomorphic function of R-charge two and  the fields $A_\mu,\sigma, D$ act on the matter fields in the appropriate representation. Here
the derivative $D_\mu$ is covariantized with respect to the spin and gauge connection and also to the R-symmetry background $A^R$.
As usual in Euclidean signature,  fields and their conjugate, $\phi$ and $\phi^\dagger$ for example, should be considered as independent variables. Notice that a vev for the scalar field $\sigma$ gives  mass  to the matter fields $\phi_i$ and $\psi_i$. This kind of coupling is typical of three dimensions and called  \emph{real mass} to distinguish it from the mass terms that can be introduced through the  superpotential $W$. One can  check that the Lagrangian is invariant under the following supersymmetry transformations
\bea
Q A_\mu &= \frac i2 \lambda^\dag \gamma_\mu \epsilon &
Q\lambda &= + \frac12 \gamma^{\mu\nu} \epsilon F_{\mu\nu} - D\epsilon + i \gamma^\mu \epsilon \, D_\mu\sigma \nonumber  \\
\wt Q A_\mu &= \frac i2 \tilde\epsilon^\dag \gamma_\mu \lambda &
\wt Q \lambda^\dag &= - \frac12 \tilde\epsilon^\dag \gamma^{\mu\nu} F_{\mu\nu} + \tilde\epsilon^\dag D + i \tilde\epsilon^\dag \gamma^\mu D_\mu\sigma  \nonumber\\
QD &= - \frac i2 D_\mu \lambda^\dag \gamma^\mu \epsilon + \frac i2 [\lambda^\dag \epsilon, \sigma ] \quad  &
Q \lambda^\dag &= 0 \qquad\quad
Q\sigma = - \frac12 \lambda^\dag \epsilon \nonumber \\
\wt Q D &= \frac i2 \tilde\epsilon^\dag \gamma^\mu D_\mu \lambda + \frac i2 [\sigma, \tilde\epsilon^\dag\lambda] &
\wt Q \lambda &= 0 \qquad\quad
\wt Q \sigma = - \frac12 \tilde\epsilon^\dag \lambda  \;.
\eea
for the vector multiplet fields and
\bea
Q\phi &= 0  & Q\psi &= \big( i\gamma^\mu D_\mu\phi + i \sigma\phi \big)\epsilon &  \wt Q\psi &= \tilde\epsilon^c F \nonumber \\
\wt Q\phi &= - \tilde\epsilon^\dag\psi \quad  & \wt Q\psi^\dag &= \tilde\epsilon^\dag \big( -i\gamma^\mu D_\mu \phi^\dag + i \phi^\dag \sigma \big) & Q\psi^\dag &= - \epsilon^{c\dag} F^\dag \\
Q\phi^\dag &= \psi^\dag\epsilon  & QF &= \epsilon^{c\dag} \big( i \gamma^\mu D_\mu \psi - i \sigma\psi - i \lambda\phi \big) & \wt QF &= 0 \\
\wt Q\phi^\dag &= 0  \quad & \wt Q F^\dag &= \big( -i D_\mu\psi^\dag \gamma^\mu - i \psi^\dag \sigma + i \phi^\dag \lambda^\dag \big) \tilde\epsilon^c \quad & QF^\dag &= 0 \;.
\eea
for the chiral multiplets. To future purposes, we also define   ${\cal Q}= Q+\wt Q$.

Notice that the Lagrangian \eqref{Lagr} and the transformations of supersymmetry are almost identical to the flat space ones with the further covariantization with respect to the metric and the background R-symmetry. This is not always the case in curved space, where extra terms should be included to maintain supersymmetry. In general, a Lagrangian is not invariant under flat space supersymmetry transformations when defined on a curved space because covariant derivatives do not commute anymore. With a topological twist, the spinor $\epsilon$   
is covariantly constant and the problem is milder. 

\subsection{The localization formula}\label{subsec:localization}

The \emph{topologically twisted index} is just the $\Sigma_\fg \times S^1$ path integral  of the  theory discussed in the previous section. We can evaluate it using 
localization. The  basic idea of localization  is simple. Let us review it briefly, referring to \cite{Pestun:2016jze,Marino:2011nm} for more details.\footnote{These references cover the developments after \cite{Pestun:2007rz}. The idea of localization in physics is much older  and it has been applied to many systems since  \cite{Witten:1982im}.}   In a theory with a fermionic symmetry squaring to zero (or to a bosonic symmetry of the theory)\footnote{In our case ${\cal Q}^2$ is a linear combination of a gauge transformation and a rotation along $S^1$.}  we can deform the action with a  ${\cal Q}$-exact term, ${\cal Q} V$, where $V$ is a fermionic functional invariant under all the symmetries. The new action is still ${\cal Q}$ invariant and the path integral 
independent of the deformation
\be  Z^{susy}(t) = \int e^{-S + t {\cal Q} V} \, ; \qquad \qquad \frac{d}{dt}Z^{susy} =  \int e^{-S + t {\cal Q} V} {\cal Q} V  =0 \, , \ee
since  ${\cal Q}$ acts as a total derivative. The path integral can be then computed for $t\rightarrow \infty$ and, if we choose $V$ cleverly and we are lucky, it reduces to a sum over saddle points of a classical contribution and a one-loop determinant,
\be Z^{susy}(t=\infty) =\sum_{{\rm saddle\, points}} e^{-S_{{\rm classical}}} \frac{\det {\rm fermions}}{\det {\rm bosons}} \, .\ee
In many supersymmetric gauge theories on Euclidean manifolds, this approach successfully reduces the path integral to the evaluation of a matrix model. 

In our theory,  $\cL_\text{YM}, \cL_\text{mat}$ and  $\cL_\text{W}$ are not only ${\cal Q}$-closed but also ${\cal Q}$-exact. For example, up to total derivatives,
\be
\tilde\epsilon^\dag\epsilon \, \cL_\text{YM} = {\cal Q}  \wt Q \Tr \Big( \tfrac12 \lambda^\dag\lambda + 2D\sigma \Big) \, .\ee
Therefore we can put arbitrary coefficients in front of the various ${\cal Q}$-exact terms in the Lagrangian
\be 
\cL=\frac{1}{g^2}\cL_\text{YM} + \cL_{\text{CS}} + \frac{1}{\lambda^2} \cL_\text{mat}+ \frac{1}{\eta^2} \cL_\text{W}\, ,\ee
and the path integral is independent of $g,\lambda,\eta$. We can then take the limit $g,\lambda,\eta\rightarrow 0$ and evaluate
the path integral in the saddle point approximation. Notice in particular that    $\cL_\text{W}$  is ${\cal Q}$-exact. This means that the partition function is independent of the precise form of the interactions in the Lagrangian. The superpotential is important nevertheless for determining the global symmetries of the theory, which enter in the partition function through chemical potentials.

We now give the localization formula for the topologically twisted index. Since the computation is complicated and subtle, we just provide the final formula,  referring to \cite{Pestun:2016jze,Marino:2011nm} for a general  introduction to localization and to \cite{Benini:2015noa,Benini:2016hjo} for the details of this particular computation. 

The path integral of the topologically twisted theory on $\Sigma_\fg\times S^1$ for a ${\cal N}=2$ supersymmetric gauge theory with gauge group $G$ can be written as a contour integral 
\be\label{locformula}  Z_{\Sigma_\fg\times S^1} =  \frac{1}{|W|} \sum_{\fm\in \Gamma} \oint_C Z_{int} (u, \fm) \, , \ee
of a meromorphic form of Cartan-valued variables $u$, summed over a lattice $\Gamma$ of magnetic fluxes. $W$ is just the order of the Weyl group of $G$. We will explain all the  ingredients in the following, referring to \cite{Benini:2015noa,Benini:2016hjo} for proofs. Notice that from now on we drop the superscript \emph{susy} from the partition functions. 

\subsubsection{The BPS locus}\label{subsubsec:BPS}

We have a family of saddle points labeled by the vev of the scalar field $\sigma$, the value of the Wilson line $A_t$ along $S^1$ and 
a quantized magnetic flux $\fm$ along $\Sigma_\fg$. As standard in localization computation, these saddle points can be found as the locus where the fermionic variations vanish.  The  gaugino BPS equations read
\be\label{BPSgaugino} Q\lambda = \left ( \frac12 \gamma^{\mu\nu}  F_{\mu\nu} - D \right ) \epsilon + i \gamma^\mu \epsilon \, D_\mu\sigma  = 0\, , \ee
and are solved by setting the two terms on the right-hand side to zero. The second term in \eqref{BPSgaugino},  $D_\mu\sigma$, vanishes for constant commuting adjoint fields $\sigma$ and $A_t$.  With a gauge transformation, we can  transform them  in the Cartan subalgebra. We can combine these fields in a complex Cartan-valued
quantity \be u=A_t+ i \beta\sigma\, . \ee The Wilson line  $A_t$ is periodic, invariant under a shift of any  element $\chi$ of the co-root lattice $\Gamma$, $A_t \sim A_t + 2 \pi \chi$,\footnote{The co-root lattice is defined by the requirement that $e^{2\pi i \chi}$ acts as the identity on any representation of the group $G$, and  defines  the weight lattice of the Langland (or S-dual) group $\check G$.} the physical object being the holonomy $e^{i A_t}$. So $u$ naturally lives on a cylinder and it is convenient to define the quantity  $x=e^{i u}$. For a $U(1)$ theory, $u$ lives on the cylinder $S^1\times \mathbb{R}$ and $x$ on the punctured plane $\mathbb{C}^*$. 
The first term in \eqref{BPSgaugino}, using $\gamma_{12}\epsilon=i \epsilon$, implies that the auxiliary field $D$ is proportional to the gauge field strength along $\Sigma_\fg$, $D= i F_{12}$, and both live in the Cartan subalgebra. The curvature along $\Sigma_\fg$ is quantized
\be \frac{1}{2\pi} \int_{\Sigma_\fg} F =  \fm \in \Gamma \, \ee
where $\Gamma$ is again the co-root lattice. For a $U(1)$ theory $\fm$ is just an integer. 

The path integral involves a sum over saddle points and is therefore given as an integral over $u$ and a sum over the magnetic fluxes $\fm$. Both variables live in the Cartan subalgebra and are only defined up to an action of the Weyl group, the surviving gauge symmetry.
This explains the factor $1/|W|$ in \eqref{locformula}. For a $U(N)$ theory the co-root lattice is just $\Gamma=\mathbb{Z}^N$ and the Weyl group is the permutation group of $N$ elements with $|W|=N!$

\subsubsection{The integrand}\label{subsubsec:integrand}

The  contribution to the saddle point of the classical action  comes only from the Chern--Simons term\footnote{This expression follows   from a holomorphic recombination of the terms $A_t\wedge F_{12}$ and $\sigma D$ in the Chern--Simons action, using $D=i F_{12}$ and $u=A_t+i \beta \sigma$.}
\be
\label{classical}
Z_\text{class}^\text{CS}(u) = x^{k \fm} \equiv \prod_{i=1}^r x_i^{k \fm_i}
\ee
where $x=e^{i u}$. 

The one-loop determinant receives contributions  from  the  vector multiplets and the chiral multiplets.  The vector multiplet contribution is 
\be
\label{vector1-loop}
Z_\text{1-loop}^\text{gauge} (u)=   \prod_{\alpha \in G} (1-x^\alpha)^{1-\fg} \; (i\, du)^r
\ee
where $\alpha$ are the roots of $G$ and, for convenience, we included the integration measure $(du)^r$ in this expression. The chiral multiplet contribution is
\be
\label{chiral1-loop}
Z_\text{1-loop}^\text{chiral} (u,\fm) = \prod_{\rho \in \fR} \Big( \frac{x^{\rho/2}}{1-x^\rho} \Big)^{\rho(\fm) +(\fg-1)(r_\phi -1)}
\ee
where $\fR$ is the representation under the  gauge group $G$, $\rho$ are the corresponding weights and $r_\phi$ is the R-charge of the field. These expressions arise by taking ratios of determinants for fermionic and bosonic fields, computed by expanding in modes on $\Sigma_\fg \times S^1$. Due to supersymmetry, most of the modes cancel between bosons and fermions and we are left with the contribution of a set of zero-modes (indeed a convenient way to perform this computation is via an index theorem). For a chiral multiplet these zero modes contribute
\be  \prod_{\rho \in \fR}\prod_{n=-\infty}^\infty  \left ( \frac{2 \pi i}{\beta} n + i \rho(\frac{A_t}{\beta}+ i  \sigma) \right)^{-(\rho(\fm) +(\fg-1)(r_\phi -1))} \, .\ee
The term in bracket represents the mass of a chiral multiplet  mode due to the coupling to $\sigma$, which acts as a real mass, to the Wilson line $A_t$ and to the KK momentum $n$ along the circle $S^1$. The exponent is the multiplicity of the zero-mode that can be easily obtained using the Riemann-Roch theorem. Notice that this multiplicity must be an integer and therefore the R-charges  $r_\phi$ must be quantized.\footnote{On $S^2\times S^1$ the R-charges $r_\phi$ must be integer. In the case of a higher genus Riemann surface it is enough to require  that the quantity $(1-\fg)(r_\phi -1)$ is an integer.}   This infinite product needs to be regularized. In  \eqref{chiral1-loop} we chose a parity invariant regularization. There are other possible ones.\footnote{Parity acts as $u\rightarrow -u$ and   \eqref{chiral1-loop} is obviously invariant. A gauge invariant regularization breaking parity is used in \cite{Closset:2016arn,Closset:2017zgf,Closset:2017bse,Closset:2018ghr}. The latter has the advantage of clarifying subtle sign issues   and simplifying the mapping of parameters between dual theories. However, even for theories with zero CS, in the gauge invariant regularization one has to introduce extra effective CS contact terms, which makes the physical interpretation less transparent.}  

The full integrand is
\be\label{integrand}  Z_{int}(u,\fm) = Z_{pert}(u,\fm) \left ( \det_{ab} \frac{\partial^2 \log Z_{pert}(u,\fm)}{\partial i u_a \partial\fm_b} \right)^g\, ,
\ee
where
\be Z_{pert}(u,\fm) = Z_\text{class}^\text{CS}(u,\fm) Z_\text{1-loop}^\text{gauge}(u) Z_\text{1-loop}^\text{chiral}(u,\fm) \, .\ee
The determinant term exists only on a Riemann surface of genus $\fg>0$ and arise from the integration of the extra $\fg$ fermionic zero-modes existing on these surfaces.

\subsubsection{The contour}\label{subsubsec:JK}

The integrand \eqref{integrand} is a meromorphic form in the Cartan variables $u$ with poles at $x^\rho=1$, the points in the BPS locus where chiral multiplets become massless, and  at the boundaries $x=0$ and $x=\infty$ of the moduli space. The partition function is  obtained by using the residue theorem. Supersymmetry will choose the correct integration contour and tell us which poles to include. One might hope that we need to integrate over some simple contour, like the unit circle in the plane $x$, but one actually discovers that the contour is highly  non-trivial and depends on the charges of the matter fields. For example, for a $U(1)$ theory with chiral fields of charge $Q_i$ and Chern--Simons level $k$, defining the effective Chern--Simons level\footnote{This is actually the  Chern--Simons level that one sees at one-loop after integrating out the matter fields (they  have mass $\sigma$ at a generic point of the BPS locus).}
\be k_{\rm eff}(\sigma) = k +\frac12 \sum_i Q_i^2 \sign (Q_i \sigma) \, ,\ee
the rule is to take the residues of the poles created by fields with positive charge $Q_i>0$, the residue at the origin $x=0$ if $k_{\rm eff}(\infty)<0$ and the residue at infinity $x=\infty$ if $k_{\rm eff}(-\infty)>0$. The  rule for a generic gauge group can be written in terms  of the so-called Jeffrey--Kirwan (JK) residue \citep{JeffreyKirwan}, a prescription for dealing with poles arising from multiple intersecting hyperplane singularities. To explain it properly will lead us too far  and we refer to \cite{Benini:2015noa,Benini:2016hjo} for details. The JK residue  also appears in localization computations for elliptic genera in two dimensions, quantum mechanics, and various other partition functions   \citep{Benini:2013nda,Hori:2014tda,Closset:2015rna}. 

The reader may object  that we are supposed to integrate over the  BPS locus, which is  the whole complex plane, and not to perform a contour integral in $u$.  Luckily again supersymmetry comes to a rescue. On $\Sigma_\fg\times S^1$ there are gaugino zero-modes that contribute an extra term in the integrand in addition to the one-loop determinant. It turns out that the full integrand is a total derivative in $\bar u$ and we can reduce the integral over the $u$ plane to a contour integral around the singularities, as discussed in details in \cite{Benini:2015noa}.

\subsubsection{Adding flavor fugacities}\label{subsubsec:flavor} 

If the theory has a flavor symmetry group $F$ acting on the chiral fields, we can introduce extra parameters in a supersymmetric way. We can just gauge the flavor symmetry and then freeze all the bosonic fields to background values that are preserved by supersymmetry. The background bosonic  fields give rise to supersymmetric couplings in the Lagrangian. The analysis of fermionic variations is identical to the one performed 
in Sect.~\ref{subsubsec:BPS} for gauge symmetries.  We need to solve \eqref{BPSgaugino} for a background multiplet, $(A^F_\mu,\lambda^F,\sigma^F,D^F)$. The result is that  we can turn on in a supersymmetric way
a constant value for $\sigma^F$ and $A_t^F$ which we combine into a complex quantity $u_F = A_t^F+i \beta \sigma^F$, and a background magnetic
flux $\fm_F$ with $D^F=i \fm_F$. $\sigma^F$ appears in the Lagrangian as a real mass for the chiral fields.  
In three dimensions, any  gauge theory  with a $U(1)$ factor with field strength $F$ has also a \emph{topological symmetry} associated with the current $J= * F$, which is automatically conserved. We can similarly introduce parameters $u_T$ and $\fm_T$ for the topological symmetry. 

The path integral is then  a function of $x_F,x_T$ and $\fm_F,\fm_T$. In the localization formula we just need to replace the one-loop determinant of a chiral field with
\be
\label{chiral1-loop2}
Z_\text{1-loop}^\text{chiral} (u,\fm; u_F,\fm_F) = \prod_{\rho \in \fR} \Big( \frac{x^{\rho/2} x_F^{\nu/2}}{1-x^\rho x_F^\nu} \Big)^{\rho(\fm) +\nu(\fm^F)+(\fg-1)(r_\phi -1)}
\ee
where $x_F=e^{i u_F}$ and $\nu$ is the weight of the chiral field under the flavor symmetry $F$. There is no modification to the vector multiplet determinant. A $U(1)$ topological symmetry just contributes a classical term 
\be x^{\fm_T} x_T^\fm \ee 
to the classical action. 

\subsubsection{The trace interpretation}\label{subsubsec:trace}

As any path integral that involves an $S^1$ factor, the topologically twisted index can be written as a trace\footnote{See footnote \ref{foot1}. The factor $e^{i A_t^G J^G}$ represents the insertion of a Wilson line $A_t^G$.} 
\be\label{trace00} Z_{\Sigma_\fg \times S^1} (x_G,\fm_G)= \Tr (-1)^F e^{i A_t^G J^G} e^{-\beta H_\fg} \, \ee
where $H_\fg$ is the Hamiltonian of the topologically twisted theory on $\Sigma_\fg$, in the presence of magnetic fluxes $\fm_G=(\fm_F,\fm_T)$ and a supersymmetric background $x_G=(x_F,x_T)$ for the global symmetries, whose conserved charges have been denoted as $J^G$.   The Hamiltonian $H_\fg$ explicitly depends on the magnetic fluxes $\fm_G$ and the real masses $\sigma^G$. If sufficiently real masses are turned on, the spectrum of $H_\fg$ is discrete and the trace is well-defined.

\subsubsection{An Example: SQED}\label{subsubsec:SQED} 
 
To explain all the ingredients,  we can give a simple example of the final formula, using supersymmetric QED. This is  a $U(1)$ theory with two chiral multiplets $Q$ and $\tilde Q$ of charges $\pm1$ (electron and positron), and no Chern--Simons couplings. Since there is no superpotential, we have many possible choices of integer R-charges for the fields. We choose to assign R-charge $+1$ to both $Q$ and $\tilde Q$. There is an axial flavor symmetry $U(1)_A$ acting on $Q$ and $\tilde Q$ with equal charges and a topological symmetry $U(1)_T$.  The charges of 
the chiral fields are 
 \be
\label{table charges SQED}
\begin{array}{c|ccc|c}
 & U(1)_g & U(1)_T & U(1)_A & U(1)_R \\
\hline
Q & 1 & 0 & 1 & 1 \\
\tilde Q & -1 & 0 & 1 & 1 \\
\end{array}
\ee
The topological symmetry acts only on non-perturbative states constructed with  monopole operators.
We introduce a gauge variable $x$, with associated magnetic flux $\fm$,  and  flavor and topological variables $y=x_F$ and $\xi=x_T$,
with associated background fluxes $\fn=\fm_F$ and $\ft=\fm_T$.  According to our rules, the partition function on $S^2\times S^1$ is
\be
Z(y,\xi,\fn,\ft) = \sum_{\fm\in\bZ} \int \frac{dx}{2\pi i\, x} x^\ft (-\xi)^\fm \Big( \frac{x^\frac12 y^\frac12}{1-xy} \Big)^{\fm+\fn} \Big( \frac{x^{-\frac12} y^\frac12}{1-x^{-1}y} \Big)^{-\fm+\fn} \;
\ee
where we included an extra $(-1)^\fm$, which can be reabsorbed in the definition of $\xi$, for later convenience.\footnote{For a more careful discussion of sign ambiguities see \cite{Closset:2017zgf}. They will not play any important role in these lectures.} 
Notice that gauge and flavor variables enter in a similar way in this formula. The main difference is that $x$ and $\fm$ are integrated and summed over, while $y,\xi$ and $\fn,\ft$ are background parameters.

Our prescription  instructs us to take the residues from the field $Q$ with positive gauge charge, whose pole is at $x = \frac1y$. By computing residues and resumming the result, one finds 
\be\label{SQEDfin}
Z (y,\xi,\fn,\ft) = \Big( \frac y{1-y^2} \Big)^{2\fn-1} \Big( \frac{\xi^\frac12 y^{-\frac12}}{ 1-\xi y^{-1} } \Big)^{\ft - \fn+1} \Big( \frac{\xi^{-\frac12} y^{-\frac12} }{1-\xi^{-1} y^{-1} } \Big)^{-\ft-\fn+1} \;.
\ee
One recognizes here the product of three factors of the form \eqref{chiral1-loop2} that we can associate with  chiral multiplets. Indeed it is well known 
that the mirror theory to SQED is a Wess-Zumino model with fields $M,T,\tilde T$ and  a cubic superpotential $W = M T\tilde T$ \citep{Aharony:1997bx}.

\subsection{Interpretation of the localization formula}\label{subsec:interpretation}

We can give an interpretation of the localization formula for the theory on $\Sigma_\fg\times S^1$ in two different ways that correspond
to two different dimensional reductions of the three-dimensional theory. Compactification on $\Sigma_\fg$ gives rise to a quantum mechanics and compactification on  $S^1$ to  a two-dimensional $(2,2)$ supersymmetric theory.

\subsubsection{Reduction to quantum mechanics}\label{subsec:QM}

Compactifying on $\Sigma_\fg$, we obtain a supersymmetric quantum mechanics describing an infinite number of KK modes on $\Sigma_\fg$. These are particles living on the Riemann surface in the presence of a magnetic field for the R-symmetry and magnetic fluxes $\fm_G$ for the global symmetries. These magnetic fields create Landau levels.  The trace \eqref{trace00} can be interpreted as the Witten index \citep{Witten:1982df} of this particular quantum mechanics. Let us understand this concept better.

The quantum mechanics in question has ${\cal N}=2$ supersymmetry. With no background for the global symmetries, the algebra of supersymmetry is simply $\{ \bar Q, Q\}=H_\fg$, where $Q$ is a complex supercharge.  The index is just
\be \Tr (-1)^F e^{-\beta H_\fg} \, ,\ee
and, according to standard arguments, is independent of $\beta$. Indeed, any state $\psi$ with $H_{\fg}\ne 0$ has a non-zero partner $Q\psi$ with the same energy and opposite statistic  and their contributions cancel in the trace. Therefore the only contribution  comes from ground states.\footnote{By the algebra of supersymmetry $H_\fg \psi=0$ is equivalent  to $Q\psi=0$ and, therefore, ground states not necessarily  have a partner.} The index  is then clearly independent of $\beta$  and it is an integer counting the number of ground states with signs (plus for bosonic ones, minus for fermionic ones). When we turn on backgrounds for the global symmetries, the supersymmetry algebra is modified to $\{ \bar Q, Q\}=H_\fg -\sigma^G J^G$, where $J^G$ is the conserved charge associated to the global symmetry.\footnote{See, for example, \cite{Hori:2014tda} or Appendix~C of \cite{Benini:2015eyy}.} Using \eqref{trace00}, we can now write the index as follows
\be \Tr (-1)^F e^{i A_t^G J^G} e^{-\beta H_\fg} = \Tr (-1)^F e^{i (A_t^G+ i \beta \sigma^G) J^G} e^{-\beta \{ \bar Q, Q\}} =\sum_n g(n) x_G^n\, ,\ee
where $g(n)$ is the number of supersymmetric states,  $Q\psi=0$, with charge $n$ under the global symmetry and, as usual,  $x_G= e^{i (A_t^G+ i \beta \sigma^G)}$. In deriving this expression, we used again the fact that states with $Q\psi\ne 0$ are paired by supersymmetry and have the same energy and charge. This time, the states that contribute to the trace are chiral states with $H_\fg=\sigma_G J^G$. In this way, we have obtained an equivariant index, where the supersymmetric states are graded according to their charge by powers of the fugacity $x_G$. Notice also that this argument shows that the topologically twisted index is  an holomorphic function of the fugacities,  as we already found using localization.

Let us also notice that the integrand of the localization formula has a simple Hamiltonian interpretation. There are two type of multiplets
in the ${\cal N}=2$ quantum mechanics we are discussing: the chiral multiplet containing a complex scalar $\phi$ and a spinor as dynamical fields, and
the Fermi multiplet containing only a spinor \citep{Hori:2014tda}. The Landau levels on $\Sigma_\fg$ give rise to zero-modes with multiplicities dictated by the Riemann--Roch theorem. One can see that the zero-modes organize themselves into $\rho(\fm) +(\fg-1)(r_\phi -1)$ chiral multiplets if   $\rho(\fm) +(\fg-1)(r_\phi -1)>0$, and $|\rho(\fm) +(\fg-1)(r_\phi -1)|$ Fermi multiplets if $\rho(\fm) +(\fg-1)(r_\phi -1)<0$, where for simplicity we set the flavor fugacities to zero. We can now compute the index for Fermi and chiral multiplets. For the Fermi multiplet the Hilbert space is a fermionic Fock space, and assigning charge $-\frac\rho2$ and fermion number $0$ to the vacuum, the index is 
\be \frac{1 - x^\rho}{x^{\frac\rho2}}\, .\ee
For the chiral multiplet  the Hilbert space is the product of a bosonic Fock space generated by $\phi,\phi^\dag$ and a fermionic Fock space; assigning fermion number $1$ to the vacuum, the index is 
\be (-x^{-\frac\rho2} + x^\frac\rho2) \sum_{n\geq 0} x^{n\rho} \sum_{m \geq 0} x^{-m\rho} = \frac{x^{\frac\rho2}}{1-x^\rho}\, .\ee
Raising these quantities to a power corresponding to the multiplicity, and taking into account the different signs for the two types of multiplets, we recover exactly the contribution \eqref{chiral1-loop} of a three-dimensional chiral multiplet to the partition function.

In particular, the localization formula  for the topologically twisted index is just the sum over many topological sectors labelled by $\fm$ of the localization formula for the Witten index of ${\cal N}=2$ quantum mechanics   found  in \cite{Hori:2014tda}.

\subsubsection{Reduction to two dimensions}\label{subsec:BE}

We can alternatively reduce our three-dimensional theory on $S^1$ and obtain a $(2,2)$ supersymmetric theory containing all the KK modes on $S^1$. At a generic point of the Coulomb branch where $\sigma\ne 0$, all the non-Cartan gauge bosons and the chiral multiplets are massive.\footnote{Due to the KK mass or their coupling to $\sigma$.} We can integrate them out  and write a Lagrangian for the Cartan modes of the vector multiplets. In two dimensions, a vector multiplet can be described using a twisted chiral multiplet  $\Sigma$ and its interactions is described by a twisted superpotential $\int d\theta_+ d\bar\theta_-  \cW$.\footnote{$\Sigma$ has a scalar  as its lowest component and it  satisfies $\bar D_+\Sigma=  D_- \Sigma =0$. See \cite{Witten:1993yc}.} 
 
It is interesting to observe that such twisted superpotential  $\cW$ enters explicitly in the integrand of the localization formula \citep{Closset:2016arn,Closset:2017zgf}. Indeed, 
the dependence on the gauge flux $\fm$ can be explicitly written as
\be\label{res} \sum_{\fm \in \Gamma} \int \frac{dx_i}{2\pi i x_i}  \, Q(x)\,  e^{i \fm_i \frac{\partial  \cW}{\partial u_i} }  \, ,\ee
where $Q(x)$ is a meromorphic function independent of $\fm$. The function $\cW$, up to an overall normalization and sign ambiguities that we fix for convenience,  is given by\footnote{\label{polylog}Polylogarithms $\Li_s (z)$ are defined by  ${\rm Li}_s (z) = \sum_{n = 1}^{\infty} \frac{z^n}{n^s}$ for $|z| < 1$ and by analytic continuation outside the disk. Notice, in particular, that
${\rm Li}_1(z)= -\log (1-z)$. For $s\ge 1$, there exists a the branch cut that we take along  $[1 , + \infty)$. They satisfy $\partial_u {\rm Li}_s (e^{iu}) =i \, {\rm Li}_{s-1} (e^{iu})$ and,  for $0< \re u < 2\pi$, ${\rm Li}_s (e^{i u}) + (-1)^s {\rm Li}_s (e^{- i u}) = - \frac{(2 i \pi)^s}{s!} B_s \left( \frac{u}{2 \pi} \right) \equiv i^{s - 2} g_s (u)$, where $B_s (u)$ are the Bernoulli polynomials. In this paper we need, in particular,
$g_2 ( u ) = \frac{u^2}{2} - \pi u + \frac{\pi^2}{3}$ and $g_3 ( u ) = \frac{u^3}{6} - \frac{\pi}{2} u^2 + \frac{\pi^2}{3} u$. One then sees that, for  $0< \re u < 2\pi$ and $\im  u\ll 0$,  ${\rm Li}_s (e^{i u}) \sim i^{s - 2} g_s (u)$.  Notice also that  $\cW$ is a multi-valued function   but, since  the action is defined up to integer multiples of $2\pi i$,  the path integral and all physical  observables are single valued.}
\be\label{twisted} \cW(u) =  \frac{k}{2} \sum_i u_i^2 + \sum_{\fR} \left ( \frac{1}{2} g_2(\rho(u)) - {\rm Li}_2(e^{i \rho(u)}) \right ) \, ,\ee
with $g_2 ( u ) = \frac{u^2}{2} - \pi u + \frac{\pi^2}{3}$. As argued in \cite{Nekrasov:2009uh, Nekrasov:2014xaa}, this can be interpreted  as the effective twisted superpotential  of the two-dimensional theory obtained by compactifying on $S^1$. The first term in \eqref{twisted} is the classical contribution coming from the  CS term, and the second  is the sum of all the perturbative contributions of massive fields, including the infinite tower of KK modes. Indeed, a one-loop diagram for a mode  of mass $m$ contributes  a term proportional to $ i (\Sigma+m) (\log (\Sigma+m)/2\pi -1)$ to $\cW$  and there are no higher order corrections \citep{Witten:1993yc}. 
The contribution of the KK modes of a chiral multiplet, whose mass depends on  $\sigma$,  the Wilson line $A_t$ and  the KK momentum $n$,  can be resummed to
\be i \sum_{n\in \mathbb{Z}} (u+2 \pi n) \left ( \log \frac{u+2 \pi n}{2 \pi} -1\right ) = -{\rm Li}_2(e^{i \rho(u)}) \, . \ee
The other term in the round bracket in \eqref{twisted} is local and it is due to our choice of a parity invariant regularization.\footnote{For a choice of a gauge invariant regularization and an extensive discussion of other issues related to definition of $\cW$ see \cite{Closset:2017zgf}. } Using the asymptotic expansion of the polylogarithms, we find that the content of the bracket in \eqref{twisted} behaves, for large $\sigma$,  as
\be \frac{\rho(u)^2}{4} \sign(\rho(\sigma)) \, . \ee
This can be  interpreted as a one-loop effective Chern--Simons term obtained by integrating out a  field of mass $\rho(\sigma)$.\footnote{For a field of mass $m$ and charge $Q_i$ the one-loop effective Chern--Simons term is $k_{\rm eff}=k+ \frac12 Q_i^2 \sign(m)$.}

The Jeffrey--Kirwan prescription  typically selects poles in the integrand of the localization formula that are  contained in a half-lattice $\fm_i \ge M$ (or $\fm_i \le M$) for some cut-off $M$.  We can then use the geometric series to resum the integrand in \eqref{res}
\be   \int \frac{dx_i}{2\pi i x_i} \frac{Q(x) e^{i M \frac{\partial  \cW}{\partial u_i}}}{\prod_i \Big(1- e^{i \frac{\partial  \cW}{\partial u_i}}\Big)} \, , \ee
and evaluate the index by taking the residues at the poles
\be\label{BE} \exp \left ( i \frac{\partial  \cW}{\partial u_i} \right ) =1 \, .\ee
The cut-off $M$ disappears in the process.
The solutions to \eqref{BE} are  the so-called \emph{Bethe vacua} of the two-dimensional theory. They  play an important role in the  Bethe/gauge correspondence \citep{Nekrasov:2009uh,Nekrasov:2009rc}.

We then find the following general characterization of the topologically twisted index as a sum over Bethe vacua
\be\label{sumBethevacua} 
Z_{\Sigma_g\times S^1} = \sum_{x^*} \frac{Q(x^*)}{\det_{ij} (-\partial^2_{u_i u_j} \cW(x^*))} \, ,  \ee
where $x^*$ are the solutions of \eqref{BE}. The expression for the topologically twisted index as a sum over Bethe vacua was first derived  by  topological field theory arguments in \cite{Okuda:2012nx,Okuda:2013fea,Nekrasov:2014xaa,Okuda:2015yea}. In the context of localization, this expression for the index has been derived and generalized  in \cite{Closset:2016arn,Closset:2017zgf,Closset:2017bse,Closset:2018ghr}.  The expression in \eqref{sumBethevacua} can be also written  as
\be\label{sumBethevacua2} 
Z_{\Sigma_g\times S^1} = \sum_{x^*}  {\cal H} (x^*)^{\fg-1}\, ,  \ee
in terms of a \emph{handle-guing operator} ${\cal H} (x) = e^{\Omega (x)} \det_{ij} \partial^2_{u_i u_j} \cW(x)$ and an \emph{effective dilaton} $\Omega(x)$
whose complete characterization  in terms of field theory data can be found in the above mentioned papers. 
Here we just notice that, for genus $\fg>0$,  the Hessian of $\cW$  enters at the power $\fg-1$. Indeed, the determinant  in \eqref{integrand} contributes $\fg$ extra powers of the Hessian  that combine with
the denominator in \eqref{sumBethevacua}. 

A very interesting result of \cite{Closset:2017zgf,Closset:2017bse,Closset:2018ghr} is the generalization of 
formula \eqref{sumBethevacua2} to three-dimensional manifolds that are not a direct product. For example, the supersymmetric partition function on a three-manifold $\cM_3$ that is an $S^1$ fibration of Chern class $p$ over a Riemann surface $\Sigma_{\fg}$  can be written as a sum over the very same set of Bethe vacua, 
\be\label{sumBethevacua3} 
Z_{\cM_3} = \sum_{x^*}  \cF(x^*)^p  {\cal H} (x^*)^{\fg-1}\, ,  \ee
with a \emph{fibering operator} $\cF(x)$ that can be expressed in terms of field theory data. There exists a similar result for the partition function on more general three-dimensional manifolds and also
for some selected four-dimensional ones  \citep{Closset:2017zgf,Closset:2017bse,Closset:2018ghr}. The particular case of the formula for the four-dimensional superconformal index plays a role
in the physics of AdS$_5$ black holes \citep{Benini:2018mlo,Benini:2018ywd}, as discussed in Sect.~\ref{sec:interpretation}. 

Let us give a couple of examples of Bethe vacua. For a pure ${\cal N}=2$ Chern--Simons theory with gauge group $SU(2)$, the expression for the partition function \eqref{integrand} is
\be
Z = \frac{(-1)^{\fg-1}}2 \sum_{\fm \in \bZ} \int_\text{JK} \frac{dx}{2\pi i x} \, (2k)^\fg x^{2k\fm} \bigg[ \frac{(1-x^2)^2}{x^2} \bigg]^{1-\fg} \;,
\ee
where we used $x_i=(x,1/x)$ and $\fm_i=(\fm,-\fm)$. The twisted superpotential receives contribution only from the classical action: $\cW=\sum_i k u_i^2/2= k u^2$. The Bethe vacua  \eqref{BE} are then $x^{2k}=1$ with solutions the $2 k$-roots of unity.  Formula \eqref{sumBethevacua}  gives, up to an ambiguous sign,
\be
Z = 
\Big( \frac{\bar k + 2}2 \Big)^{\fg-1} \sum_{j=1}^{\bar k + 1} \Big( \sin \frac{\pi i}{\bar k+2} j \Big)^{2-2\fg} \;,
\ee
where $\bar k =k-2$. This is the well-known Verlinde formula for the CS partition function on $\Sigma_\fg\times S^1$.\footnote{Since $\sigma$ and $\lambda$ are massive and free, they can be integrated out leading to a shift in the CS coupling. An ${\cal N}=2$ Chern--Simons theory is thus equivalent to a bosonic CS theory with level $\bar k =k-2$.} Notice that the root $x=1$ is not  included in the sum: as a general rule, the Bethe vacua that are also  zeros of the Vandermonde determinant are not physical.  

In the presence of matter, the Bethe equations \eqref{BE} are more complicated. In the SQED example discussed in Sect.~\ref{subsubsec:SQED}, the Bethe equation is  
\be \frac{\xi (y-x)}{1- xy}=1 \, ,\ee
with solution $x= (1- \xi y)/(y-\xi)$. It is easy to see that \eqref{sumBethevacua} correctly reproduces \eqref{SQEDfin}.
For a general theory with gauge group $G$, the Bethe equations \eqref{BE} cannot be analytically solved.

\section{The entropy of dyonic AdS$_4$ black holes}\label{sec:lecture3}

In this section we will derive microscopically the entropy of a family of BPS static dyonic black holes   in AdS$_4\times S^7$ \citep{Benini:2015eyy,Benini:2016rke}. 
These solutions have been found  in ${\cal N}=2$ gauged supergravity in four-dimensions  \citep{Cacciatori:2009iz,DallAgata:2010ejj,Hristov:2010ri,Katmadas:2014faa,Halmagyi:2014qza} and later uplifted to $M$-theory. We then start by briefly discussing the main features of ${\cal N}=2$ gauged supergravity in four dimensions. This will be also useful to write an entropy functional.  We then consider the large $N$ limit of the topologically twisted index for the ABJM theory in three dimensions and show that it reproduces the entropy of the dual black holes.

We will only consider the case of static black holes in this section. A field theory derivation for the entropy of dyonic rotating black holes in AdS$_4\times S^7$ \citep{Hristov:2018spe} is still missing.

\subsection{AdS$_4$ dyonic static black holes}\label{subsec:BH}
BPS  black holes in AdS$_4$ can be found by studying an effective four-dimensional  ${\cal N}=2$ gauged supergravity. We start discussing the main features of the effective theory. 

The ${\cal N}=2$  supergravity  multiplets are:

\begin{itemize}
\item the graviton multiplet, whose bosonic components are the metric $g_{\mu\nu}$  and a vector field $A_\mu^0$, called graviphoton;
\item the vector multiplet,   whose bosonic components are a vector $A_{\mu}^i$  and a complex scalar $z$;
\item the hypermultiplet, whose bosonic components are four real scalars $q^\alpha$.

\end{itemize}

For simplicity, we will restrict to ${\cal N}=2$  gauged supergravities with $n_V$ vector multiplets and no hypermultiplets.\footnote{Theory with hypermultiplets have been considered for matching the entropy of black holes in massive type IIA and other models \citep{Hosseini:2017fjo,Benini:2017oxt,Bobev:2018uxk}.} This will be enough to describe the black holes in AdS$_4\times S^7$. The theory contains  $n_V+1$ vector multiplets $A_\mu^\Lambda$ and $n_V$ complex scalar fields $z_i$, where  $\Lambda=0,1,\ldots, n_V$ and $i=1,\ldots , n_V$. The Lagrangian can be written
in terms of  a holomorphic prepotential $\cF(X^\Lambda)$, which is a homogeneous function of degree two,  and a vector of magnetic and electric Fayet--Iliopoulos (FI) parameters $(g^\Lambda, g_\Lambda)$.  $X^\Lambda(z_i)$ are a set of $n_V+1$ homogeneous coordinates on the scalar manifold. The theory is invariant under rescaling of the $X^\Lambda$ and one can identify the physical scalar fields with $z_i=X^i/X^0$.\footnote{Other choices of gauge fixing for the rescaling symmetry are possible,  corresponding to  field redefinitions.}  It is also convenient to define $\cF_\Lambda \equiv \partial_\Lambda \cF$. The theory is fully covariant under a $Sp(2 n_V+2)$ group of electric/magnetic dualities acting on $(X^\Lambda,{\cal F}_\Lambda)$ and $(g^\Lambda, g_\Lambda)$ as symplectic vectors.

The action of the bosonic part of the theory reads \citep{Andrianopoli:1996cm}
\bea
 S^{(4)} =\frac{1}{8 \pi G_{{\text N}}^{(4)} } \int_{\bR^{3,1}} & \bigg[ \frac{1}{2} R^{(4)} \star_4 1 + \frac12 \im \cN_{\Lambda \Sigma} F^\Lambda \wedge \star_4 F^{\Sigma}
 + \frac12 \re \cN_{\Lambda \Sigma} F^{\Lambda} \wedge F^{\Sigma} \\
 & - g_{i \bar{j}} D z^{i} \wedge \star_4 D \bar{z}^{\bar{j}}
 - V(z, \bar{z}) \star_4 1 \bigg] \, . \nonumber
\eea
The  metric on the scalar manifold is given by
\bea
 \label{Kahler:metric:Kahler}
 g_{i \bar{j}} = \partial_i \partial_{\bar{j}} \cK(z , \bar{z}) \, .
\eea
Here, $\cK(z , \bar{z})$ is the K\"ahler potential and it reads
\bea
 \label{4d:Kahler:prepotential}
 e^{- \cK (z , \bar{z})} = i \left( \bar{X}^\Lambda \cF_\Lambda - X^\Lambda \bar{\cF}_\Lambda \right) \, .
\eea
The matrix $\cN_{\Lambda \Sigma}$ of the gauge kinetic term is a
function of the vector multiplet scalars and is given by
\bea
\label{period:matrix:4d}
 \cN_{\Lambda \Sigma} = \bar{\cF}_{\Lambda \Sigma} + 2 i \frac{\im \cF_{\Lambda \Delta} \im \cF_{\Sigma \Theta} X^\Delta X^\Theta}{\im \cF_{\Delta \Theta} X^\Delta X^\Theta} \, .
\eea
Finally, the scalar potential reads
\bea
 V(z, \bar{z}) =  g^{i\bar j}  D_i {\cal L} \bar D_{\bar j} \bar {\cal L} - 3 | {\cal L}|^2  \, ,\eea
where ${\cal L}= e^{{\cal K}/2}\left ( X^\Lambda g_\Lambda - \cF_\Lambda g^\Lambda \right )$ and $D_i{\cal L}= \partial_i {\cal L} + \partial_i {\cal K} {\cal L}/2$.   

The ansatz  for a static dyonic black hole with horizon $\Sigma_\fg$ is of the form\footnote{We normalize the metric on $\Sigma_\fg$ such that the  scalar curvature is $2\kappa$, where 
$\kappa = 1$ for $S^2$, $\kappa = 0$ for $T^2$, and $\kappa = -1$ for $\Sigma_\fg$ with $\fg>1$. The volume is then $\Vol(\Sigma_\fg) = 2\pi \eta$ where $\eta =  2 |\fg-1|$  for $\fg \neq 1$ and $ \eta=1$ for  $\fg = 1$.} 
\bea
ds^2 &= - e^{2U(r)} dt^2 + e^{-2U(r)} \big( dr^2 + V(r)^2 ds^2_{\Sigma_\fg} \big) \, \nonumber \\
A^\Lambda &= a_0(r) dt + a_1(r) A_{\Sigma_\fg} \, ,  
\eea
where  $A_{\Sigma_\fg}$ is the gauge potential for a magnetic flux on $\Sigma_\fg$. For example, for $\Sigma_\fg=S^2$ we can take $A_{S^2}= -\cos \theta d \phi$.
We assume that the scalar fields $z^i$ have only radial dependence.  We are interested in solutions that are asymptotic to AdS$_4$ for large values of the radial coordinate, 
\be e^{ U(r)}  \sim r\, , \qquad  V(r) \sim r^2\, ,\qquad r \gg 1\, , \ee
and approach a regular horizon AdS$_2\times \Sigma_\fg$ at some fixed value  $r=r_0$, 
\be e^{ U(r)}  \sim r-r_0\, , \qquad  V(r) \sim e^{U(r)}\, ,\qquad r \sim r_0 \, . \ee
Notice that we can also interpret these black holes as domain walls interpolating between AdS$_4$ and AdS$_2\times \Sigma_\fg$. The AdS$_2$ factor suggests the existence of a superconformal quantum mechanics describing the horizon microstates. We expect that this is the IR limit of the quantum mechanics discussed in Sect.~\ref{subsec:QM}.

There are two conserved
quantities
\be
\int_{\Sigma_\fg} \! F^\Lambda = \Vol(\Sigma_\fg) \, p^\Lambda \;,\quad \int_{\Sigma_\fg} \! G_\Lambda = \Vol(\Sigma_\fg) \, q_\Lambda \;,
\ee
where $G_\Lambda = 8 \pi G_\text{N}\, \delta (\mathscr{L} \dvol_4)/\delta F^\Lambda$,  corresponding to the magnetic and electric charges of the black hole. Under $Sp(2 n_V+2)$ they transform as  a symplectic  vector $(p^\Lambda,  q_\Lambda)$.  In a frame with purely electric gauging $g_\Lambda$, the magnetic and electric charges are quantized as follows
\be
\label{charge quantization}
 \Vol(\Sigma_\fg)   \, p^\Lambda \, g_\Lambda  \in 2 \pi \bZ \;,\qquad \frac{\Vol(\Sigma_\fg)  \, q_\Lambda}{4G_{{\text{N}}}^{(4)} g_\Lambda} \in 2 \pi \bZ\, ,
\ee
not summed over $\Lambda$. 

As discussed in Sect.~\ref{subsubsec:magnetic}, supersymmetry is realized with a topological twist. In particular, the Killing spinors $\epsilon_A$, $A=1,2$,  only depend on the radial coordinate.   The BPS equations give a set of ordinary differential equations for the  functions $U,V,a_0,a_1,z_i, \epsilon_A$ that are explicitly given in \cite{Cacciatori:2009iz,DallAgata:2010ejj,Hristov:2010ri,Katmadas:2014faa,Halmagyi:2014qza}.
For our purposes, the only important point is that  the gravitino variation contains, among other pieces,
\be \delta  \psi_{\mu A} =\partial_\mu \epsilon_A +\frac{1}{4} \omega_\mu^{ab} \Gamma_{ab} \epsilon_A + \frac i2  g_\Lambda A_\mu^\Lambda (\sigma^3)_A^{\,\,\,\,  B} \epsilon_B +\dots \ .\ee
The vanishing of this variation, when the index $\mu$ is restricted to $\Sigma_\fg$, requires that $A_\mu^\Lambda$ cancels the spin connection and one obtains
\be\label{twist} \sum_\Lambda g_\Lambda p^\Lambda = -\kappa \, ,\ee
where, with standard notations, $\kappa=1$ for horizon $S^2$, $\kappa =0 $ for $T^2$ and $\kappa=-1$ for $\fg >1$. We see that a linear combination of the magnetic charges  is fixed by the twist. In a general theory with also magnetic FI the previous condition
is replaced by
\be \sum_\Lambda \left (g_\Lambda p^\Lambda  -  g^\Lambda q_\Lambda \right ) = -\kappa \, \ee
which is manifestly symplectic invariant. The previous  discussion closely parallels the field theory analysis of the supersymmetries preserved by a topological twist in Sect.~\ref{subsec:loc}.\footnote{This is not a coincidence \citep{Klare:2012gn}: when holography applies, solving   the  Killing spinor equations in bulk near the AdS boundary gives a set of constraints on the boundary theory that are equivalent to those obtained with the approach proposed in \cite{Festuccia:2011ws}.} 

It has been noticed in \cite{DallAgata:2010ejj} that the BPS equations of gauged supergravity for the near-horizon geometry can be put in the form of \emph{attractor equations}.\footnote{The attractor mechanism  for AdS$_4$ static black holes in ${\cal N}=2$ gauged supergravity is discussed in \cite{Cacciatori:2009iz,DallAgata:2010ejj,Chimento:2015rra}. For some recent progress  for dyonic rotating black holes see \cite{Hristov:2018spe}.} 
The BPS equations  are indeed  equivalent to the extremization of the quantity 
\bea
 \label{attractor:mechanism:4D}
 \cI_{\text{sugra}} (X^\Lambda) = -i \frac{\Vol(\Sigma_\fg)}{4 G_{{\text{N}}}^{(4)}} \frac{ q_\Lambda X^\Lambda - p^\Lambda \cF_\Lambda}{g_\Lambda X^\Lambda - g^\Lambda \cF_\Lambda} \, ,
\eea
with respect to the horizon-value of the symplectic sections $X^\Lambda$,  combined with  the requirement that the  value of $\cI_{\text{sugra}}$  at the critical point   $\bar X^\Lambda$ is \emph{real}. 
In general, in gauged supergravity, $\cF (X^\Lambda)$ is a homogeneous function of degree two, so we can equivalently define $\hat Y^\Lambda \equiv X^\Lambda/(g_\Sigma X^\Sigma - g^\Sigma \cF_\Sigma)$ and extremize
\bea
 \label{attractor}
 \cI_{\text{sugra}} (\hat Y^\Lambda) =  i \frac{ \Vol(\Sigma_\fg)}{4 G_{{\text{N}}}^{(4)}}  \left ( p^\Lambda \cF_\Lambda (\hat Y) - q_\Lambda \hat Y^\Lambda \right ) \, .
\eea
The extremization of \eqref{attractor:mechanism:4D} gives a set of algebraic equations for  the value of the physical scalars $z^i$ at the horizon,
and the entropy of the black hole is  given by evaluating the functional \eqref{attractor:mechanism:4D} at its extremum
\be \label{attractor2}
 S_{\text{BH}} (p^\Lambda , q_\Lambda) = \cI_{\text{sugra}} (\bar X^\Lambda) \, .
\ee

\subsubsection{Black holes in AdS$_4\times S^7$}\label{subsubsec:BHS7}

M-theory on AdS$_4\times S^7$ can be consistently truncated to an  ${\cal N}=2$ gauged supergravity containing the four vectors parameterizing the Cartan subgroup of the $SO(8)$ isometry of $S^7$.
In the language of gauged supergravity, one is the graviphoton and the other three give rise to a model with three vector multiplets,  $n_V=3$. By explicitly reducing M-theory on AdS$_4\times S^7$, one can determine the prepotential\footnote{See for example \cite{Hristov:2012bk}.}
\be \label{ABJMgauged} \cF = -2 i \sqrt{X^0 X^1 X^2 X^3} \, ,\ee
and the FI, $g_\Lambda\equiv g, \, g^\Lambda=0$, that are purely electric.  With these notations the AdS$_4$ vacuum has radius $L^2=1/2g^2$.  

We can introduce four magnetic and electric charges  $(p^\Lambda,  q_\Lambda)$. However,  two of these charges are determined by supersymmetry. Indeed, one is fixed by the twisting condition \eqref{twist} that gives a linear constraint among the magnetic charges. For  purely magnetic black holes  \citep{Cacciatori:2009iz}, this is the only constraint and we find a three-dimensional family of solutions. They are particularly simple since all the scalars $z_i$ are real. We can write, for example,  the solution for a black hole with $S^2$ horizon
\bea
\label{BH_metric}
ds^2 &= - \frac12 e^{\mathcal{K} (X)} \Big(r - \frac cr \Big)^2 dt^2 + 2 \frac{ e^{-\mathcal{K} (X)} \, dr^2}{\big( r - \frac cr \big)^2} + 2 e^{-\mathcal{K} (X)} \, r^2 \big( d\theta^2 + \sin^2 \theta\, d\phi^2 \big) \, , \\
F^\Lambda_{\theta \phi} &=  p^\Lambda \, \sin \theta 
\eea
where the real sections  are given by $X^\Lambda = \frac{1}{4} - \frac{\beta_\Lambda}{r}$ 
and the parameters $\beta^\Lambda$ and $c$ are determined in terms of the magnetic charges by
\bea
\label{r_h}
c = 4 \big( \beta_0^2 + \beta_1^2 + \beta_2^2 + \beta_3^2 \big) - \frac12 \, , \,\,  -\sqrt{2} p^\Lambda - \frac12 = 16 \beta_\Lambda^2 - 4 \sum\nolimits_\Sigma \beta_\Sigma^2 \, , \,\, \sum_\Lambda \beta_\Lambda = 0 \, \nonumber 
\eea
where we  also set $L=1$ or, equivalently, $g = 1/\sqrt{2}$. The generic dyonic black holes found in \cite{Katmadas:2014faa,Halmagyi:2014qza}  are more complicated and we will not report here the form of the solution. For dyonic black holes there is an extra constraint on the charges that  follows from the requirement that the entropy computed through the attractor mechanism \eqref{attractor2} is a real number. This constraint is highly non linear in the charges and leaves a six-dimensional family of black holes. 

The entropy can be written  by using  \eqref{attractor2}, even without knowing the explicit form of the metric. The final expression can be written in a symplectic invariant  form 
\bea
S_{\text{BH}} (p^\Lambda , q_\Lambda) =\frac{\Vol(\Sigma_\fg)}{8 \sqrt{2} G_{{\text N}}^{(4)}} \frac{\sqrt{ I_4(\Gamma,\Gamma,G,G) \pm \sqrt{ I_4(\Gamma,\Gamma,G,G)^2 - 64 I_4(\Gamma) I_4(G)}}}{ I_4(G)}  \nonumber
\eea
where $\Gamma=(p^\Lambda, q_\Lambda)$ and $G=(g^\Lambda,g_\Lambda)$ are symplectic vectors containing the charges and the FI parameters, and $I_4$ is a quartic polynomial, known as the quartic invariant, whose   
explicit expression can be found in \cite{Katmadas:2014faa,Halmagyi:2014qza}. In the simplest case of a purely magnetic black hole with $p^1=p^2=p^3\equiv -p/(2g)$, $p^0=(3 p-2)/(2g)$ and horizon $S^2$, we find an expression of the form
\be S_{\text{BH}} \sim \sqrt{-1+6 p -6 p^2 + \sqrt{(6 p -1)(-1+2 p)^{3}}} \, .\ee
 Notice  that this expression is quite complicated, especially if compared with simple forms of the entropy as a function of charges that one can find for some asymptotically flat black holes.

The solutions with spherical horizon can be generalized by adding rotation \citep{Hristov:2018spe}. For completeness we report the form of the entropy  
\bea
S_{\text{BH}} (p^\Lambda , q_\Lambda , j) =\pi \frac{\sqrt{ I_4(\Gamma,\Gamma,G,G) \pm \sqrt{ I_4(\Gamma,\Gamma,G,G)^2 - 64 I_4(G)  (I_4(\Gamma) + j^2)}} }{\sqrt{8} G_{{\text N}}^{(4)} I_4(G)} \nonumber
\eea
where $j$ is the angular momentum. This time supersymmetry imposes  three constraints on the charges 
and leaves again a six-dimensional family of rotating black holes.

\subsection{The dual field theory}

The CFT dual to AdS$_4\times S^7$ is the so-called ABJM theory \citep{Aharony:2008ug}. We briefly discuss its properties and then we write the corresponding  topological twisted index using the rules discussed in Sect.~\ref{sec:lecture2}.

The ABJM theory describes the low-energy dynamics of $N$ M2-branes on $\bC^4/\bZ_k$ \citep{Aharony:2008ug}. It is a three-dimensional supersymmetric Chern--Simons-matter theory with gauge group $U(N)_k \times U(N)_{-k}$ (the subscripts are the CS levels) and matter in bifundamental representation.  The matter content, in ${\cal N}=2$ notations, is described by the quiver diagram
\begin{center}
\begin{tikzpicture}
\draw (-2,0) circle [radius=.4]; \node at (-2,0) {$N$}; \node at (-1.5,-.5) {\footnotesize{$k$}};
\draw (2,0) circle [radius=.4]; \node at (2,0) {$N$}; \node at (2.5,-.5) {\footnotesize{$-k$}};
\draw [decoration={markings, mark=at position 1 with {\arrow[scale=2.5]{>>}}}, postaction={decorate}, shorten >=0.4pt] (-1.5,.5) arc [radius=4.5, start angle = 110, end angle = 70];
\draw [decoration={markings, mark=at position 1 with {\arrow[scale=2.5]{>>}}}, postaction={decorate}, shorten >=0.4pt] (1.5, -.5) arc [radius=4.5, start angle = -70, end angle = -106];
\node at (0,1.2) {$A_i$}; \node at (0,-1.2) {$B_j$};
\end{tikzpicture}
\end{center}
where $i,j=1,2$ and nodes represent gauge groups and arrows represent bifundamental chiral multiplets. The theory has a quartic superpotential
\be\label{ABJMsuper}
W = \Tr \big( A_1B_1A_2B_2 - A_1B_2A_2B_1 \big) \;.
\ee
The ABJM theory has a number of interesting properties:
\begin{itemize}
\item the theory has $\cN=6$ superconformal symmetry, non-perturbatively enhanced to $\cN=8$
for $k=1,2$;
\item it has an $SU(4)$ R-symmetry,  enhanced to $SO(8)$ for $k=1,2$;
\item for $N\gg k^5$ the theory is well-described by a weakly coupled M-theory background, AdS$_4\times S^7/\mathbb{Z}_k$;
\item the free energy on $S^3$ can be computed using localization and scales as $O(N^{3/2})$ in the M-theory limit \citep{Drukker:2010nc}
\be F_{S^3}=\log Z_{S^3} = \frac{\pi \sqrt{2}}{3} \sqrt{k} N^{3/2} \, .\ee
\end{itemize}
For a review of these properties see \cite{Klebanov:2009sg,Marino:2011nm}. 

We will consider the case $k=1$ where the theory has maximal supersymmetry, $SO(8)$ R-symmetry  and is dual to AdS$_4\times S^7$. The four abelian symmetries of the theory, $U(1)^4\subset SO(8)$ correspond, in ${\cal N}=2$ notation, to an R-symmetry and three global symmetries. There are many choices of $U(1)$ R-symmetry corresponding to  different decompositions $SO(8)\rightarrow U(1)_R\times U(1)^3$. In order to write the index we need to select one with integer charges. Introducing a  natural basis of $U(1)$ R-symmetries, 
\be
\begin{array}{c|cccc}
 &  R_1 & R_2 & R_3 & R_4 \\
 \hline
A_1 & 2 & 0 & 0 & 0 \\
A_2 & 0 & 2 & 0 & 0  \\
B_1 & 0 & 0 & 2 & 0  \\
B_2 & 0 & 0 & 0 & 2  \\
\end{array}
\ee
we can for example choose the R-symmetry $\sum_a R_a/2$ that has integer charges. The remaining three $U(1)$s  combine to give three flavor symmetries, say $(R_a-R_4)/2$ for $a=1,2,3$.\footnote{We are cheating a little bit here. The ABJM theory has  also two topological symmetries, $T_1$ and $T_2$, associated with the two $U(1)$ gauge groups. This apparently makes a total of five $U(1)$ global symmetries. However $T_1+T_2$ is decoupled, and  the baryonic symmetry that rotates $A_i$ and $B_i$ with opposite charges is actually gauged. More precisely, due the  CS term, a linear combination of $T_1-T_2$ and the baryonic symmetry differ by a gauge transformation and are therefore equivalent. In the index we could introduce extra fluxes and fugacities for $T_1$ and $T_2$ but these can be re-absorbed by a shift of the fluxes and a rescaling of the integration variables. See \cite{Benini:2015eyy} for more details.}  

Our general rules for the index allow to introduce a number of independent fluxes and fugacities equal to the number of global  symmetries.   We then introduce three magnetic fluxes $\fp$ and three fugacities $y$ for the  three flavor symmetries of ABJM. It will be convenient to choose a redundant but democratic parameterization of these quantities.
We assign a flux and a fugacity, $\fp_a$ and $y_a$ with $a=1,2,3,4$, to each of the fields $A_1,A_2,B_1,B_2$ in the order indicated. The  index is given by 
\begin{multline}
\label{initial Z}
Z = \frac1{(N!)^2} \sum_{\fm, \wt\fm \in \bZ^N} \int_\cC \; \prod_{i=1}^N \frac{dx_i}{2\pi i x_i} \, \frac{d\tilde x_i}{2\pi i \tilde x_i} \, x_i^{k \fm_i} \, \tilde x_i^{-k  \wt\fm_i} 
\prod_{i\neq j}^N \Big [\Big( 1 - \frac{x_i}{x_j} \Big)\, \Big( 1 - \frac{\tilde x_i}{\tilde x_j} \Big)\Big ]^{1-\fg}\\
\hskip 0.1truecm \prod_{i,j=1}^N \prod_{a=1,2}
\bigg( \frac{ \sqrt{ \frac{x_i}{\tilde x_j} \, y_a} }{ 1- \frac{x_i}{\tilde x_j} \, y_a } \bigg)^{\fm_i - \wt\fm_j - \fp_a +1-\fg}
\prod _{a=3,4} \bigg( \frac{ \sqrt{ \frac{\tilde x_j}{x_i} \, y_a} }{ 1- \frac{\tilde x_j}{x_i} \, y_a } \bigg)^{\wt\fm_j - \fm_i - \fp_a +1-\fg}  \times \\
\hskip 0.1truecm
  \left(\det_{AB} \frac{\partial^2 \cW}{\partial u_A\partial u_B}\right )^\fg \, , \end{multline}
where we used the rules of Sect.~\ref{subsubsec:integrand}. Notice that the Hessian of $\cW$ should be computed using the $2N$ variables $u_A=(u_i, \tilde u_i)$. We have also  included the CS term $k$ in order to make clear where the terms come from but soon we will set  $k=1$. 
 As already said, the fugacities are not independent. Since the superpotential 
\eqref{ABJMsuper} must have charge zero under a global symmetry we must set
\be \prod_{a=1}^4 y_a=1 \, .\ee
This translates into a constraint for the corresponding (complexified) chemical potentials  $\Delta_a$, $y_a=e^{i \Delta_a}$,
\be\label{per} \sum_a \Delta_a \in 2\pi \mathbb{Z} \, ,\ee
since the $\Delta_a$ are only defined modulo $2 \pi$.\footnote{In the notation of section  \ref{subsubsec:flavor}, $\Delta_a= A_{t\, a}^{F} + i \beta \sigma_a^{F}$, where $A_{t\, a}^{F}$ and $\sigma_a^{F}$ are the backgrounds for the $a$-th symmetry. The periodicity of $\Delta_a$ is due to the periodicity of the Wilson line $A_{t\, a}^{F}$.}  Similarly, the four fluxes $\fp_a$ are not independent. To understand our parameterization, let us compare the chiral fields contributions in \eqref{initial Z} with \eqref{chiral1-loop2}.
Identifying exponents we have
\be -\fp_a +1 -\fg = \fm_a^{F} +(\fg -1)( r_a -1) \, ,\ee
where $ \fm_a^{F}$ is an assignment of  background fluxes  for the global symmetry and $r_a$ the R-charge of the $a$-th field. Since $W$ has charge zero under global symmetries and charge two under R-symmetries, we have $\sum_{a=1}^4 \fm_a^{F}=0$ and  $\sum_{a=1}^4 r_a=2$, so that 
\be\label{fluxtwist}  \sum_{a=1}^4 \fp_a = 2(1-\fg) \, .\ee
 
The dependence of our index  on three magnetic fluxes and three fugacities  fits well with the family of  black holes discussed in Sect.~\ref{subsubsec:BHS7} that have three  magnetic and three electric charges.  \eqref{fluxtwist} is clearly the analog of \eqref{twist} and already suggests the following identification between parameters $p^\Lambda \rightarrow - \kappa \fp_a /(2 g (1-\fg))$.

The index can be written as a sum over Bethe vacua \eqref{sumBethevacua}. The twisted superpotential \eqref{twisted} 
reads\footnote{We used the of polylogarithm identities  given in footnote \ref{polylog} in order to  recombine the terms in \eqref{twisted}, and discarded terms that do not contribute to the Bethe equations \eqref{BE}. We also introduced an extra minus sign in the definition of $\cW$ in order to match the original conventions in \cite{Benini:2015eyy,Hosseini:2016tor}. It is easy to check directly that the equations \eqref{BAE} give the position of the poles of the integrand after we sum the geometric series in  $m_i$ and $\tilde m_i$ and that,  with the given definition of $\cW$, \eqref{BAE} are equivalent to \eqref{BE}.} 
 \be
\label{twistedABJM}
\cW = \sum_{i=1}^N  \frac k2 \big( \tilde u_i^2 -  u_i^2 \big)  + \sum_{i,j=1}^N \bigg[ \sum_{a=3,4} \Li_2\big( e^{i(\tilde u_j - u_i + \Delta_a)} \big) - \sum_{a=1,2} \Li_2 \big( e^{i( \tilde u_j - u_i - \Delta_a)} \big) \bigg]
\ee
and the Bethe vacua equations are
\be
\label{BAE}
x_i^k \prod_{j=1}^N \frac{ \big( 1- y_3 \frac{\tilde x_j}{x_i} \big) \big( 1- y_4 \frac{\tilde x_j}{x_i} \big) }{ \big( 1- y_1^{-1} \frac{\tilde x_j}{x_i} \big) \big( 1- y_2^{-1} \frac{\tilde x_j}{x_i} \big) }  = \tilde x_j^k \prod_{i=1}^N \frac{ \big( 1- y_3 \frac{\tilde x_j}{x_i} \big) \big( 1- y_4 \frac{\tilde x_j}{x_i} \big) }{ \big( 1- y_1^{-1} \frac{\tilde x_j}{x_i} \big) \big( 1- y_2^{-1} \frac{\tilde x_j}{x_i} \big) } =1  \;.
\ee
In the large $N$ limit we expect that just one Bethe vacuum dominates the partition function. 

\subsection{The ABJM Bethe vacua in the large $N$ limit}\label{sec:LargeNW}

We want to study the solutions of \eqref{BAE} in the large $N$ limit \citep{Benini:2015eyy}. By running numerics, one discovers that the imaginary parts of the solutions $u_i$ and $\tilde u_i$ grow with $N$,
\be\label{scaling} u_i = i N^\alpha t_i  + v_i \, \qquad \tilde u_i = i N^\alpha t_i  + \tilde v_i \, \ee
and are equal for the two sets, while the real parts remain bounded. As usual, in the large $N$ limit, the distributions   of $u_i$ and $\tilde u_i$ become almost continuous and we introduce a parameter  $t(i/N)=t_i$, defined in an interval $[t_-,t_+]$. We also introduce two functions of $t$, $v(t)$ and $\tilde v(t)$, defined implicitly by $v(i/N)=v_i, \, \tilde v(i/N)=\tilde v_i$, and a normalized density
\be\label{normc} \rho(t)=\frac1N \frac{di}{dt} \, , \qquad \int_{t_-}^{t_+} \rho(t) d t =1 \, .\ee

The interesting feature of this model is that $\cW$ becomes a \emph{local} functional,\footnote{This is similar to other matrix models solved using localization in three and five dimensions \citep{Herzog:2010hf,Jafferis:2011zi,Jafferis:2012iv,Minahan:2013jwa}.} 
\be\label{BP}  \cW[\rho(t),\delta v(t)] =  i N^{1+\alpha} \int dt \, t \rho(t) \delta v(t)  +   i N^{2-\alpha} \int dt   \rho(t)^2 \sum_{a=1}^4 g_3 (- \epsilon_a \delta v(t) + \Delta_a) \, ,\ee
where $\delta v(t)=\tilde v(t) - v(t)$, $g_3(u)=  \frac16 u^3- \frac12 \pi u^2 + \frac{\pi^2}{3} u$ and $\epsilon_a=1$ for $a=1,2$ and $\epsilon_a=-1$ for $a=3,4$. We also assumed that $\Delta_a$ are real and \be\label{ineq}  0<  -\epsilon_a \delta v(t) + \Delta_a < 2\pi \ee for all $a$. The first term in $\cW$ comes from the Chern--Simons interaction and the second is the contribution of matter fields.  The derivation of \eqref{BP} is given in \cite{Benini:2015eyy}. Here we just mention few facts. 

\begin{itemize}
\item  $\cW$  is local because of the exponential terms $e^{i( \tilde u_j - u_i \pm \Delta_a)}$ in the arguments of polylogs in \eqref{twistedABJM}. Due to \eqref{scaling}, for $j>i$ the polylogs are exponentially suppressed in the large $N$ limit. 
%
For $i>j$ the exponential is large but we can use the identity $\Li_2(e^{iu})+\Li_2(e^{-iu}) =\frac12 u^2- \pi u + \frac{\pi^2}{3}$, valid for $\re u\in [0,2\pi]$, (see  footnote \ref{polylog}) to transform it into a polynomial plus exponentially suppressed terms. As a consequence, up to polynomial terms, the main contribution comes for values of the indices $i\sim j$ and makes the functional local. 
\item Terms with higher powers of $N$ cancel. For more general ${\cal N}=2$ theories this is not automatic and imposes conditions on the matter content of the theories for which this method works \citep{Hosseini:2016tor}.
\item  Polynomial terms coming from this manipulation or Chern--Simons terms that are not in \eqref{BP} happily combine into a contribution $\sum_{i=1}^N 2\pi n_i u_i + 2 \pi \tilde n_i \tilde u_i$ to $\cW$, where $n_i$ and $\tilde n_i$ are integers. These angular ambiguities disappear in the  Bethe equations \eqref{BE}.
\end{itemize}

In general, the two contributions in \eqref{BP} have different powers of $N$. They  compete and give a sensible functional with a minimum only for $\alpha=1/2$. We then see that $\cW$ scales as $N^{3/2}$ as predicted by holography for  AdS$_4$ black holes. We will then set $\alpha=1/2$ from now on.

\begin{figure}[t]
\centering
\includegraphics[scale=.6]{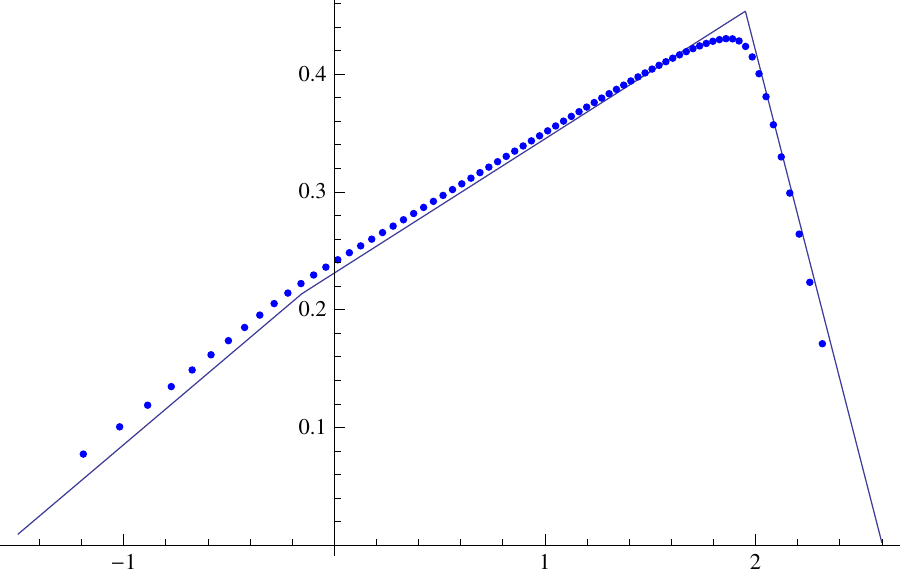}
\hspace{1cm}
\includegraphics[scale=.6]{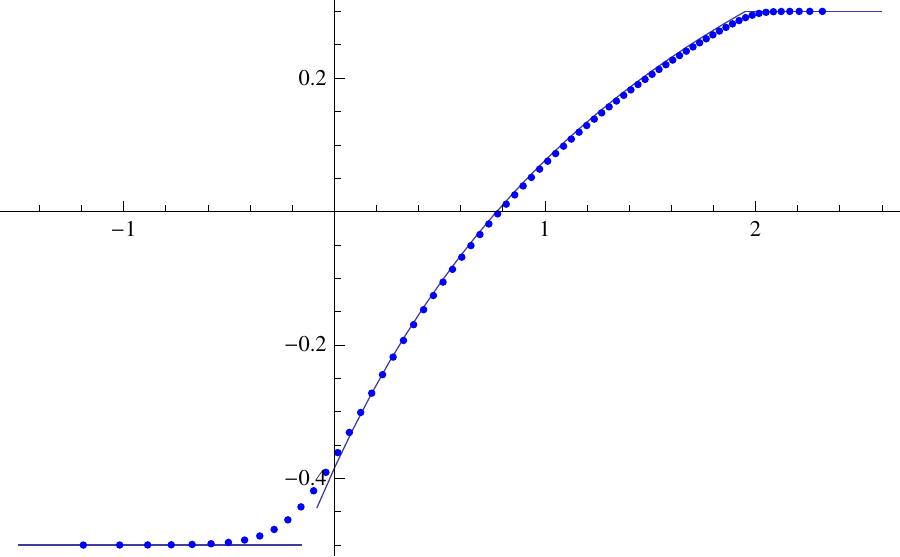}
\caption{Plots of the density of eigenvalues $\rho(t)$ and the function $\delta v(t)$ for $N=75$, $\Delta_1=0.3$, $\Delta_2=0.4$, $\Delta_3=0.5$ with $\sum_a\Delta_a=2\pi$ and $k=1$. The blue dots represent the numerical simulation, while the solid grey line is the analytical result.
\label{fig: comparisonplots}}
\end{figure}

In order to extremize \eqref{BP} we add to $\cW$ a Lagrange multiplier term 
\be -i N^{3/2}\mu \left ( \int  \rho(t) dt - 1\right ) \ee
 that enforces the normalization condition \eqref{normc}. Differentiating $\cW$ with respect to $\delta v(t)$ and $\rho(t)$, we obtain a pair of algebraic equations 
\begin{eqnarray}
& t   -   \rho(t) \sum_{a=1}^4 \epsilon_a g_3^\prime (- \epsilon_a  \delta v(t) +\Delta_a) = 0 \, ,  \label{eqsLargeN}\\
& t \delta v(t) + 2 \rho(t) \sum_{a=1}^4 g_3(- \epsilon_a  \delta v(t) +\Delta_a) = \mu \, , \label{eqsLargeN2}
\end{eqnarray}
which can be easily solved in terms of rational functions of $t$. The solution is depicted in Fig.~\ref{fig: comparisonplots} together with the numerical solution for large $N$.

From the figure we see that  $\rho(t)$ and $\delta v(t)$    are piece-wise continuous functions of $t$. The solution in \eqref{eqsLargeN} and \eqref{eqsLargeN2} only covers the central part of these functions. Numerics suggest that there are two external intervals, which we dub \emph{tails}, where $\delta v(t)$ is actually constant in the large $N$ limit.  It turns out that such constant value corresponds to the saturation of the inequality \eqref{ineq} for some value of $a$, 
$\delta v(t)= \epsilon_a \Delta_a$. The inequalities \eqref{ineq} are necessary to restrict   to a particular determination of the multi-valued polylog functions  and the saturation corresponds to  the position of the cuts. The numerics suggest that, once  $v_i$ and $\tilde v_i$ hit the cut, their value is frozen. The value for $\rho(t)$ in the tails can be obtained by \emph{solving} its equation of motion, \eqref{eqsLargeN2}, setting $\delta v(t)$ to the  constant value $\epsilon_a \Delta_a$ and \emph{ignoring} the equation of motion for $\delta v(t)$, \eqref{eqsLargeN}, which would be inconsistent. The end-points of the interval, $t_-$ and $t_+$, are  finally determined by $\rho(t_{\pm})=0$. 

Obviously, the equation of motion for $\delta v(t)$, \eqref{eqsLargeN}, must be satisfied at finite $N$.   The main correction to $\delta v(t)$ and to its equation comes from the terms with $i=j$ in \eqref{twistedABJM}. Such terms  contribute 
\be \delta \cW = N \int dt \rho(t) \bigg[ \sum_{a=3,4} \Li_2\big( e^{i(\delta v(t) + \Delta_a)} \big) - \sum_{a=1,2} \Li_2 \big( e^{i( \delta v(t)- \Delta_a)} \big) \bigg] \, .\ee
Notice that these terms are  suppressed  compared to $\cW$.\footnote{This is a standard argument in the context of matrix models: $\sum_{i\ne j}^N =O(N^2)$ while  $\sum_{i=1}^N =O(N)$. Notice that the contribution to $\cW$ comes from terms with $i$ almost equal to $j$ and contributions to $\delta \cW$ from terms with $i=j$.} 
They contribute a term
\be\label{corr}  -   \frac{1 }{\sqrt{N}}  \bigg[ \sum_{a=1}^4 \epsilon_a \log\big( 1- e^{i( \delta v(t) -\epsilon_a \Delta_a)} \big)  \bigg] \, ,\ee
to the right-hand side of the equation \eqref{eqsLargeN}  for $\delta v(t)$. Such a correction is generically of order  $1/\sqrt{N}$.  However, on the tails, since $\delta v(t)=\epsilon_b \Delta_b$ for some $b$, one of the  logarithms blows up and the correction can be effectively of order one. Indeed, the equation of motion for $\delta v(t)$ can be satisfied if
\be \delta v(t) =\epsilon_b \left (\Delta_b - e^{- \sqrt{N} Y_b(t)}\right )\, , \ee
where $Y_b(t)$ is a quantity of order one. In this case, on the tail, the  equation becomes
\be t    -   \rho(t) \sum_{a=1}^4 \epsilon_a g_3^\prime (- \epsilon_a  \delta v(t) +\Delta_a)  = \epsilon_b Y_b(t)  \, ,\ee
not summed over $b$, which determines the value of $Y_b(t)$. Notice that, quite remarkably, the correction to $\delta v(t)$ is not power-like but exponentially small. The equations for $\rho(t)$ and the  value of $\cW$  on the solution are not affected by these corrections in the large $N$ limit since $\Li_2(z)$ is finite for $z\rightarrow 1$.  However, these corrections are important for evaluating the index. 
  
  The explicit solution is as follows \citep{Benini:2015eyy}. Let us first take $\Delta_a$ real. Using the periodicity of $\Delta_a$, we can always restrict to the case where $0\le \Delta_a\le 2 \pi$. We will also assume that $\Delta_1\leq \Delta_2$, $\Delta_3\leq \Delta_4$. The constraint \eqref{per}  can be satisfied only for $\sum_{a} \Delta_a =0,2\pi, 4 \pi, 6\pi, 8\pi$ and we need to consider all possible cases. We find a  solution for $\sum_a \Delta_a = 2 \pi$. We have a central region where 
\be
\begin{aligned}
\rho &= \frac{2\pi \mu + t(\Delta_3 \Delta_4 - \Delta_1 \Delta_2)}{(\Delta_1 + \Delta_3)(\Delta_2 + \Delta_3)(\Delta_1 + \Delta_4)(\Delta_2 + \Delta_4)} \\[.5em]
\delta v &= \frac{\mu(\Delta_1 \Delta_2 - \Delta_3 \Delta_4) + t \sum_{a<b<c} \Delta_a \Delta_b \Delta_c }{ 2\pi \mu + t ( \Delta_3 \Delta_4 - \Delta_1 \Delta_2) }
\end{aligned}
\qquad -\frac{\mu}{\Delta_4}   < t < \frac{\mu}{\Delta_2} \, .
\ee
When $\delta v$ hits $-\Delta_3$ on the left the solution becomes
\be
\rho = \frac{\mu + t\Delta_3}{(\Delta_1 + \Delta_3)(\Delta_2 + \Delta_3)(\Delta_4 - \Delta_3)} \, , \,\,   \delta v = - \Delta_3 \;, \,\,   \,\,\,\,  -\frac{\mu}{\Delta_3}  < t <-\frac{\mu}{\Delta_4}  \, ,
\ee
with the exponentially small correction  $Y_3 = (- t\Delta_4 -\mu)/(\Delta_4 - \Delta_3)$, while when  $\delta v$ hits $\Delta_1$ on the right the solution becomes
\be
\rho = \frac{\mu - t \Delta_1}{(\Delta_1 + \Delta_3)(\Delta_1 + \Delta_4)(\Delta_2 - \Delta_1)} \, ,\,\,\,\,\,\,\,\,\,\,\,\,
\delta v = \Delta_1 \;,\,\, \, \frac{\mu}{\Delta_2}  < t < \frac{\mu}{\Delta_1}  \, ,
\ee
with $Y_1 =(t\Delta_2 - \mu)/(\Delta_2 - \Delta_1)$.
It turns out that, for $\sum_{a=1}^4 \Delta_a =0, 4 \pi, 8\pi$, equations \eqref{eqsLargeN} and \eqref{eqsLargeN2} have no regular solutions.  There is also a solution for $\sum_a \Delta_a = 6 \pi$ which, however, is obtained by the previous one by a discrete symmetry of the index: $\Delta_a \rightarrow 2 \pi -\Delta_a$ $\left(y_a\rightarrow y_a^{-1}\right)$. 

We can also evaluate the twisted superpotential on the solution and find
\be
\label{Vsolution sum 2pi -- end}
\wt \cW (\Delta) = \frac {2i N^{3/2}}{3} \sqrt{ 2 \Delta_1 \Delta_2 \Delta_3 \Delta_4} \, , \qquad  \Delta_a \in [0,2\pi]\, ,  \qquad  \sum_{a=1}^4 \Delta_a = 2\pi \, .
\ee
The result for a different determination of the $\Delta_a$ is obtained by periodicity: just replace $\Delta_a$ with $[\Delta_a]= (\Delta_a \, {\rm mod} \, 2 \pi)$. We can also extend by holomorphicity the result to complex $\Delta_a$.

It is interesting to observe that  $\wt \cW(\Delta)$ has the same functional dependence of  two important physical  quantities appearing in the study of ABJM and its dual AdS$_4\times S^7$.  One is of purely field theory origin. It is known  that, for any ${\cal N}=2$ theory, there exists a family of supersymmetric Lagrangians on $S^3$ parameterized by arbitrary R-charges of the chiral fields \citep{Hama:2011ea,Jafferis:2010un}. When the theory is superconformal, the resulting partition function has an extremum precisely at the exact R-symmetry of the theory \citep{Jafferis:2010un}. For ABJM at $k=1$, the $S^3$ free energy reads \citep{Jafferis:2011zi}
\be\label{freeS3ABJM} F_{S^3}(r) =  \frac {4\pi N^{3/2}}{3} \sqrt{ 2 r_1 r_2  r_3 r_4} \, ,\ee
where $r_a$ are a general assignment of R-charges for the fields $A_1,A_2,B_1,B_2$ satisfying $\sum_{a=1}^4 r_a=2$.
We see that \citep{Hosseini:2016tor}
\be \wt \cW (\Delta_a) = \frac{ \pi i}{2} F_{S^3} \left (\frac{\Delta_a}{\pi}\right ) \, .\ee  
The second quantity, of supergravity origin, is the prepotential of the ${\cal N}=2$ gauged supergravity
describing the low energy theory of the holographic dual.  Restricted to the $U(1)^4$ gauge sector, the prepotential is given by \eqref{ABJMgauged} and we see that
\be \wt \cW (\Delta_a) = -\frac{ \sqrt{2} N^{3/2}}{3} {\cal F} (\Delta_a) \, .\ee 
This will be important for comparison with the attractor mechanism.

\subsection{The large $N$ limit of the ABJM index}

Using \eqref{sumBethevacua}, up to an irrelevant overall factor, the index is given by 
\bea
\label{final Z}
& \left(\det_{AB} \frac{\partial^2 \cW}{\partial u_A\partial u_B}\right )^{\fg-1} \prod_{i\neq j}^N \Big( 1 - x_i^*/x_j^* \Big)^{1-\fg} \, \Big( 1 - \tilde x_i^*/\tilde x_j^* \Big)^{1-\fg} \times \\
& \prod_{i,j=1}^N \prod_{a=1,2}
\left( \frac{ \sqrt{ x_i^* \, y_a/\tilde x_j^*} }{ 1- x_i^* \, y_a/\tilde x_j^* } \right )^{- \fp_a +1-\fg}
\prod _{b=3,4} \left ( \frac{ \sqrt{ \tilde x_j^*\, y_b/x_i^*} }{ 1- \tilde x_j^*\, y_b/x_i^* } \right )^{- \fp_b +1-\fg}   \, , \eea
where $x_i^*$ and $\tilde x_i^*$ is the large $N$ Bethe vacuum found in the previous section. By taking the large $N$ limit of \eqref{final Z}, after some manipulations, one finds 
\begin{eqnarray} \label{Z large N functional} &  \log Z  =- N^\frac32 \int dt\, \rho(t)^2 \bigg[ (1-\fg) \frac{2\pi^2}3 + \sum_{a=1}^4  \big( \fp_a-1 + \fg \big)\, g_3'\big(-\epsilon_a \delta v(t) + \Delta_a \big) \bigg] \quad  \nonumber \\
 & \hskip -1truecm  - N^\frac32 \sum_{a=1}^4 \fp_a \int_{\delta v \,\approx\, \varepsilon_a \Delta_a}  dt\, \rho(t) \, Y_a(t)  \, , 
\end{eqnarray}
up to corrections of order $N\log N$.  The first contribution in the first line of \eqref{Z large N functional} comes from the Vandermonde determinant and the second from the matter contribution. The second line in \eqref{Z large N functional} comes from the tails. Since the logarithm of the one-loop determinant of the chiral fields is singular on such regions, we need to take into account the exponentially small corrections $Y_a$ to the tails. The exponent $-\fp_a +1-\fg$ of the one-loop determinant is corrected to $-\fp_a$ by an analogous and subtle contribution from the determinant of the Hessian of $\cW$.\footnote{See \cite{Benini:2015eyy} for details.}

By plugging in the explicit solution for $\rho(t), \delta v(t)$ and $Y_a(t)$ we find
\be
\label{Z large N}
\log Z  = - \frac{2\sqrt2}3 N^\frac32  \sum_{a=1}^4 \fp_a \parfrac{}{\Delta_a} \sqrt{\Delta_1 \Delta_2 \Delta_3 \Delta_4} \;.
\ee

Notice that 
\be
\label{indexTh}
\log Z  =   i  \sum_{a=1}^4 \fp_a \parfrac{}{\Delta_a} \wt \cW (\Delta) \;.
\ee
One can show that this remarkable identity can be extended to other ${\cal N}=2$ quiver theories. Indeed it can be proved using only the equations of motion for $\rho$ and $\delta v$,  and taking into account the extra terms on the tails \citep{Hosseini:2016tor}. 

\subsection{Matching  index and entropy for ABJM}\label{sec:matching}

We can now extract the degeneracy of states from the index using \eqref{entropyind}. We introduce four electric charges $\fq_a$, adapted to the democratic  basis of charges we are using, and we set $j_i=0$ since we are interested in static black holes. The degeneracy of states with given electric charge  is then obtain by extremizing the functional
\be\label{II} \cI(\Delta) = \log Z(\Delta) - i \sum_{a=1}^4 \fq_a \Delta_a \, ,\ee
which implicitly depends on the magnetic charges $\fp_a$ through $Z$. 
This is the $\cI$- extremization principle introduced in \cite{Benini:2015eyy,Benini:2016rke}. 
For purely magnetic black holes we just extremize $\log Z$. For a generic dyonic static black hole,
we obtain
\be \label{degeneracyABJM} S_{{\rm micro}}(\fp_a,\fq_a) = \log Z - i \sum_{a=1}^4 \fq_a \Delta_a = i \sum_{a=1}^4 \left ( \fp_a \parfrac{}{\Delta_a} \wt \cW (\Delta)  - \fq_a \Delta_a \right )\, ,\ee
evaluated at its critical point in $\Delta_a$ with the constraints  $\sum_{a=1}^4 \Delta_a = 2\pi$  and $\re \Delta_a \in [0,2\pi]$.

By an explicit computation one can see that this expression correctly reproduces the entropy of the black holes in \cite{Cacciatori:2009iz,DallAgata:2010ejj,Hristov:2010ri,Katmadas:2014faa,Halmagyi:2014qza}. This can be checked more easily by comparing with the attractor equations \eqref{attractor2}. Using \eqref{attractor:mechanism:4D}  we find 
\be \label{entropyatrrABJM}  S_{{\rm BH}}(\fp^\Lambda,\fq_\Lambda) =  -\frac{i \Vol(\Sigma_\fg)}{4 G_{\text{N}}^{(4)}} \frac{ q_\Lambda X^\Lambda - p^\Lambda F_\Lambda}{g_\Lambda X^\Lambda - g^\Lambda F_\Lambda} = i \sum_{\Lambda=0}^3 \left ( \hat \fp^\Lambda \parfrac{}{\hat X_\Lambda} \wt \cW (\hat X)  - \hat \fq_\Lambda \hat X^\Lambda \right )\, .\ee
Here we used the data of the relevant gauged supergravity (see \eqref{ABJMgauged})
\be   g_\Lambda \equiv g\, , \qquad g^\Lambda=0\, ,\qquad   \cF = -2 i \sqrt{X^0 X^1 X^2 X^3} \, ,\ee
we defined adapted scalar fields 
\be \hat X^\Lambda = \frac{2 \pi X^\Lambda}{\sum_{\Lambda=0}^3 X^\Lambda} \, ,\ee
satisfying $\sum_{\Lambda=0}^3 \hat X^\Lambda = 2\pi$, and enforced the Dirac quantization condition of fluxes \eqref{charge quantization} by defining the integers $\hat \fp^\Lambda$ and $\hat \fq_\Lambda$,
\be
\label{charge quantization2}
 \Vol(\Sigma_\fg)   \, p^\Lambda \, g_\Lambda  = -2 \pi \hat \fp^\Lambda \;,\qquad \frac{\Vol(\Sigma_\fg)  \, q_\Lambda}{4G_\text{N} g_\Lambda} = 2 \pi \hat \fq_\Lambda \, .
\ee Finally, we used the known holographic dictionary (see for example \citealt{Marino:2011nm})
\be \frac{1}{2 g^2 G_{N}^{(4)}} = \frac{2 \sqrt{2}}{3} N^{3/2} \, .\ee
Using the identification\footnote{Notice that the identification is consistent with the twisting condition \eqref{twist} and the constraint $\sum_{a=1}^4 \Delta_a = 2\pi$.}
\bea  
 a=1,2,3,4 & \rightarrow  \Lambda =0,1,2,3 \, , \\
 \Delta_a &\rightarrow \hat X^\Lambda \, , \\
(\fp_a,\fq_a) & \rightarrow (\hat \fp^\Lambda, \hat \fq_\Lambda) \, ,
 \eea  
 we see that $S_{{\rm micro}}(\fp_a,\fq_a) = S_{{\rm BH}}(\fp^\Lambda,\fq_\Lambda)$. The extremization of \eqref{degeneracyABJM} is thus completely equivalent to the attractor mechanism in gauged supergravity. The correspondence, up to constants,  is
\bea & \wt \cW(\Delta) \, \rightarrow \cF (X^\Lambda) \, , \\
& \cI(\Delta) \, \rightarrow \cI_{{\rm sugra}} (X^\Lambda) \, . \\
\eea

Notice that the functional $\cI(\Delta)$ 
is only defined up to integer multiples of $2\pi i$, due to the presence of $\log Z$. So the microscopical entropy should be properly defined
as $S_{{\rm micro}}(\fp_a,\fq_a) = \cI \, ({\rm mod} \, 2\pi i)$.  

In field theory, $\cI$ only depends on three independent chemical potentials, which we can choose to be $\Delta_a$ with $a=1,2,3$. The extremization  of $\cI$ determines them in terms of  three electric charges, $q_a-q_4$,
\be\label{legendre} \frac{\partial \log Z}{\partial \Delta_a} = i (\fq_a - \fq_4) \, ,\qquad   a=1,2,3 \, . \ee
The entropy can be then expressed in terms of $\fq_a-\fq_4$ as  
 \be\label{entropyI}  \log Z - i \sum_{a=1}^3 (\fq_a -\fq_4) \Delta_a \, \,   ({\rm mod} \, 2\pi i)  \, .\ee
In these derivations we used that  $\fq_a$ are integers and $\sum_{a=1}^4 \Delta_a = 2 \pi$. 

Correspondingly, in gravity, the family of black holes depends only on three independent electric charges, as discussed in Sect.~\ref{subsec:attractor}. Indeed, the requirement that $S_{{\rm BH}}(\fp^\Lambda,\fq_\Lambda)$ is real gives a constraints on the charges. For given magnetic charges $\fp_a$ and flavor electric charges $\fq_a-\fq_4$, there is at most  one value of $\fq_4$ that leads to  black hole with regular horizon. 

We can  see the left hand side of \eqref{legendre} as determining the average electric charge $\fq_a-\fq_4$ in our statistical ensemble at large $N$.
The index only depends on the global symmetries of the theory and there are three of them. Correspondingly, with our method,
we cannot determine  the average electric charge associated with the R-symmetry. However, this value is fixed by the BPS equations
in gravity.   It would be interesting to find a purely field theoretical method for testing this prediction for the fourth charge.  Notice that, with the permutationally invariant definition of $\cI$ given in \eqref{II}, the value of $q_4$ predicted by gravity  is precisely the one that makes  the critical value of $\cI$ real.

Let us conclude this section with a comment. We computed the entropy as a Legendre transform of the grand canonical partition function, according to a natural microscopic point of view. However, in holography, the  partition function of the boundary field theory can  be also identified  with the (holographically  renormalized) on-shell action of the bulk theory. By consistency,  we should be able to extract the same information from the gravitational on-shell action. The agreement of these points of view has been discussed in \cite{Azzurli:2017kxo,Halmagyi:2017hmw,Cabo-Bizet:2017xdr}. In particular, the authors of \cite{Bobev:2020pjk} found a family of smooth Euclidean solutions with boundary $\Sigma_g\times S^1$ whose on-shell action coincides  with the topologically twisted index of ABJM for general values of the fugacities and that contains one element with an AdS$_2$ geometry that can be wick-rotated to the original Lorentzian black hole.  

\subsection{Other examples and generalizations}

The large $N$ limit of the topologically twisted index can be evaluated for other ${\cal N}=2$ quiver Chern--Simons theories with an AdS$_4$ dual.  There is a large class of Yang--Mills--Chern--Simons theories with fundamental and bi-fundamental chiral fields that have been proposed as duals of M-theory and massive type IIA compactifications. Most of these are obtained by dimensionally reducing a \emph{parent} four-dimensional quiver gauge theory  with an AdS$_5\times SE_5$ dual, where $SE_5$ is a five-dimensional Sasaki--Einstein manifold,  adding Chern--Simons terms and flavoring with fundamentals.   Holography predicts that the  twisted index  scales as $N^{3/2}$ for  theories  dual to M-theory on seven-dimensional Sasaki--Einstein manifolds, and as $N^{5/3}$ for a class of theories  dual to massive type IIA  on warped six-manifolds.\footnote{See \cite{Hanany:2008fj, Hanany:2008cd, Martelli:2008si,Gaiotto:2009tk,Benini:2009qs,Jafferis:2008qz,Herzog:2010hf,Gulotta:2011vp,Crichigno:2012sk,Martelli:2009ga}  for examples of theories with M-theory dual and \cite{Guarino:2015jca,Fluder:2015eoa} for theories with massive type IIA dual.} For a large class of these theories the large $N$ method discussed in  Sect.~\ref{sec:LargeNW} applies and the twisted index has been evaluated in \cite{Hosseini:2016tor,Hosseini:2016ume,Jain:2019lqb,Jain:2019euv}.  Quite remarkably, in all these theories  the on-shell twisted superpotential coincides with the corresponding $S^3$ free energy \citep{Hosseini:2016tor}
\be\label{S3} \wt \cW (\Delta_a) = \frac{ i \pi}{d}F_{S^3} \left (\frac{\Delta_a}{\pi}\right ) \, ,\ee
where $d=2$ for M-theory examples and $d=3$ in massive type IIA ones.\footnote{The twisted index depends on chemical potentials for the global symmetries. As for ABJM,  we can use a redundant basis where we assign a chemical potential $\Delta_a$ to each field $\phi_a$. For each term $W_I$ in the superpotential we must require $\sum_{a \in W_I} \Delta _a = 2 \pi p$ with $p\in \bZ$
for all the fields $\phi_a$ appearing in $W_I$. It turns out that, as for ABJM, a large $N$ solution exists only for $p=1$ (up to equivalent solutions).  On the other hand the $S^3$ free energy is a function of the R-charges of the fields, and these are constrained to satisfy  $\sum_{a \in W_I} r _a = 2$. The identity \eqref{S3} makes sense  because  $p=1$.} The partition function is given by the following \emph{index theorem} \citep{Hosseini:2016tor}\footnote{See \cite{Hosseini:2018qsx} for more details and applications.} 
\be\label{Z large N conjecture}
 {\quad \rule[-1.4em]{0pt}{3.4em}
 \log Z =  (1-\fg) \left ( \frac{d i}{\pi} \, \wt{\mathcal{W}}(\Delta_I) \,
 +i  \sum_{I}\, \left[ \left(\frac{\fp_I}{1-\fg} - \frac{\Delta_I}{\pi}\right) \frac{\partial \wt{\mathcal{W}}(\Delta_I)}{\partial \Delta_I}
  \right] \right ) \, .
 \quad}
\ee
It is often possible, as for ABJM, to choose a convenient redundant parameterization for  the chemical potentials such that $\wt \cW$ becomes a homogeneous functions of degree $d$. For this choice the twisted index is given again by
\be
\label{indexTh2}
\log Z  =  i  \sum_{a} \fp_a \parfrac{}{\Delta_a} \wt \cW (\Delta) \;.
\ee
%

The entropy of dyonic static black holes in such theories is obtained by taking  the Legendre transform of \eqref{indexTh2}. Comparison with the attractor mechanism in the form \eqref{attractor} then requires that
\be\label{Watt}  \wt \cW (\Delta) \sim \cF (X^\Lambda) \, ,\ee
under some map  $\Delta_a \rightarrow X^\Lambda$. \eqref{Watt} is however oversimplified. The effective low energy theory for M or type IIA compactifications on general manifolds is an ${\cal N}=2$ gauged supergravity containing massless vector multiplets associated with the global symmetries of the CFT. In general, such gauged supergravity contains also hypermultiplets and a number of vector multiplet that is larger than the number of global symmetries. Therefore, in \eqref{Watt}, the number of $\Delta_a$ does not match the number of sections $X^\Lambda$. In general,   the hypermultiplets  have a VEV at the horizon and give a mass to some of the vector fields. If $n_V$ is the number of vector multiplets and $n_H$ is the number of hypermultiplets,  $n_V - n_H$ is the number of massless vectors,  corresponding to the number of global symmetries and therefore to the number of independent $\Delta_a$. 
The BPS equations for the hyperinos are typically algebraic and give linear constraints among the sections $X^\Lambda$.
By solving these constraints we can obtain an effective prepotential $\cF_{{\rm eff}}(X^\Lambda)$ for the massless vector multiplets only. If we believe that this further truncated theory correctly describes the horizon of the black hole, \eqref{Watt} should hold 
with $\cF(X^\Lambda)$ replaced by $\cF_{{\rm eff}}(X^\Lambda)$.\footnote{This happens for example for the mass deformed ABJM model  \citep{Bobev:2018uxk},
where the value of one $\Delta_a$  is fixed by the mass deformation and, in gravity, one $X^\Lambda$ is fixed by the hyperino conditions.} 

Two notable applications where this works perfectly are the following.  A well-known example of massive type IIA background is the warped AdS$_4\times Y_6$  flux vacua of massive type IIA dual  to the  $\mathcal{N}=2$ $U(N)$  gauge theory with three adjoint
multiplets  and a Chern--Simons coupling $k$  \citep{Guarino:2015jca}.  It corresponds to 
an internal manifold $Y_6$ with the topology of $S^6$. The theory has two global symmetries and there exists a family of 
black holes depending on two magnetic charges \citep{Guarino:2017pkw}. The entropy of such black holes has been matched with the prediction of the twisted index in \cite{Hosseini:2017fjo,Benini:2017oxt}. For this example
\be \wt \cW(\Delta_a) \sim F_{S^3} (\Delta_a) \sim \cF_{{\rm eff}}(\Delta_a) \sim (\Delta_1\Delta_2\Delta_3)^{2/3} \, , \qquad \sum_{a=1}^3 \Delta_a = 2\pi\, .\ee
The second example involves the so-called \emph{universal black hole} \citep{Azzurli:2017kxo}. This is dual  to the \emph{universal twist} \citep{Benini:2015bwz,Bobev:2017uzs}  defined by a set of magnetic fluxes $\fp_a$ proportional to the exact R-symmetry of the CFT.  This black hole is a solution of minimal gauged supergravity and, as such, can be embedded in all M-theory and massive type IIA compactifications, thus explaining the name universal. Hence it  provides an infinite family of examples. Since 
$\wt \cW$ is proportional to the $S^3$ free energy (see  \eqref{S3}) and the latter is extremized at the exact R-symmetry, it follows easily from \eqref{Z large N conjecture} that 
\be S_{\text {BH}}(\fp_a) = (\fg-1) F_{S^3} \, .\ee
This relation agrees with the gravitational prediction based on  minimal gauged supergravity \citep{Azzurli:2017kxo}.

There have been progresses in various directions and also open problems:
\begin{itemize}
\item For theories with M-theory dual the large $N$ method discussed above  works only when  the bi-fundamental fields  transform in a real representation of the gauge group and the total number of fundamentals is equal to the total number of anti-fundamentals. The same restriction has been found in \cite{Jafferis:2011zi} for the large $N$ evaluation of the $S^3$ free energy. It is not known yet how to take the large $N$ limit for the $S^3$ free energy or the twisted index in the case of chiral quivers. Also in the case of vector-like quivers, where the method works, there is an extra complication since baryonic symmetries are invisible in the large $N$ limit (accidental flat directions of the matrix model). It is still not clear how to introduce and study baryonic fluxes in the large $N$ limit of the twisted index. 
\item A gravitational dual of $\cI$-extremization was proposed in \cite{Couzens:2018wnk}, generalizing similar results for $a$- and $c$-extremization \citep{Martelli:2005tp,Gauntlett:2018dpc}. In this approach, the entropy of an M theory  black hole with AdS$_2$ horizon is obtained by extremizing a geometric quantity associated with a supersymmetric background  where some of the equations of motion have been relaxed. As a result, the entropy can be expressed in terms of simple geometrical data of the internal manifold and can be computed also in the cases where the explicit solution is not known. In particular, this provides an explicit formula for the entropy and the exact R-symmetry of the associated superconformal mechanics for an infinite class of models based on toric Sasaki-Einstein manifolds. The agreement of the gravitational picture with the available field theory results at large $N$ has been discussed  in \cite{Gauntlett:2019roi,Hosseini:2019ddy,Kim:2019umc}.
\item The microstate counting for dyonic rotating black holes \citep{Hristov:2018spe} is still an open problem. The chemical potential dual to rotation in the bulk is associated with an Omega-background for the boundary theory on $S^2\times S^1$. The topologically twisted index in an Omega-background is known \citep{Benini:2015noa} but it gives rise to a complicated matrix model. For particular values of the $\Omega$-deformation, it has been  written as a sum over Bethe vacua \citep{Closset:2018ghr}. This could be useful to solve the matrix model in the large $N$ limit. The result is expected to match the entropy functional proposed in \cite{Hosseini:2019iad} by gluing gravitational blocks.
\item For  asymptotically flat black holes, there is a huge literature including  higher derivative corrections  and highly detailed precision tests. For asymptotically AdS black holes, the story is just begun. At the moment, $1/N$ corrections to the twisted index for ABJM and the above massive type IIA background have been computed numerically focusing  on  the universal logarithmic correction  \citep{Liu:2017vbl,Liu:2017vll,Jeon:2017aif,Liu:2018bac,PandoZayas:2019hdb}. The ABJM matrix model provides a quantum corrected entropy functional that would be interesting to study further. In particular, it would be interesting to find the analog of  standard results and conjectures about the quantum entropy of asymptotically flat back holes, like the OSV conjecture \citep{Ooguri:2004zv}, the Sen's quantum entropy functional \citep{Sen:2008vm,Sen:2009vz} and localization in supergravity \citep{Dabholkar:2010uh,Dabholkar:2011ec},\footnote{See \cite{Iqbal:2003ds} for an earlier localization computation in gravity.}   to cite only few of them. In particular, some interesting results about  localization in AdS$_2\times S^2$ have been already  found in \cite{Hristov:2018lod,Hristov:2019xku}.
\item The topologically twisted index has been extended to five-dimensions \citep{Hosseini:2018uzp,Crichigno:2018adf}  and successfully compared with the entropy of 
AdS$_6$ black holes with horizon AdS$_2\times \Sigma_{\fg_1}\times \Sigma_{\fg_2}$ \citep{Suh:2018tul,Hosseini:2018usu,Suh:2018szn,Suh:2018qyv,Fluder:2019szh}. The expression for the entropy is  given by a  generalization of the index theorem \eqref{indexTh2} and \eqref{Z large N conjecture}.
   \end{itemize}
Other interesting related results and solutions can be found in \cite{Cabo-Bizet:2017jsl,Toldo:2017qsh,Gang:2018hjd,Gang:2019uay,Hosseini:2019and,Bae:2019poj}. 

\section{Black strings in AdS$_5$}\label{sec:lecture4}

The methods introduced in  Sect.~\ref{sec:lecture3} can be generalized to other dimensions and can be used to provide  general tests of holography. In particular, 
they can be applied to  domain wall solutions interpolating between AdS$_{d+n}$ and AdS$_d \times  \cM_n$, with a topological twist along the $n$-dimensional compact manifold $\cM_n$.
In this section,  making a brief digression from  the main subject of these notes,  we discuss the example of black strings in AdS$_5$. 

More precisely, we consider a family of black strings in AdS$_5$ with near horizon geometry AdS$_3\times \Sigma_\fg$. They correspond holographically to a twisted compactification of a four-dimensional theory flowing  to a two-dimensional  CFT in the IR. The properties of this CFT, and, in particular, its elliptic genus, can be computed using a topologically twisted index for four-dimensional theories on $\Sigma_\fg \times T^2$.  
We can make contact with the physics of black holes by compactifying the black string on a circle, as in the standard example  \citep{Strominger:1996sh}. In this way we can also make contact with a Cardy formula approach to the microstate counting. Other examples of similar tests of holography related to twisted compactifications can be found in \cite{Toldo:2017qsh,Crichigno:2018adf,Hosseini:2018uzp}.

\subsection{Black string solutions in AdS$_5\times S^5$}\label{sec:blackstrings}
We  consider  solutions of a five-dimensional ${\cal N}=2$ effective gauged supergravity with abelian vector multiplets   of the form  
\bea
ds^2 &= e^{2f(r)} (- dt^2  + dz^2) + e^{-2f(r)} dr^2 + e^{2g(r)} ds^2_{\Sigma_\fg}  \, \nonumber \\
A^\Lambda &= p^\Lambda A_{\Sigma_\fg} \, ,  
\eea
where  $A_{\Sigma_\fg}$ is the gauge potential for a magnetic flux on $\Sigma_\fg$ and supersymmetry  is preserved with a twist along $\Sigma_{\fg}$.  We are interested in  solutions that are asymptotic to AdS$_5$ for large values of the radial coordinate, 
\be e^{ f(r)}  \sim r\, , \qquad  e^{g(r)} \sim r \, ,\qquad r \gg 1 \ee
and approach a regular horizon AdS$_3\times \Sigma_\fg$ at some fixed value  $r=r_0$, 
\be e^{ f(r)}  \sim r-r_0\, , \qquad  e^{g(r)}  \sim {\rm constant} \, ,\qquad r \sim r_0 \, . \ee
These solutions can be interpreted as black strings extended in the direction $z$ or, equivalently,  as domain walls interpolating between AdS$_5$ and AdS$_3\times \Sigma_\fg$. They are holographically dual to a twisted compactification 
of a SCFT$_4$ on $\Sigma_\fg$ that flows in the IR to a two-dimensional $(0,2)$ SCFT$_2$ associated with the horizon factor AdS$_3$. 

Solutions  that can be embedded in AdS$_5\times S^5$ have been found in \cite{Benini:2013cda}, using a five-dimensional gauged supergravity with three abelian gauge fields associated with the isometries $U(1)^3\subset SO(6)$ of $S^5$. The solution depends on three fluxes $\fp_a$ constrained by the twisting condition
\be \fp_1+\fp_2+\fp_3 = 2 - 2 \fg \, ,\ee
and  corresponds to a twisted compactification of ${\cal N}=4$  SYM  on $\Sigma_\fg$. Holography suggests that this theory flows to an IR two-dimensional SCFT. The value of the central charge of the IR SCFT$_2$ in the large $N$ limit can be extracted from the solution using standard arguments \citep{Brown:1986nw,Henningson:1998gx} and reads \citep{Benini:2013cda}
\be\label{cr} c = \frac{ 3 R_{{\rm AdS}_3}}{2 G_N^{(3)}} = 
  12 N^2 \frac{ \fp_1 \fp_2 \fp_3}{ \fp_1^2+\fp_2^2+\fp_3^2 - 2 \fp_1\fp_2 - 2 \fp_2 \fp_3 -2 \fp_3\fp_1} \, . \ee
  
This result can be successfully compared with field theory \citep{Benini:2013cda}, where central charges can be computed  using extremization techniques.  
Let us briefly explain how this works. Consider first the case of ${\cal N}=1$ four-dimensional SCFTs and  ${\cal N}=4$  SYM  as an example. In ${\cal N}=1$ language, the  theory contains three chiral multiplets $\phi_i$ with the superpotential
\bea
 \label{SYM:superpotential}
 W = \Tr \left( \phi_3 \left[ \phi_1, \phi_2 \right] \right)  \, .
\eea
We introduce  a generic R-charge assignment $\Delta_a$ for the three chiral fields $\phi_i$. Since the superpotential has R-charge $2$, we must have  $\Delta_1+\Delta_2+\Delta_3=2$. The exact R-symmetry of an ${\cal N}=1$ superconformal theory can be obtained by extremizing  a trial $a$-charge
\be\label{acharge0}  a(\Delta_a) = \frac{9}{32} \Tr R(\Delta_a) ^3 - \frac{3}{32} \Tr R(\Delta_a)  \, ,\ee
where $R(\Delta_a)$ is the matrix of R-charges of the fermionic fields. This construction is known as $a$-maximization \citep{Intriligator:2003jj}. For ${\cal N}=4$ SYM  at large $N$ we find\footnote{For ${\cal N}=4$ SYM $\Tr R$ is identically zero. For theories with an AdS dual, $\Tr R=0$ in the large $N$ limit.}
\be\label{acharge}  a(\Delta_a) =   \frac{9}{32} \, N^2 \left ( 1   + \sum_{a=1}^3 (\Delta_a -1)^3 \right ) = \frac{27}{32} N^2 \Delta_1\Delta_2 \Delta_3  \, ,\ee
where the first contribution  in the bracket comes from the gauginos and the second from the fermions in the chiral multiplets. This expression is trivially extremized for $\Delta_1=\Delta_2=\Delta_3=2/3$, the exact R-charges of the fields $\phi_i$.
The critical value of $a(\Delta)$ is the central charge $a$ of the SCFT$_4$. Similarly, the exact R-symmetry and the right-moving central charge of a two-dimensional $(0,2)$ SCFT 
are obtained  by extremizing the trial quantity  \citep{Benini:2012cz}
\be c_r(\Delta_a) =  3 \Tr \gamma_3 R(\Delta_a)^2 \, , \ee
where $\gamma_3$ is the two-dimensional chirality operator and $R(\Delta_a)$ the matrix of R-charges of the massless fermionic fields. This construction is known as $c$-extremization \citep{Benini:2012cz}. For  the twisted compactification of ${\cal N}=4$ SYM, the quantity $c_r$  can be computed using topological arguments. We just need to know the difference between the number of fermionic zero modes with positive and negative chirality. This is easily computed from the Riemann-Roch theorem as in section  \ref{subsec:QM}. We then find at large $N$ \citep{Benini:2013cda}
\bea\label{trialcr} c_r(\Delta_a) &= - 3\, N^2 \left ( 1-\fg  + \sum_{a=1}^3 ( \fp_a -1 + \fg) (\Delta_a -1)^2 \right ) \\ & = -3 N^2(\Delta_1\Delta_2 \fp_3 + \Delta_2\Delta_3 \fp_1 + \Delta_3\Delta_1 \fp_2 ) \, , \eea
where again  the first contribution in the first line bracket comes from gauginos and the second from matter fields. 
One can easily check that the extremization of \eqref{trialcr} with respect to $\Delta_a$  reproduces \eqref{cr}. The agreement is valid in the large $N$ limit where $c=c_l=c_r$.\footnote{$c_r-c_l$ is equal to the gravitational anomaly $k={\rm Tr} \gamma_3$ which is of order one in the large $N$ limit  \citep{Benini:2012cz}.} 

Notice  that, in the large $N$ limit, the $c_r$ trial central charge can be written as  \citep{Hosseini:2016cyf}
\be\label{ccharge} c_r(\Delta_a) = -\frac{32 }{9} \sum_{a=1}^3 \fp_a \frac{\partial a(\Delta_a)}{\partial \Delta_a} \, . \ee
It is interesting to observe that formula \eqref{ccharge}, with a suitable parameterization for the fluxes and the R-charges, holds for all the twisted compactifications of ${\cal N}=1$ SCFT$_4$ dual to AdS$_5\times SE_5$, where $SE_5$ is a toric Sasaki--Einstein manifold \citep{Hosseini:2016cyf}.\footnote{We refer to \cite{Hosseini:2016cyf} for a proof, to \cite{Amariti:2017iuz} for more detailed expressions, and to the appendices of \cite{Hosseini:2018uzp,Hosseini:2019use} for an alternative derivation based on the anomaly polynomial.}

In the case of black strings the information about states is encoded in the \emph{elliptic genus} of the two-dimensional $(0,2)$ CFT
\bea
\label{trace0}
Z (y,q) =  {\rm Tr}   (-1)^F q^{L_0} \prod_I y_I^{J_I} \, ,
\eea
where $q = e^{2 \pi i \tau}$, with $\tau$  the modular parameter of $T^2$, $y_I$ are fugacities for the global symmetries and $L_0$ the left-moving Virasoro generator. Since the index is independent of the scale, we can evaluate it in the UV, where it becomes the topologically twisted index,  defined as the partition function on $\Sigma_\fg \times T^2$, with a topological twist along $\Sigma_\fg$. 

\subsection{The topologically twisted index on $T^2\times \Sigma_\fg$}\label{sec:topT2}

For an $\cN = 1$ four-dimensional gauge theory with  a non-anomalous $\U(1)$ R-symmetry, the topologically twisted index is a function of $q = e^{2 \pi i \tau}$, where $\tau$ is the modular parameter of $T^2$, fugacities $y_I$ for the global symmetries and flavor magnetic fluxes $\fm^F_I$ on $\Sigma_\fg$ parameterizing the twist.
It can be computed using localization and it is given by an elliptic generalization of the formulae in Sect.~\ref{subsec:localization}  \citep{Closset:2013sxa,Benini:2015noa}.   
Explicitly, for a theory with gauge group $G$ and a set of chiral multiplets transforming in representations $\fR_I$ of $G$ with R-charge $r_I$,
the topologically twisted index is given by a contour integral of a meromorphic form
\bea
 \label{path integral index}
 Z (\fp, y) & =  \frac1{|W|} \; \sum_{\fm \,\in\, \Gamma_\fh} \; \oint_\cC \;   \prod_{\text{Cartan}} \left (\frac{dx}{2\pi i x}  \eta(q)^{2(1-\fg)} \right )
 \prod_{\alpha \in G} \left[ \frac{\theta_1(x^\alpha ; q)}{i \eta(q)} \right]^{1-\fg} \times \\
  &  \prod_I \prod_{\rho_I \in \fR_I} \bigg[ \frac{i \eta(q)}{\theta_1(x^{\rho_I} y_I ; q)} \bigg]^{\rho_I(\fm) + (\fg -1)(r_I-1) +\fm^F_I} \left ( \det_{ij} \frac{\partial^2 \log Z_{pert}(u,\fm)}{\partial i u_i \partial\fm_j} \right)^\fg \, , 
\eea
where $\alpha$ are the roots of $G$, $\rho_I$ the weights of the representation $\fR_I$ and $|W|$ denotes the order of the Weyl group.
In this formula, $\theta_1(x; q)$ is a Jacobi theta function and $\eta(q)$ is the Dedekind eta function. The zero-mode gauge variables  $x = e^{i u}$  parameterize the Wilson lines on the two directions of the torus 
\be
u = 2 \pi \oint_{\textmd{A-cycle}} A - 2 \pi \tau \oint_{\textmd{B-cycle}} A \, , 
\ee
and are defined modulo 
\be u_i \sim u_i + 2 \pi n + 2 \pi m \tau\, ,\qquad\qquad  n\, ,m \in \mathbb{Z} \, . \ee
As in three dimensions, the result is summed over a lattice of gauge magnetic fluxes $\fm$  living in the co-root lattice $\Gamma_{\fh}$ of the gauge group $G$ and the contour of integration selects the Jeffrey--Kirwan prescription for taking the residues. In four dimensions there is a one-loop contribution from 
the Cartan components of the vector multiplets.
One can show that the integrand in \eqref{path integral index} is a well-defined meromorphic function of $x$ on the torus provided that the gauge and the gauge-flavor anomalies vanish. The index has a trace interpretation as a sum over a Hilbert space of states on $\Sigma_\fg \times S^1$
\bea
\label{trace}
Z (\fn, y, q) =  {\rm Tr}_{\Sigma_\fg \times S^1}  (-1)^F q^{H_L} \prod_I y_I^{J_I} \, .
\eea
This trace reduces to the elliptic genus of the two-dimensional theory obtained by the twisted compactification on $\Sigma_\fg$.

We now consider the index for ${\cal N}=4$ SYM. The superpotential \eqref{SYM:superpotential}
imposes the following constraints on the chemical potentials $\Delta_a$ and the flavor magnetic fluxes $-\fp_a+1-\fg=(\fg -1)(r_a-1) +\fm^F_a$ associated with the fields $\phi_a$,
\be
 \label{SYM:constraints}
 \sum_{a = 1}^{3} \Delta_a \in 2 \pi \mathbb{Z} \, , \qquad \qquad \sum_{a = 1}^{3} \fp_a = 2 -2 \fg \, .
 \ee 
The topologically twisted index for the  SYM theory with gauge group $SU(N)$ is then given by
\bea
 \label{SYM path integral_constraint}
 Z = & \frac{1}{N!} \;
 \sum_{\substack{\fm \,\in\, \mathbb{Z}^N \, , \\ \sum_i \fm_i = 0}} \; \int_\cC \;
 \prod_{i=1}^{N - 1} \frac{d x_i}{2 \pi i x_i} \eta(q)^{2(N-1)(1-\fg)} \prod_{i,j=1}^{N} \left( \frac{\theta_1\left( \frac{x_i}{x_j} ; q\right)}{i \eta(q)}\right)^{1-\fg} \\
 &  \prod_{i,j=1}^{N} \prod_{a=1}^{3} \left[ \frac{i \eta(q)}{\theta_1\left( \frac{x_i}{x_j} y_a ; q\right)} \right]^{\fm_i - \fm_j - \fp_a + 1-\fg}   \left ( \det_{ab} \frac{\partial^2 \log Z_{pert}(u,\fm)}{\partial i u_a \partial\fm_b} \right)^g \, ,
\eea
with $y_a=e^{i \Delta_a}$. The Bethe vacua are determined by
\bea\label{BAN=4}
e^{i \frac{\partial \cW}{\partial u_i}} =  \prod_{j=1}^{N} \prod_{a=1}^{3} \frac{ \theta_1\left( \frac{x_j}{x_i} y_a ; q\right) }{\theta_1\left( \frac{x_i}{x_j} y_a ; q\right)} =1\, .\eea
The Bethe equations \eqref{BAN=4} have a remarkably simple solution \citep{Hosseini:2016cyf,Hong:2018viz}\footnote{This solution was found in the high temperature limit in \cite{Hosseini:2016cyf} and proved to be exact for all $\tau$ in \cite{Hong:2018viz}. Using $SL(2,\mathbb{Z})$, the authors of \cite{Hong:2018viz} have found many other solutions of the Bethe equations for ${\cal N}=4$ SYM at finite $\tau$. Non-standard solutions corresponding to continua of Bethe vacua have been found in \cite{ArabiArdehali:2019orz}.  The solution \eqref{solell}  extends to more general quivers \citep{Hosseini:2016cyf,Lezcano:2019pae,Lanir:2019abx}.}
\be \label{solell}  u_k =  -\frac{2 \pi \tau}{N} \left (k - \frac{N+1}{2}  \right )  \, .\ee
These Bethe vacua will play a role also in  the physics of AdS$_5$ black holes \citep{Benini:2018mlo,Benini:2018ywd}, as discussed in Sect.~\ref{sec:interpretation}. 

This time the index is too hard to solve in the large $N$ limit. We can study instead the \emph{high temperature limit} corresponding to a shrinking of the torus given by $\tau= i\beta/(2\pi)$ with $\beta \rightarrow 0^+$ \citep{Hosseini:2016cyf}.\footnote{Notice that $\beta$ is not really a temperature, but rather a parameterization of the modular parameter of the torus.} This limit is also particularly interesting from the field theory point of view because it controls the asymptotic growing
of the number of states with the dimension \citep{Cardy:1986ie}.  
Using \eqref{solell} it is easy to compute, at leading order in $\beta$, \citep{Hosseini:2016cyf}
\bea\label{bstwisted} 
\wt \cW (\Delta) &= \frac{i (N^2-1)}{2\beta} \Delta_1\Delta_2\Delta_3 \, ,\\
\log Z (\fp, \Delta) &= i \sum_{a=1}^3 \fp_a \frac{\partial \wt \cW}{\partial \Delta_a} \, ,
\eea 
where $\wt \cW(\Delta)$ is the on-shell value of the twisted superpotential. As in the three-dimensional case, the result is valid for $\sum_a \Delta=2\pi$.
We see a striking similarity with the black hole case, in particular equation \eqref{indexTh2}. Moreover, the twisted superpotential is proportional to
the trial $a$-central charge \eqref{acharge} of the four-dimensional SCFT. This statement is the four-dimensional analog of \eqref{S3}. 

Comparing with \eqref{ccharge} we find the \emph{Cardy formula}  
\be\label{Cardy} \log Z (\fp, \Delta) = \frac{\pi^2}{6 \beta} c_r(\Delta) = \frac{\pi i }{12 \tau} c_r(\Delta) \, ,\ee
valid at leading order in $\beta$  and at large $N$.

This result can be extended to all ${\cal N}=1$ SCFT$_4$ dual to AdS$_5\times SE_5$ vacua \citep{Hosseini:2016cyf}. It can be also generalized to the case of the \emph{refined} topologically twisted index on $S^2\times T^2$
\bea
\label{trace2}
Z (\fn, y, q) =  {\rm Tr}_{S^2 \times S^1}  (-1)^F q^{H_L} \zeta^{2 J} \prod_I y_I^{J_I} \, ,
\eea
where $\zeta= e^{i \omega/2}$ is a fugacity for the angular momentum.  The result is \citep{Hosseini:2019lkt}
\be\label{Cardy2} \log Z (\fp, \Delta) = \frac{\pi^2}{6 \beta} \left( c_r(\Delta) -  \frac{8 \omega^2}{9 \pi^2} a(\fp)\right ) \, ,\ee
valid at leading order in $\beta$  and at large $N$.

The example discussed in this section can be actually 
generalized to many other  flows interpolating between AdS$_{3+n}$ and AdS$_3\times \cM_n$ where supersymmetry is preserved along the compact manifold $\cM_n$ with a topological twist. Examples of compactifications of the $(2,0)$ theory in six dimensions on the product of two Riemann surfaces $\Sigma_{\fg_1}\times \Sigma_{\fg_2}$ are discussed
for example in \cite{Hosseini:2018uzp}.

\subsection{Back to Cardy}\label{sec:Cardy0}

Similarly to what is done for generic CFT$_2$ \citep{Cardy:1986ie}, we can extract information on the growing of the number of supersymmetric states with the energy from the asymptotic behavior \eqref{Cardy} of the elliptic genus.  

From the definition of the index as a trace \eqref{trace2}, we see that  the number of supersymmetric  states with momentum $n$, electric charge $\fq_a$ under the Cartan subgroup of  $SO(6)$ and angular momentum $j$  can be extracted as a Fourier coefficient 
\be
 \label{ch:1:micro:density}
 d ( \fp ,n, j, \fq) =  -i \int_{i \mathbb{R}}   \frac{d \beta}{2\pi} \int_{0}^{2\pi} \frac{d \Delta_a}{2\pi} \, Z (\fp, \Delta) \, e^{  \beta n - i  \sum_{a=1}^3 \Delta_a \fq_a - i \omega j}  \delta \left( \sum_{a=1}^3 \Delta=2\pi\right)\, ,
\ee
where $\beta=-2 \pi i \tau$ and the corresponding integration is over the imaginary axis.

In the  limit of large charges, we can use the saddle point approximation. Consider first $\fq=0$ and $j=0$.
The number of supersymmetric  states  with charges $(\fp , n)$ can be obtained by extremizing
\be
 \label{SBH}
 \cI ( \beta , \Delta) \equiv \log Z (\fp, \Delta) +  \beta n \ee
with respect to $\Delta$ and $\beta$. Given \eqref{Cardy},  we see that the extremization with respect to $\Delta$  is the $c$-extremization principle \citep{Benini:2012cz,Benini:2013cda}
and sets the trial right-moving central charge $c_r (\Delta)$ to its \emph{exact} value $c_{{\rm CFT}}(\fp)$ given in \eqref{cr}. Extremizing $\cI (\beta , \Delta)$ with respect to $\beta$ yields
\be
 \label{crit:tbeta}
 { \beta} (\fp  , n)
 = \pi \sqrt{\frac{c_{\text{CFT}} (\fp)}{6 n} } \, .
\ee
Plugging back \eqref{crit:tbeta} into $\cI ( \beta , \Delta)$, we find for the entropy of states
\bea\label{Cardys3}
S(\fp, n)\equiv  \log d (\fp ,  n, 0, 0) =   \cI \big|_{\text{crit}} (\beta , \Delta)  = 2 \pi \sqrt{ \frac{ n\, c_{\text{CFT}}(\fp)}{6}} \, .
\eea
This is obviously nothing else than Cardy formula \citep{Cardy:1986ie}.\footnote{For a further discussion about the asymptotic behavior see \cite{Hosseini:2018qsx}.}

One can generalize the previous computation to the case $\fq_a\ne 0$ and $j\ne 0$ by extremizing
\be
 \label{SBH2}
 \cI ( \beta , \Delta) \equiv \log Z (\fp, \Delta) + \beta n  - i \sum_a \fq_a \Delta_a - i \omega j  \, .\ee
After some manipulations, the result can be expressed in the form  \citep{Hosseini:2019lkt,Hosseini:2020vgl}
\bea\label{Cardyelectric}
S(\fp, \fq, n) = 2\pi  \sqrt{ \frac{c_{\text{CFT}}}{6} \left( n +\frac12 \sum_{I,K} q_I A_{IK}^{-1}(\fp) q_K  - \frac{27}{16 a(\fp)} j^2 \right) }   \, ,
\eea
where the index $I$ runs over a set of independent global symmetries $J_K$ and $A_{IK}= \Tr \gamma_3 J_I J_K$ is the 't Hooft  anomaly matrix of the two-dimensional theory.  This result holds for a generic ${\cal N}=1$ quiver with a Sasaki-Einstein dual. The particular combination of $n$, electric charges and angular momentum appearing in \eqref{Cardyelectric} is related to the properties of the elliptic genus and is familiar from
the physics of asymptotically flat black holes \citep{Dijkgraaf:2000fq,Kraus:2006nb,Manschot:2007ha}.\footnote{Our elliptic genus  is actually a meromorphic function of $\Delta_a$ and this leads to a more complicated structure of the corresponding modular forms, related to wall-crossing phenomena. For asymptotically flat black holes this leads to interesting behaviours (see for example \citealt{Dabholkar:2012nd}). It would be interesting to see if the same happens for the black holes discussed in Sect.~\ref{sec:Cardy} obtained by compactifying the black string.}

\subsection{Cardy formula and black holes}\label{sec:Cardy}

There is standard argument to obtain a black holes from a black string:  compactify  along the circle inside AdS$_3$ and add a momentum $n$ along it. More precisely, one  replaces the near-horizon geometry of the five-dimensional  black string with $\text{BTZ} \times \Sigma_{\fg_1}$, where the metric for the extremal BTZ reads \citep{Banados:1992wn}
\bea
 \label{BTZ}
 d s_3^2= \frac14 \left (\frac{ - d t^2+ d r^2}{r^2} \right ) + \rho \left[ d z +\left( - \frac14 + \frac{1}{2 \rho r} \right) d t \right]^2 \, .
\eea
Here, the parameter $\rho$ is related to the electric charge $n$.
This solution is \emph{locally} equivalent to AdS$_3$, since there exists locally only one constant curvature metric in three dimensions, and solves the same BPS equations; however, BTZ and AdS$_3$ are inequivalent globally.
Compactifying the full five-dimensional  black string of \cite{Benini:2013cda} on the circle with the extra momentum we obtain a static BPS black hole in four dimensions,
with magnetic charges $\fp_a$ and  electric charge $n$.
This can be thought as a domain wall that interpolates between an AdS$_2 \times \Sigma_{\fg_1}$ near-horizon region and a complicated asymptotic \emph{non-AdS$_4$} vacuum \citep{Hristov:2014eza}. By another standard argument, the entropy of  such a black hole is given by the number of states of the CFT with momentum $n$, and is therefore given by the Cardy formula \eqref{Cardys3}
\bea\label{Cardys32}
S_{\text{BH}} (\fp ,  n) = \cI \big|_{\text{crit}} (\beta , \Delta)  = 2 \pi \sqrt{ \frac{ n\, c_{\text{CFT}}(\fp )}{6}} \, .
\eea
One can see that this prediction matches the black holes entropy computed from supergravity  \citep{Hristov:2014eza}. Moreover, a family of rotating  dyonic black strings in AdS$_5\times S^5$ was found in \cite{Hosseini:2019lkt} and the entropy of the compactified black hole successfully compared with \eqref{Cardyelectric}. An analogous matching for black strings in AdS$_7\times S^4$ was discussed in \cite{Hosseini:2020vgl}.

It is also interesting to observe that,  for static black holes, the field theory extremization of $\cI$ becomes again equivalent to the attractor mechanism
in the reduced four-dimensional gauged supergravity. The dimensional reduced supergravity has one more vector coming from the reduction on the circle, prepotential $\cF \sim  \frac{X^1 X^2 X^3}{X^0}$, and purely electric FI parameters $g_0=0$ and $g_i=g$ \citep{Hristov:2014eza}.  Under the identifications $\hat X^0 \rightarrow \beta\, , \hat X^i \rightarrow i \Delta_a \, 
, (\fq_\Lambda , \fp^\Lambda) \rightarrow (-n,\fq_1,\fq_2,\fq_3, 0, \fp_1,\fp_2,\fp_3)$ and $\cF(\hat X)\rightarrow \wt \cW (\Delta)$,  the attractor functional \eqref{attractor}
can be identified with the $\cI$-functional \eqref{SBH}.

\section{The superconformal index}\label{sec:lectureSCI}

As we discussed in Sect.~\ref{sec:lecture1}, we should be able to obtain the entropy of supersymmetric Kerr--Newman black holes in AdS$_{d+1}$ by counting  states in the dual CFT on $S^d\times \mathbb{R}$. The relevant supersymmetric quantity to consider is the \emph{superconformal index}, which enumerates BPS states on $S^d\times \mathbb{R}$  \citep{Romelsberger:2005eg,Kinney:2005ej}.
  
We start by considering ${\cal N}=1$ supersymmetric field theories on $S^3\times \mathbb{R}$. We refer to \cite{Festuccia:2011ws} for an explicit description of the supersymmetric Lagrangian  on $S^3\times \mathbb{R}$ and the corresponding supersymmetry transformations. The superconformal index is defined as the trace
\bea\label{supindex}    I(p,q,u) = \rm{Tr}_{S^3} (-1)^F e^{-\beta \{ \cQ, \cQ^\dagger\}} p^{J_1+\frac{R}{2}}  q^{J_2+\frac{R}{2}}  u^{J}  \eea
on the Hilbert space of states on $S^3$.  Here $J_1$ and $J_2$ generate rotations on $S^3$, $R$ is the R-symmetry generator, and $J$ denotes collectively  the global symmetries. 
 $\cQ$ is a supercharge with $J_1=J_2=-1/2$ and $R=1$ and satisfies the algebra
\bea \{ \cQ, \cQ^\dagger \} = \Delta- J_1- J_2 - \frac32 R \, ,\eea
where $\Delta$ generates translation along $\mathbb{R}$. We introduced fugacities, $p$, $q$, and $u$,  for all the generators that commute with $\cQ$, which are $J_1+R/2$, $J_2+R/2$ and the global symmetries $J$.
The quantity \eqref{supindex} is a Witten index in the sense discussed  in Sect.~\ref{index} and it is therefore independent of $\beta$ and invariant under continuous small deformations of the Lagrangian.
Despite the name, the supersymmetric index \eqref{supindex} makes sense also for non-conformal theories. Since we are interested in holography, we will just consider the case of conformal theories where 
$\Delta$ can be identified with the dilatation operator. 

The index \eqref{supindex} can be also written as the Euclidean supersymmetric partition function on $S^3\times S^1$, as discussed in Sect.~\ref{index}, and computed using localization. The precise relation among the two quantities involves a prefactor
\bea \cZ^{susy}_{S^3\times S^1} (p,q,u) = e^{-\beta E_{c.e.}} I(p, q, u) \, ,\eea
where the quantity $E_{c.e}$, called  supersymmetric Casimir energy \citep{Assel:2014paa,Assel:2015nca,Lorenzen:2014pna,Genolini:2016sxe,Martelli:2015kuk,Closset:2019ucb}, is due to the regularization of the one-loop  determinants and
it can be interpreted as the vacuum expectation value of the Hamiltonian.

The superconformal index can be computed explicitly through localization or, being invariant under continuous deformations, just by enumerating the gauge invariant states annihilated by $\cQ$ and $\cQ^\dagger$ in the weakly coupled UV theory.
The result for an ${\cal N}=1$ theory with gauge group $G$ and chiral matter in the representation $\cR_a$ and R-charge $r_a$ is \citep{Romelsberger:2005eg,Kinney:2005ej,Dolan:2008qi}
\bea\label{siloc}  I(p,q,u) = \frac{(p,p)_\infty^r (q,q)^r_\infty}{|W|} \prod_{i=1}^r \oint \frac{d z_i}{2\pi i z_i}  \frac{\prod_a \prod _{\rho \in \cR_a} \Gamma ( (pq)^{r_a/2} z^{\rho_a} u^{\nu_a} ; p, q)}{\prod _{\alpha} \Gamma ( z^{\rho_a};p, q)} \, ,\eea
where $r$ is the rank of the gauge group, $\alpha$ the roots, $\rho_a$ are the weights of the representation $\cR_a$, $\nu_a$ the flavor weights,  and $|W|$ the order of the gauge group. The integration is over the Cartan subgroup of $G$ and is taken over the unit circle for all variables $z_i$. The special functions appearing in the previous formula are the elliptic Gamma function \citep{Felder_2000}
\bea \Gamma( z; p,q) = \prod_{n,m=0}^\infty \frac{1 - p^{n+1} q^{m+1}/z}{1- p^n q^n z}\, ,\qquad  |p|<1\, , |q|<1\, , \eea
and the q-Pochhammer symbol
\bea (z; q)_\infty = \prod_{n}^\infty (1- q^n z)\, ,\qquad   |q|<1\, . \eea
In the  localization approach, numerator and denominator of \eqref{siloc} arise as  one-loop determinants for chiral matter multiplets and  vector multiplets, respectively. The reader can compare the formula for the superconformal index  \eqref{siloc} with the analogous one for the topologically twisted one \eqref{locformula} and look for analogies and differences.\footnote{One main difference is the sum over magnetic fluxes in the \eqref{locformula}. From a more technical point of view, another big difference is the integration contour.}

The superconformal index can be defined and computed through localization also in other dimensions \citep{Bhattacharya:2008zy,Pestun:2016jze}. For three-dimensional theories the elliptic Gamma functions are replaced by hyperbolic ones and there is an extra sum over magnetic fluxes, as in \eqref{locformula}, due to the existence of local BPS monopole operators in three dimensions \citep{Kim:2009wb,Imamura:2011su,Kapustin:2011jm,Dimofte:2011py}.   

\section{Electrically charged  rotating black holes}\label{sec:lecture5}

In this section we discuss the case of Kerr--Newman black holes in various dimensions. These are electrically charged  rotating black holes without a twist. As argued in Sect.~\ref{sec:lecture1}, they are qualitatively different from the magnetically charged black holes  discussed 
in Sect.~\ref{sec:lecture3}.

In general, we should be able to recover the entropy of the electrically charged  rotating   black holes in AdS$_d$ from the BPS partition function \eqref{BPSpf} that counts supersymmetric states of the dual CFT on $S^{d-2}\times \mathbb{R}$. Since the black holes preserve just two real supercharges, we need to count 1/16 BPS states and this is a hard problem. Old attempts to evaluate  the BPS partition function of ${\cal N}=4$ SYM \citep{Grant:2008sk,Chang:2013fba,Yokoyama:2014qwa} reached the somehow disappointing conclusion that there is a subset of states whose number grows with $N$ but slower  than the entropy of the black holes. 
Alternatively,  one may try to replace the BPS partition function with the corresponding index \eqref{BPSindex}. The appropriate index is the superconformal one, defined in the previous section,  that can be expressed and computed as a supersymmetric Euclidean path integral on $S^{d-2}\times S^1$.  It is known that, for generic real fugacities,  the superconformal index is a quantity of order one in the large $N$ limit \citep{Kinney:2005ej} and, as such, it does not reproduce the entropy which grows with powers of $N$. As a difference with the twisted index, already in the large $N$ limit, there is a large cancellation between bosonic and fermionic supersymmetric states  and $\cZ_{\rm index}(\Delta_a, \omega_i)\ne \cZ(\Delta_a, \omega_i)$.  All these (partial) results have stood for long time as puzzles about supersymmetric electrically charged   rotating black holes.  

However, the entropy functionals for  many Kerr--Newman black holes in different dimensions can be expressed in terms of quantities with a clear field theory interpretation  \citep{Hosseini:2017mds,Hosseini:2018dob,Choi:2018fdc}, thus suggesting that the entropy can be always reproduced by a field theory computation. These entropy functionals also suggest complex value for the chemical potentials $\Delta_a$ and $\omega_i$. As stressed in \cite{Choi:2018hmj,Benini:2018ywd},  the computation in \cite{Kinney:2005ej} is valid only for \emph{real fugacities} and the introduction of phases in the fugacities can obstruct the cancellation   at large $N$  and lead to an enhancement of the entropy. This is also in the spirit of the $\cI$-extremization principle discussed in Sect.~\ref{subsec:Iextremization}. Recent results for AdS$_5$ and other dimensions, started with the work of  \cite{Cabo-Bizet:2018ehj,Choi:2018hmj,Benini:2018ywd},  confirm this point of view  and lead to various derivations of the entropy using the index, 
as we discuss in this section. 

\subsection{The entropy functional for electrically charged rotating black holes
}\label{sec:entropyfunct}

For many Kerr--Newman black holes, the entropy, as a function of electric charges $q_a$ and angular momenta $j_i$, can be written as a Legendre transform
\be\label{entropyind2} S(q_a,j_i) = \log \cZ(\Delta_a, \omega_i) - i (\Delta_a q_a +\omega_i j_i) \, \ee
of a  quantity  $\log \cZ(\Delta_a, \omega_i)$   related to anomalies or free energies of the  dual CFT. This was derived in \cite{Hosseini:2017mds} for five-dimensional black holes and generalized to other dimensions in \cite{Hosseini:2018dob,Choi:2018fdc}.\footnote{A proposal for non BPS black holes can be found in \cite{Larsen:2019oll}.} 
We consider first the example of AdS$_5\times S^5$ and we come back later to the general case. 
\subsubsection{The entropy functional for Kerr--Newman black holes in AdS$_5\times S^5$
}\label{sec:entropyfunct1}

As discussed in Sect.~\ref{subsubsec:electric}, the Kerr--Newman black holes in AdS$_5\times S^5$ depend on three charges $q_1,q_2,q_3$ associated with $U(1)^3\subset SO(6)$, the Cartan subgroup of the isometry of $S^5$,
and two angular momenta $j_1$ and $j_2$ in AdS$_5$. Supersymmetry requires a non-linear constraint among these conserved charges, whose form we discuss below, and leaves a four-dimensional family of BPS black holes \citep{Gutowski:2004yv,Gutowski:2004ez,Chong:2005da,Chong:2005hr,Kunduri:2006ek}.
The entropy can be compactly written as \citep{Kim:2006he}
\bea\label{KN5} S_{\text{BH}}(q_a,j_i)= 2 \pi \sqrt{q_1 q_2 + q_1 q_3 + q_2 q_3 -\frac{\pi}{4 G_{{\text{N}}}^{(5)} g^3}(j_1+j_2)} \, ,\eea
where $G_{{\text{N}}}^{(5)}$ is the five-dimensional Newton constant and $g$ the gauge coupling of the five-dimensional effective supergravity. Holography relates these quantities to the number of colors of the dual field theory, ${\cal N}=4$ SYM in four dimensions,
through $ N^2= \pi/(2 G_{{\text{N}}}^{(5)} g^3)$. We see that black holes with charges and angular momenta of order $O(N^2)$ have an entropy of order $O(N^2)$.

Quite remarkably, the entropy \eqref{KN5} can be  written as the Legendre transform of a very simple quantity  \citep{Hosseini:2017mds}
\bea\label{entropyind0} S_{\text{BH}}(q_a,j_i) =  -  i \frac{ N^2}{2}  \frac{\Delta_1\Delta_2\Delta_3}{\omega_1\omega_2} -i \left(\sum_{a=1}^3\Delta_a q_a +\sum_{i=1}^2\omega_i j_i\right ) \Big |_{{\rm extremum} \, \bar \Delta_a,\bar \omega_i} \, ,\eea
where the chemical potentials are constrained by
\bea\label{con1} \Delta_1+\Delta_2+\Delta_3 -\omega_1-\omega_2=\pm 2\pi \, .\eea
This constraint resembles the analogous one, \eqref{Vsolution sum 2pi -- end},  for black holes in ABJM.
The quantity  in \eqref{entropyind0} is  extremized for \emph{complex values} of the chemical potentials $\Delta_a$ and $\omega_i$. However, quite remarkably, the on-shell value \eqref{entropyind0} becomes real once we impose the  non-linear constraint among charges imposed by supersymmetry. In fact, a simple way of characterize the constraint on charges is to identify it with the imaginary part of entropy functional  \eqref{entropyind0} at its extremum.  The fact that the critical value for $\Delta_a$ and $\omega_i$ are complex will also play an important for the field theory interpretation of the result. The two choice of signs in \eqref{con1} lead to the same final result. Formally, this is due to the fact that \eqref{entropyind0} is a holomorphic homogeneous function of the chemical potentials of degree one.\footnote{Consider the two functionals $S_{\pm} =S_{\text{BH}} - i \Lambda (\sum_{a=1}^3 \Delta_a -\sum_{i=1}^2 \omega_i \mp 2\pi)$, where the constraint is enforced by the Lagrange multiplier $\Lambda$. Given the homogeneity of  $S_{\text{BH}} $, the extremal value is $S_{\pm}=\pm 2\pi i \Lambda$. Since $q_a$ and $j_i$ are real, it is immediate to see that, if $(\Delta_a,\omega_i,\Lambda)$ is an extremum of  $S_+$, then  $(-\bar\Delta_a,-\bar\omega_i, \bar \Lambda)$ is an extremum of $S_-$ and the extremal values are related by $S_+=\bar S_-$. Therefore the extremization with different constraints \eqref{con1} give the same results for the real part of the entropy.}

The quantity 
\bea\label{EE} \log \cZ(\Delta_a, \omega_i) =-  i \frac{ N^2}{2}  \frac{\Delta_1\Delta_2\Delta_3}{\omega_1\omega_2} \, \eea
must be interpreted as a grand canonical partition function, and, as discussed in \ref{subsec:attractor}, should be related to the on-shell action  of the Euclidean black hole. 
This has been proved in \cite{Cabo-Bizet:2018ehj} by taking the  zero-temperature limit of the on-shell action of a family of supersymmetric complexified  non-extremal Euclidean solutions. In this approach, one can also derive the  complex value for the chemical potentials at the saddle point.\footnote{The chemical potentials are extracted following the logic discussed at the end of Sect.~\ref{grancan}.} The  constraints  $\sum_a \Delta_a - \sum_i \omega_i=\pm 2 \pi$ arise due to regularity conditions to be  imposed on  the Killing spinors. 

\subsubsection{The entropy functional for Kerr--Newman black holes in diverse dimensions
}\label{sec:entropyfunct2}

The expression for the entropy functionals for Kerr--Newman black holes in diverse dimensions
is schematically given in Table~\ref{RotEntropy}.

\begin{table}[h!!!!]
\caption{Entropy functionals for  electrically charged  rotating black holes. For simplicity of notations, we opted for  a uniform notation for all dimensions, involving some sign redefinitions in comparison to \cite{Hosseini:2017mds,Hosseini:2018dob,Choi:2018fdc}, to which we refer for more precise statements.}
\label{RotEntropy}
\vskip 0.5truecm
\centering
\begin{tabular}{lc l  c | c |} \toprule \toprule \noalign{\smallskip}
 \hskip -0.4truecm AdS$_4\times S^7$ & \hskip -0.6truecm $F(\Delta_a)=\sqrt{\Delta_1\Delta_2\Delta_3\Delta_4}$ &  \hskip -0.2truecm $\log \cZ(\Delta_a, \omega_i)= -\frac{4\sqrt{2}N^{3/2}}{3}\frac{\sqrt{\Delta_1\Delta_2\Delta_3\Delta_4}}{\omega_1}$ 
\\  \noalign{\smallskip}
& \hskip -0.6truecm $\Delta_1+\Delta_2+\Delta_3+\Delta_4=2$ & \hskip -0.2truecm  $\Delta_1+\Delta_2+\Delta_3 +\Delta_4-\omega_1=2\pi$ \\ \noalign{\smallskip}
 \toprule \noalign{\smallskip}
  \hskip -0.4truecm AdS$_5\times S^5$  & \hskip -0.6truecm $F(\Delta_a)=\Delta_1\Delta_2\Delta_3$ &  \hskip -0.3truecm $\log \cZ(\Delta_a, \omega_i)=  -  i \frac{ N^2}{2}  \frac{\Delta_1\Delta_2\Delta_3}{\omega_1\omega_2}$ \\\noalign{\smallskip}
 & \hskip -0.6truecm $\Delta_1+\Delta_2+\Delta_3=2$ & \hskip -0.2truecm $\Delta_1+\Delta_2+\Delta_3 -\omega_1-\omega_2=2\pi$ \\\noalign{\smallskip}
 \toprule \noalign{\smallskip}
 \hskip -0.4truecm  AdS$_6\times_W S^4$  & \hskip -0.6truecm $F(\Delta_a)=(\Delta_1\Delta_2)^{3/2}$ & \hskip -0.3truecm  $\log \cZ(\Delta_a, \omega_i) \sim  N^{5/2} \frac{(\Delta_1\Delta_2)^{3/2}}{\omega_1\omega_2}$ \\\noalign{\smallskip}
 & \hskip -0.6truecm $\Delta_1+\Delta_2=2$ & \hskip -0.2truecm $\Delta_1+\Delta_2- \omega_1-\omega_2=2\pi$ \\
 \noalign{\smallskip}
 \toprule \noalign{\smallskip}
  \hskip -0.4truecm AdS$_7\times S^4$  & \hskip -0.6truecm $F(\Delta_a)=(\Delta_1\Delta_2)^{3}$ & \hskip -0.2truecm  $\log \cZ(\Delta_a, \omega_i)=   i \frac{ N^3}{24} \frac{(\Delta_1\Delta_2)^{3}}{\omega_1\omega_2\omega_3}$ \\
 \noalign{\smallskip}
 & \hskip -0.6truecm $\Delta_1+\Delta_2=2$ & \hskip -0.2truecm $\Delta_1+\Delta_2- \omega_1-\omega_2-\omega_3=2\pi$ \\
 \noalign{\smallskip}
 \toprule
 \toprule
\end{tabular}
\end{table}

 The content of the table refers to   electrically charged  rotating black holes  in each dimension that can be embedded in a maximally supersymmetric string theory or M-theory background. In addition to  the  AdS$_5\times S^5$  black holes   in type IIB, there are  analogous AdS$_4\times S^7$ and AdS$_7\times S^4$   black holes  in M-theory \citep{Chow:2007ts,Cvetic:2005zi,Chong:2004dy}. The dual field theories are well known: the  ABJM theory in three dimensions   and the $(2,0)$ theory in six dimensions. In addition, there are black holes  in the warped AdS$_6\times_W S^4$ background of massive type IIA \citep{Brandhuber:1999np,Chow:2008ip}.  The dual CFT is the ${\cal N}=1$  five-dimensional fixed point associated with D4-D8-O8 branes in type IIA found in \cite{Seiberg:1996bd}. Notice that, in six dimensions, the maximal superconformal algebra has only sixteen supercharges instead of thirty-two and the superconformal theory with such an algebra is not unique.  Among the theories with an AdS dual, the D4-D8-O8 system is somehow the simplest and most studied.  

The chemical potentials in the table refer to the isometries of the  internal manifold.\footnote{In the case of the D4-D8-O8 system, the field theory has $SU(2)_R\times SU(2)\times E_{n_f+1}$ symmetry, where $SU(2)_R \times SU(2)$ is realized by the isometry of the warped $S^4$ and $E_{n_f+1}$ by the theory on the $N_f$ physical D8 branes. We introduced a symmetric notation for the  chemical potentials  associated with $SU(2)_R \times SU(2)$ following  the notations of \cite{Hosseini:2018uzp}. The entropy functional discussed in \cite{Choi:2018fdc} refers to the case $\Delta_1=\Delta_2$.}
Notice that the chemical potentials  are always subject to a constraint. Indeed, as already discussed in Sect.~\ref{sec:lecture1}, in all dimensions,  we expect the existence of a family of supersymmetric black holes depending on the possible electric charges and spins with a constraint among them.\footnote{Supersymmetric hairy AdS$_5$ black holes  depending on all the charges has been recently found in \cite{Markeviciute:2018yal,Markeviciute:2018cqs}. Their entropy seems to be subleading compared to the Kerr--Newman black hole.} This explains the constraint among chemical potentials. The family of black holes with generic charges and spins allowed by the  constraint is  known only for AdS$_5\times S^5$ (and AdS$_4\times S^7$) and \eqref{entropyind2} has been fully checked  only in these cases \citep{Hosseini:2017mds}. In all other dimensions  the relation \eqref{entropyind2} has been checked for the solutions available in the literature  \citep{Hosseini:2018dob,Choi:2018fdc}. 

In the  second column of the table, there is  a quantity, $F(\Delta_a)$, with a clear field theory interpretation. The reader can recognize the $S^3$ free energy of ABJM in the first row   and the trial $a$-charge of ${\cal N}=4$ SYM in the second row, see  \eqref{freeS3ABJM} and  \eqref{acharge},  which are both functions of trial R-charges satisfying 
\bea \label{c1} \sum_a \Delta_a=2\, .\eea  
In general, $F(\Delta_a)$ is the sphere free energy for  odd-dimensional CFTs, and a particular combination of t'Hooft anomaly coefficients in even dimensions.\footnote{ It is also curious to observe that $F(\Delta)$ is the on-shell value of the twisted superpotential  of the CFT in three and four-dimensions, as discussed in Sects.~\ref{sec:lecture3} and \ref{sec:lecture4}, and  the on-shell Seiberg--Witten prepotential in  five- and six-dimensional computations \citep{Hosseini:2018uzp}.} In all cases, the quantity $\log \cZ(\Delta_a, \omega_i)$ can be obtained by taking the quotient  of $F(\Delta)$  by the product of all angular momentum chemical potentials and by replacing the R-charge constraint \eqref{c1}  with \bea\label{c2} \sum_a \Delta_a -\sum_i \omega_i=2 \pi\, .\eea 
This constraint is strongly reminiscent of the analogous constraint \eqref{Vsolution sum 2pi -- end} for magnetically charged black holes. Notice that $\log \cZ(\Delta_a, \omega_i)$  is a sort of equivariant generalization of $F(\Delta)$ with respect to rotations. Indeed, the expression for $\log \cZ(\Delta_a, \omega_i)$ for AdS$_5$ and AdS$_7$ can be also directly obtained by an equivariant integration of the six-dimensional and eight-dimensional anomaly polynomial for ${\cal N}=4$ SYM and the $(2,0)$ theory in six dimensions, respectively \citep{Bobev:2015kza}.\footnote{The expression for $\log \cZ(\Delta_a, \omega_i)$ for AdS$_4$ and AdS$_6$ can be instead related to a small $\epsilon$ limit of the partition functions on $\mathbb{R}^2_{\epsilon_1} \times S^1$  and $\mathbb{R}^2_{\epsilon_1}\times \mathbb{R}^2_{\epsilon_2} \times S^1$, respectively, where $\epsilon_i \propto \omega_i$ are equivariant parameters for rotations in the $\mathbb{R}^2$ planes. The sphere and twisted partitions functions in three and five dimensions can be obtained by gluing together
these basic building blocks in the spirit of \cite{Pasquetti:2011fj,Beem:2012mb, Nieri:2013yra,Gukov:2016gkn, Pasquetti:2016dyl, Nekrasov:2003vi, Bershtein:2016mxz, Kim:2013nva, Hosseini:2018uzp, Qiu:2016dyj, Festuccia:2018rew}. This point of view has been applied to the physics of black holes in \cite{Hosseini:2018uzp,Hosseini:2019iad} and \cite{Choi:2019zpz,Choi:2019dfu}.}

Various observations made for AdS$_5\times S^5$ generalize to the other dimensions. First, the quantity  in \eqref{entropyind2} is  extremized for complex values of the chemical potentials $\Delta_a$ and $\omega_i$ but  the on-shell value \eqref{entropyind2} is real, as an entropy must be. Secondly, there is a sign ambiguity in the constraint \eqref{c2} that can be replaced by  $\sum_a \Delta_a +\sum_i \omega_i=-2 \pi$ with no differences. 
Thirdly, the expression of $\log \cZ(\Delta_a, \omega_i)$  can be explicitly derived in gravity by taking the  zero-temperature limit of the on-shell action of a family of supersymmetric  Euclidean solutions \citep{Cassani:2019mms}.

Finally, we expect that \eqref{entropyind2} corresponds to the attractor mechanism in the relevant gauged supergravity. Unfortunately, the attractor mechanism in generic dimensions and, specifically, for electrically charged  rotating black holes is not known.  AdS$_5$ black holes with equal angular momenta can be dimensionally reduced to static black holes in  four dimensions and, in this case,  one can show that \eqref{entropyind2} corresponds to the attractor mechanism in four-dimensional gauged supergravity \citep{Hosseini:2017mds}. This was actually the observation that led to write  the entropy functional \eqref{entropyind2}.

\subsection{Results on the quantum field theory side}\label{sec:interpretation}
 Various field theory  derivations of the extremization principles \eqref{entropyind0} have been recently proposed for AdS$_5$.  All these results are valid in overlapping limits  and the connection between different approaches still to be understood, but all seems to indicate that the entropy is correctly accounted by the large $N$ limit of the superconformal index. Partial results in other dimensions also confirm the  content of table \ref{RotEntropy}.

Consider first the case of AdS$_5\times S^5$. The dual field theory is ${\cal N}=4$ SYM. In the language of ${\cal N}=1$ supersymmetry, it contains a vector multiplet $W_\alpha$ and three chiral multiplets $\phi_a$ subject to the superpotential \eqref{SYM:superpotential}. We introduce three R-symmetries $\cR_a$ associated with $U(1)^3\subset SO(6)$. $\cR_a$ assign charge $2$ to $\phi_a$  and zero  to the  $\phi_b$ with $b\ne a$. The exact R-symmetry is $R= (\cR_1+\cR_2+\cR_3)/3$ and the global symmetries are $q_a= (\cR_a-R)/2$, with associated  fugacities $u_a$. Only two global symmetries are independent since $\sum_{a=1}^3 q_a=0$ and, as a consequence, $\prod_{a=1}^3 u_a=1$. Defining $y_a=(pq)^{1/3} u_a$ we can write the superconformal index \eqref{supindex} as
\bea\label{BPSindex0}  I (\Delta_a, \omega_i)  = \Tr \Big |_{{\cal Q}=0} (-1)^F p^{J_1} q^{J_2} y_1^{Q_1} y_2^{Q_2} y_3^{Q_3}   = \Tr \Big |_{{\cal Q}=0} (-1)^F e^{i (\Delta_a Q_a +\omega_i J_i)} \, , \eea
with $y_a=e^{i \Delta_a}$, $p=e^{i \omega_1}$, $ q= e^{i \omega_2}$ and $Q_a=\cR_a/2$.  The fugacities are constrained by $\prod_{a=1}^3 y_a= p q$, and the index depends only on four independent parameters, as the family of BPS black holes. 

Due to  cancellations between bosonic and fermionic supersymmetric states,  the result obtained from the index can only be a lower bound on the number of BPS states. However, we may expect that, as for magnetically charged black holes,  for large $N$, the result saturates the entropy.  Most of the computations for the superconformal index  in the old literature has been performed for \emph{real fugacities}  and give results of order $O(1)$ for large $N$.  However, the extremization principle discussed above strongly suggests that we should look at the behavior of the  index  as a function of \emph{complex} chemical potentials.  

Agreement with the gravity  result \eqref{EE}
\bea\label{EE2} \log I (\Delta_a, \omega_i) =-  i \frac{ N^2}{2}  \frac{\Delta_1\Delta_2\Delta_3}{\omega_1\omega_2} \, ,\eea
with $\Delta_1+\Delta_2+\Delta_3 -\omega_1-\omega_2=\pm 2 \pi$ has been obtained analytically, up to now,  in two partially overlapping limits:

\begin{itemize}
\item
Large $N$ and equal angular momenta \citep{Benini:2018mlo,Benini:2018ywd}. In the large $N$ limit, the index has a  Stokes behavior as a functions of the chemical potentials, and it  can give a contribution to the entropy of order $O(N^2)$ along the right direction in the complex plane. 
 A crucial technical ingredient in this  approach involves  writing the superconformal index as a sum over Bethe vacua. As shown in \cite{Closset:2017zgf,Closset:2017bse,Closset:2018ghr}, the supersymmetric partition function of many three- and four-dimensional manifolds  can be expressed as a sum over two-dimensional Bethe vacua  using the formalism that we briefly discussed in Sect.~\ref{subsec:BE}.  A formula for the superconformal index was obtained in \cite{Closset:2017bse} and generalized to unequal fugacities for the angular momenta in \cite{Benini:2018mlo}. Schematically, it allows to write the index as in \eqref{sumBethevacua} 
\bea \label{BR} \cI = \sum_{x^*}\frac{Q(x^*)}{\det_{ij} (-\partial^2_{u_i u_j} W(x^*))}  \, ,\eea
where $x^*$ are the Bethe vacua of the two-dimensional theory obtained by reduction on $T^2$,  and $Q(x)$ is a suitable function whose expression can be found in \cite{Closset:2017bse,Benini:2018mlo}.\footnote{The four-manifold $S^1\times S^3$ can be considered as a torus fibration over a two-dimensional manifold using  $S^1$ and a circle inside $S^3$ for the $T^2$ fiber. For details see \cite{Closset:2017bse,Benini:2018mlo}.}   The Bethe vacua of ${\cal N}=4$ SYM on $T^2$ have been already  discussed 
in Sect.~\ref{sec:topT2} for a different purpose and are explicitly given by solutions of \eqref{BAN=4}.\footnote{As argued in \cite{ArabiArdehali:2019orz}, there can exist non-standard solutions corresponding to continua of Bethe vacua and the formula \eqref{BR} must be accordingly generalized.}  It is argued in \cite{Benini:2018ywd} that, in the large $N$ limit, 
the  particular Bethe vacuum \eqref{solell} dominates for sufficiently large charges and  reproduces   the entropy of the AdS$_5$ black holes. The same result has been reproduced by directly analysing the saddle point of the integrand \eqref{siloc} in \cite{Cabo-Bizet:2019eaf}. The extension to the case of unequal angular momenta is discussed in \cite{Benini:2020gjh}. Although it is difficult to evaluate and compare the contribution of all the Bethe vacua, one can show that there is a natural choice that leads precisely to the gravity  result \eqref{EE}.
\item
The Cardy limit, corresponding to $\omega_i \ll 1$ at fixed complex valued of $\Delta_a$ \citep{Choi:2018hmj,Choi:2018vbz}. This limit corresponds to large black holes with electric charges and angular momenta scaling as 
\bea q_a\sim \frac{1}{\omega^2} \, , \qquad  j_i \sim \frac{1}{\omega^3} \, ,\qquad \qquad  \omega_1\sim \omega_2\sim \omega \rightarrow 0\, .\eea
It is crucial that the chemical potentials are complex. As argued  in \cite{Choi:2018hmj,Choi:2018vbz}, the imaginary parts of the fugacities at the saddle point  introduce phases that optimally obstruct the cancellation between bosonic and fermionic states.  The number of states accounted by the index  in the Cardy limit correctly reproduces the entropy of large AdS$_5$ black holes and the $\omega_i\ll 1$ limit of the extremization formula  \eqref{entropyind0}. This approach has been further refined and generalized in \cite{Honda:2019cio,ArabiArdehali:2019tdm,ArabiArdehali:2019orz}.
\end{itemize}
Numerical analysis confirming the $O(N^2)$ behavior of the index for complex chemical potential  has been performed in \cite{Murthy:2020rbd,Agarwal:2020zwm}.

In all these approaches, there seems to  exist instabilities when decreasing the charges, which might  suggest the contribution of other types of black holes. Given also the recently found supersymmetric hairy black holes in AdS$_5$ \citep{Markeviciute:2018yal,Markeviciute:2018cqs},  we may expect a rich structure of the index/partition function still to be uncovered.

It was observed  in \cite{Hosseini:2017mds,Hosseini:2018dob} that the quantity \eqref{EE2} and its analogous for AdS$_7$ given in   table \ref{RotEntropy}  matches the expression for the supersymmetric Casimir energy $E_{c.e}$ quoted in the literature  with the \emph{precise} coefficient  for both ${\cal N}=4$ SYM and the $(2,0)$ theory (see for example \citealt{Bobev:2015kza}). This observation was strengthened in \cite{Cabo-Bizet:2018ehj} by considering a modified supersymmetric partition function on $S^3\times S^1$  implementing  the constraint \eqref{con1} and showing that the corresponding  supersymmetric Casimir energy $E_{c.e}$ has still the expression \eqref{EE2}.\footnote{The motivation for using this  partition function comes from holography, since  the constraint \eqref{con1} explicitly arises in the Euclidean description   of the AdS$_5$ black holes \citep{Cabo-Bizet:2018ehj}.} It is not completely clear why the supersymmetric Casimir energy, which corresponds to the energy of the vacuum of the CFT \citep{Assel:2015nca}, should be related to the entropy of the black hole. It would be intriguing if this a consequence of some modular properties of the partition function, still to be understood. 

The large $N$ holographic expectation for theories with a Sasaki--Einstein dual is \citep{Hosseini:2018dob}
\bea\label{EE3} \log I (\Delta_a, \omega_i) =-  i \frac{ 16}{27}  \frac{a(\Delta)}{\omega_1\omega_2} =-  i \frac{N^2 }{12} \sum_{a,b,c=1}^d f_{abc}   \frac{  \Delta_a \Delta_b \Delta_c}{\omega_1\omega_2} \, ,\eea
with the constraint $\sum_{a=1}^d \Delta-\omega_1-\omega_2= \pm 2\pi$, where $a(\Delta)$ is the trial central charge defined in section  \ref{sec:blackstrings} and $f_{abc}$ are  the cubic t'Hooft anomaly coefficients for a basis of $d$ independent R-charges. The coefficients $f_{abc}$ have a natural dual gravitational interpretation. They arise as intersection numbers of cycles in the internal manifold \citep{Benvenuti:2006xg}, and, from an effective field theory perspective, as Chern--Simons terms in the corresponding  gauged supergravity in five dimensions. In particular, they determine completely the structure of ${\cal N}=1$ gauged supergravity including the prepotential. 
The field theory computation of the index has been extended to other ${\cal N}=1$ superconformal theories and gives results consistent with \eqref{EE3}. In particular, the Cardy limit for a generic ${\cal N}=1$ superconformal theory has been studied 
in \cite{Kim:2019yrz,Cabo-Bizet:2019osg} with the result\footnote{The authors use a slightly modified index where $(-1)^F$ is replaced by $(-1)^R$. This replacement has the same effect as introducing complex fugacities for  the flavor symmetries. The prescription is the same introduced in \cite{Cabo-Bizet:2018ehj}  for the modified supersymmetric partition function on $S^3\times S^1$.}
\bea\label{cardygen}  \log I ( \omega_i) \underset{\omega_i\rightarrow 0}{=} \, 4 \pi^2 i \frac{3\omega_1+3\omega_2\pm 2 \pi}{ 27 \omega_1\omega_2} (3 c -5a) + 4 \pi^2 i  \frac{\omega_1+\omega_2\pm 2 \pi}{  \omega_1\omega_2} (a- c) + O(1) \nonumber \eea 
when all flavor symmetries are turned off. This formula generalize a previous result by \cite{DiPietro:2014bca} for the standard index with real fugacities. Flavor fugacities have been introduced in \cite{Amariti:2019mgp} and consistency with \eqref{EE3} in the large $N$ limit, where $c=a$,  checked for many toric models.  The large $N$ limit of the index for  equal angular momenta
has been studied in \cite{Lanir:2019abx,Cabo-Bizet:2020nkr} with results again consistent with \eqref{EE3}.\footnote{See also \cite{Lezcano:2019pae}.} The  case of unequal angular momenta is discussed in \cite{Benini:2020gjh}.

These results have been generalized and extended to other dimensions. In particular, the entropy of Kerr--Newman black holes in AdS$_4$ has been reproduced in the Cardy limit in \cite{Choi:2019zpz,Nian:2019pxj,Choi:2019dfu}, one of the method involving factorization of the partition function.
The case of AdS$_6$  and AdS$_7$ have been analysed in \cite{Choi:2019miv,Kantor:2019lfo,Nahmgoong:2019hko}. Other interesting developments can be found in \cite{Benini:2019dyp,Bobev:2019zmz,Goldstein:2019gpz}.
 
\section{Conclusions and comments}\label{sec:conclusions}

At the end of our journey, it is time to recapitulate. We have seen that, using holography,   the entropy of supersymmetric AdS black holes and black objects with large charges can be correctly accounted by the evaluation of the relevant index in the dual conformal field theory, which just enumerates the corresponding microstates.  This solves a long standing puzzle about supersymmetric  black holes in AdS and  their holographic interpretation. Many results have been obtained for most of the existing black objects in maximally supersymmetric string theory backgrounds in diverse dimensions. One can reasonably expect that the agreement will persist for the more technically involved case of black objects with arbitrary rotations and magnetic charges and of black objects in string backgrounds with reduced supersymmetry.

It is interesting to observe that, in all dimensions, a single function $F(\Delta)$ controls the entropy of most of the black holes and black objects asymptotic to a maximally supersymmetric AdS vacuum, with or without magnetic charges or rotation. For comparison, the partition function for 
Kerr--Newman black holes and magnetically charged black objects with a twist in various dimensions is reported in Table~\ref{RotEntropy} and Table~\ref{MagEntropy}, respectively.

\begin{table}[h!!!!]
  \caption{Entropy functionals for  magnetically charged  static spherically symmetric black objects with a twist. The cases of AdS$_5\times S^5$ and AdS$_7\times S^4$ correspond to black strings. The Legendre transform of $\log \cZ$ reproduces the entropy of the static black hole obtained by reduction on a circle, as discussed in Sect.~\ref{sec:lecture4}. Details and normalizations for AdS$_6$ and AdS$_7$ can be found in \cite{Hosseini:2018uzp,Crichigno:2018adf,Hosseini:2018usu,Suh:2018szn}.}
  \vskip 0.5truecm
\label{MagEntropy}
\centering
\begin{tabular}{lc l  c | c |} \toprule \toprule \noalign{\smallskip}
 \hskip -0.0truecm AdS$_4\times S^7$ & \hskip -0.0truecm $F(\Delta_a)=\sqrt{\Delta_1\Delta_2\Delta_3\Delta_4}$ &  \hskip -0.0truecm $\log \cZ= -\frac{2\sqrt{2}N^{3/2}}{3} \sum_{a=1}^4  \fp_a \frac{\partial  {\cal F} (\Delta)}{\partial \Delta_a}$ 
\\  \noalign{\smallskip}
& \hskip -0.0truecm $\Delta_1+\Delta_2+\Delta_3+\Delta_4=2$ & \hskip -0.0truecm  $\Delta_1+\Delta_2+\Delta_3 +\Delta_4=2\pi$ \\ \noalign{\smallskip}
 \toprule \noalign{\smallskip}  
  \hskip -0.0truecm AdS$_5\times S^5$  & \hskip -0.0truecm  $F(\Delta_a)=\Delta_1\Delta_2\Delta_3$ &  \hskip -0.0truecm  $\log \cZ=  -   \frac{ N^2}{2\beta} \sum_{a=1}^3  \fp_a \frac{\partial  {\cal F} (\Delta)}{\partial \Delta_a}$ \\\noalign{\smallskip}
 &  \hskip -0.6truecm $\Delta_1+\Delta_2+\Delta_3=2$ & \hskip -0.0truecm   $\Delta_1+\Delta_2+\Delta_3 =2\pi$ \\\noalign{\smallskip}
 \toprule \noalign{\smallskip} 
 \hskip -0.0truecm  AdS$_6\times_W S^4$  & \hskip -0.0truecm $F(\Delta_a)=(\Delta_1\Delta_2)^{3/2}$ & \hskip -0.0truecm  $\log \cZ \sim  N^{5/2}  \sum_{a,b =1}^2 \fp_a \tilde \fp_b\frac{\partial^2  {\cal F} (\Delta)}{\partial \Delta_a \partial \Delta_b}$ \\\noalign{\smallskip}
 & \hskip -0.0truecm  $\Delta_1+\Delta_2=2$ & \hskip -0.0truecm $\Delta_1+\Delta_2=2\pi$ \\
 \noalign{\smallskip}
 \toprule \noalign{\smallskip} 
  \hskip -0.0truecm AdS$_7\times S^4$  & \hskip -0.0truecm $F(\Delta_a)=(\Delta_1\Delta_2)^{2}$ & \hskip -0.0truecm  $\log \cZ\sim   \frac{N^3}{\beta} \sum_{a,b =1}^2 \fp_a \tilde \fp_b\frac{\partial^2  {\cal F} (\Delta)}{\partial \Delta_a \partial \Delta_b}$ \\
 \noalign{\smallskip}
 & \hskip -0.0truecm  $\Delta_1+\Delta_2=2$ & \hskip -0.0truecm  $\Delta_1+\Delta_2=2\pi$ \\
 \noalign{\smallskip}
 \toprule
 \toprule
\end{tabular}
\end{table}

 We see that the function $F(\Delta)$ determines the entropy of all such black holes. A general entropy functional built out of $F(\Delta)$ that covers all existing black holes and generalises the content of table \ref{RotEntropy} and \ref{MagEntropy} to the case of arbitrary rotations and magnetic charges has been discussed in \cite{Hosseini:2019iad} in analogy with the factorization properties of supersymmetric partition functions.

The function $F(\Delta)$ has various interpretations. In gravity, it determines the effective gauged supergravity action for the massless vectors, being the prepotential in four dimensions. In field theory, it is related to the anomalies of the dual CFT in even dimensions, and the round sphere partition function in odd dimensions. On a more technical side, there is a third interpretation in terms of the twisted superpotential of the two-dimensional theory obtained by reducing the CFT on circle or tori, as discussed at length in sections \ref{sec:lecture2} and \ref{sec:lecture4}.\footnote{For five and six-dimensional CFTs, this should be replaced with the Seiberg--Witten prepotential  \citep{Hosseini:2018uzp,Crichigno:2018adf}.} 


In these lectures we discussed black holes in the supergravity approximation, where the charges are large in units of the number of colors of the dual CFT. For asymptotically flat black holes impressive computations and precision tests have been made beyond the supergravity limit. The story for AdS black hole has just begun. 

\section*{Acknowledgements}

These notes grow out of lectures given at the CERN Winter School on Strings and Fields 2017 and the School on Supersymmetric Localization, Holography and Related Topics at ICTP in 2018, for which I heartily thank the organizers.  I would also like to thank Francesco Azzurli,  Nicolay Bobev, Marcos Crichigno, Noppadol Mekareeya, Vincent Min, Anton Nedelin, Achilleas Passias, Alessandro Tomasiello and especially Francesco Benini, Seyed Morteza Hosseini and Kiril Hristov,  for collaboration and discussions on related topics. I am partially supported by the INFN, the ERC-STG grant 637844-HBQFTNCER and the MIUR-PRIN contract 2017CC72MK003.

\bibliographystyle{ytphys}

\bibliography{refs}

\providecommand{\href}[2]{#2}\begingroup\raggedright\begin{thebibliography}{100}

\bibitem{Bekenstein:1972tm}
J.~D. Bekenstein, ``{Black holes and the second law},''
\href{http://dx.doi.org/10.1007/BF02757029}{{\em Lett. Nuovo Cim.} {\bfseries
  4} (1972) 737--740}.

\bibitem{Bardeen:1973gs}
J.~M. Bardeen, B.~Carter, and S.~W. Hawking, ``{The Four laws of black hole
  mechanics},''
\href{http://dx.doi.org/10.1007/BF01645742}{{\em Commun. Math. Phys.}
  {\bfseries 31} (1973) 161--170}.

\bibitem{Hawking:1974sw}
S.~W. Hawking, ``{Particle Creation by Black Holes},''
  \href{http://dx.doi.org/10.1007/BF02345020, 10.1007/BF01608497}{{\em Commun.
  Math. Phys.} {\bfseries 43} (1975) 199--220}.
[,167(1975)].

\bibitem{tHooft:1993dmi}
G.~{'t Hooft}, ``{Dimensional reduction in quantum gravity},'' {\em Conf.
  Proc.} {\bfseries C930308} (1993) 284--296,
\href{http://arxiv.org/abs/gr-qc/9310026}{{\ttfamily arXiv:gr-qc/9310026
  [gr-qc]}}.

\bibitem{Susskind:1994vu}
L.~Susskind, ``{The World as a hologram},''
  \href{http://dx.doi.org/10.1063/1.531249}{{\em J. Math. Phys.} {\bfseries 36}
  (1995) 6377--6396},
\href{http://arxiv.org/abs/hep-th/9409089}{{\ttfamily arXiv:hep-th/9409089
  [hep-th]}}.

\bibitem{Maldacena:1997re}
J.~M. Maldacena, ``{The Large N limit of superconformal field theories and
  supergravity},'' \href{http://dx.doi.org/10.1023/A:1026654312961,
  10.4310/ATMP.1998.v2.n2.a1}{{\em Int. J. Theor. Phys.} {\bfseries 38} (1999)
  1113--1133}, \href{http://arxiv.org/abs/hep-th/9711200}{{\ttfamily
  arXiv:hep-th/9711200 [hep-th]}}.
[Adv. Theor. Math. Phys.2,231(1998)].

\bibitem{Strominger:1996sh}
A.~Strominger and C.~Vafa, ``{Microscopic origin of the Bekenstein-Hawking
  entropy},'' \href{http://dx.doi.org/10.1016/0370-2693(96)00345-0}{{\em Phys.
  Lett. B} {\bfseries 379} (1996) 99--104},
\href{http://arxiv.org/abs/hep-th/9601029}{{\ttfamily arXiv:hep-th/9601029
  [hep-th]}}.

\bibitem{Nekrasov:2002qd}
N.~A. Nekrasov, ``{Seiberg-Witten prepotential from instanton counting},''
  \href{http://dx.doi.org/10.4310/ATMP.2003.v7.n5.a4}{{\em Adv. Theor. Math.
  Phys.} {\bfseries 7} no.~5, (2003) 831--864},
\href{http://arxiv.org/abs/hep-th/0206161}{{\ttfamily arXiv:hep-th/0206161
  [hep-th]}}.

\bibitem{Pestun:2007rz}
V.~Pestun, ``{Localization of gauge theory on a four-sphere and supersymmetric
  Wilson loops},'' \href{http://dx.doi.org/10.1007/s00220-012-1485-0}{{\em
  Commun. Math. Phys.} {\bfseries 313} (2012) 71--129},
\href{http://arxiv.org/abs/0712.2824}{{\ttfamily arXiv:0712.2824 [hep-th]}}.

\bibitem{Benini:2015eyy}
F.~Benini, K.~Hristov, and A.~Zaffaroni, ``{Black hole microstates in AdS$_{4}$
  from supersymmetric localization},''
  \href{http://dx.doi.org/10.1007/JHEP05(2016)054}{{\em JHEP} {\bfseries 05}
  (2016) 054},
\href{http://arxiv.org/abs/1511.04085}{{\ttfamily arXiv:1511.04085 [hep-th]}}.

\bibitem{Aharony:2008ug}
O.~Aharony, O.~Bergman, D.~L. Jafferis, and J.~Maldacena, ``{$\mathcal{N}=6$
  superconformal {Chern-Simons}-matter theories, M2-branes and their gravity
  duals},'' \href{http://dx.doi.org/10.1088/1126-6708/2008/10/091}{{\em JHEP}
  {\bfseries 10} (2008) 091},
\href{http://arxiv.org/abs/0806.1218}{{\ttfamily arXiv:0806.1218 [hep-th]}}.

\bibitem{Cabo-Bizet:2018ehj}
A.~Cabo-Bizet, D.~Cassani, D.~Martelli, and S.~Murthy, ``{Microscopic origin of
  the Bekenstein-Hawking entropy of supersymmetric AdS$_{5}$ black holes},''
  \href{http://dx.doi.org/10.1007/JHEP10(2019)062}{{\em JHEP} {\bfseries 10}
  (2019) 062},
\href{http://arxiv.org/abs/1810.11442}{{\ttfamily arXiv:1810.11442 [hep-th]}}.

\bibitem{Choi:2018hmj}
S.~Choi, J.~Kim, S.~Kim, and J.~Nahmgoong, ``{Large AdS black holes from
  QFT},'' {\em arXiv e-prints} (2018) ,
\href{http://arxiv.org/abs/1810.12067}{{\ttfamily arXiv:1810.12067 [hep-th]}}.

\bibitem{Benini:2018ywd}
F.~Benini and P.~Milan, ``{Black holes in 4d $\mathcal{N}=4$
  Super-Yang-Mills},'' \href{http://dx.doi.org/10.1103/PhysRevX.10.021037}{{\em
  Phys. Rev. X} {\bfseries 10} no.~2, (2020) 021037},
  \href{http://arxiv.org/abs/1812.09613}{{\ttfamily arXiv:1812.09613
  [hep-th]}}.

\bibitem{Hosseini:2016cyf}
S.~M. Hosseini, A.~Nedelin, and A.~Zaffaroni, ``{The Cardy limit of the
  topologically twisted index and black strings in AdS$_{5}$},''
  \href{http://dx.doi.org/10.1007/JHEP04(2017)014}{{\em JHEP} {\bfseries 04}
  (2017) 014},
\href{http://arxiv.org/abs/1611.09374}{{\ttfamily arXiv:1611.09374 [hep-th]}}.

\bibitem{Nekrasov:2009uh}
N.~A. Nekrasov and S.~L. Shatashvili, ``{Supersymmetric vacua and Bethe
  ansatz},'' \href{http://dx.doi.org/10.1016/j.nuclphysbps.2009.07.047}{{\em
  Nucl. Phys. Proc. Suppl.} {\bfseries 192-193} (2009) 91--112},
\href{http://arxiv.org/abs/0901.4744}{{\ttfamily arXiv:0901.4744 [hep-th]}}.

\bibitem{Marino:2011nm}
M.~Marino, ``{Lectures on localization and matrix models in supersymmetric
  {Chern-Simons}-matter theories},''
  \href{http://dx.doi.org/10.1088/1751-8113/44/46/463001}{{\em J. Phys. A}
  {\bfseries 44} (2011) 463001},
\href{http://arxiv.org/abs/1104.0783}{{\ttfamily arXiv:1104.0783 [hep-th]}}.

\bibitem{Pestun:2016jze}
V.~Pestun and M.~Zabzine, ``{Introduction to localization in quantum field
  theory},'' \href{http://dx.doi.org/10.1088/1751-8121/aa5704}{{\em J. Phys.}
  {\bfseries A50} no.~44, (2017) 443001},
\href{http://arxiv.org/abs/1608.02953}{{\ttfamily arXiv:1608.02953 [hep-th]}}.

\bibitem{Gutowski:2004yv}
J.~B. Gutowski and H.~S. Reall, ``{General supersymmetric AdS(5) black
  holes},'' \href{http://dx.doi.org/10.1088/1126-6708/2004/04/048}{{\em JHEP}
  {\bfseries 04} (2004) 048},
\href{http://arxiv.org/abs/hep-th/0401129}{{\ttfamily arXiv:hep-th/0401129
  [hep-th]}}.

\bibitem{Gutowski:2004ez}
J.~B. Gutowski and H.~S. Reall, ``{Supersymmetric AdS(5) black holes},''
  \href{http://dx.doi.org/10.1088/1126-6708/2004/02/006}{{\em JHEP} {\bfseries
  02} (2004) 006},
\href{http://arxiv.org/abs/hep-th/0401042}{{\ttfamily arXiv:hep-th/0401042
  [hep-th]}}.

\bibitem{Chong:2005da}
Z.~W. Chong, M.~Cveti\v{c}, H.~Lu, and C.~N. Pope, ``{Five-dimensional gauged
  supergravity black holes with independent rotation parameters},''
  \href{http://dx.doi.org/10.1103/PhysRevD.72.041901}{{\em Phys. Rev. D}
  {\bfseries 72} (2005) 041901},
\href{http://arxiv.org/abs/hep-th/0505112}{{\ttfamily arXiv:hep-th/0505112
  [hep-th]}}.

\bibitem{Chong:2005hr}
Z.-W. Chong, M.~Cveti\v{c}, H.~Lu, and C.~N. Pope, ``{General non-extremal
  rotating black holes in minimal five-dimensional gauged supergravity},''
  \href{http://dx.doi.org/10.1103/PhysRevLett.95.161301}{{\em Phys. Rev. Lett.}
  {\bfseries 95} (2005) 161301},
\href{http://arxiv.org/abs/hep-th/0506029}{{\ttfamily arXiv:hep-th/0506029
  [hep-th]}}.

\bibitem{Kunduri:2006ek}
H.~K. Kunduri, J.~Lucietti, and H.~S. Reall, ``{Supersymmetric multi-charge
  AdS(5) black holes},''
  \href{http://dx.doi.org/10.1088/1126-6708/2006/04/036}{{\em JHEP} {\bfseries
  04} (2006) 036},
\href{http://arxiv.org/abs/hep-th/0601156}{{\ttfamily arXiv:hep-th/0601156
  [hep-th]}}.

\bibitem{Markeviciute:2018yal}
J.~Markeviciute and J.~E. Santos, ``{Evidence for the existence of a novel
  class of supersymmetric black holes with AdS$_5\times$S$^5$ asymptotics},''
  \href{http://dx.doi.org/10.1088/1361-6382/aaf680}{{\em Class. Quant. Grav.}
  {\bfseries 36} no.~2, (2019) 02LT01},
\href{http://arxiv.org/abs/1806.01849}{{\ttfamily arXiv:1806.01849 [hep-th]}}.

\bibitem{Markeviciute:2018cqs}
J.~Markeviciute, ``{Rotating Hairy Black Holes in AdS$_5\times$S$^5$},''
  \href{http://dx.doi.org/10.1007/JHEP03(2019)110}{{\em JHEP} {\bfseries 03}
  (2019) 110},
\href{http://arxiv.org/abs/1809.04084}{{\ttfamily arXiv:1809.04084 [hep-th]}}.

\bibitem{Chong:2004dy}
Z.~W. Chong, M.~Cveti\v{c}, H.~Lu, and C.~N. Pope, ``{Non-extremal charged
  rotating black holes in seven-dimensional gauged supergravity},''
  \href{http://dx.doi.org/10.1016/j.physletb.2005.07.054}{{\em Phys. Lett. B}
  {\bfseries 626} (2005) 215--222},
\href{http://arxiv.org/abs/hep-th/0412094}{{\ttfamily arXiv:hep-th/0412094
  [hep-th]}}.

\bibitem{Cvetic:2005zi}
M.~Cveti\v{c}, G.~W. Gibbons, H.~Lu, and C.~N. Pope, ``{Rotating black holes in
  gauged supergravities: Thermodynamics, supersymmetric limits, topological
  solitons and time machines},'' {\em arXiv e-prints} (2005) ,
\href{http://arxiv.org/abs/hep-th/0504080}{{\ttfamily arXiv:hep-th/0504080
  [hep-th]}}.

\bibitem{Chow:2007ts}
D.~D.~K. Chow, ``{Equal charge black holes and seven dimensional gauged
  supergravity},'' \href{http://dx.doi.org/10.1088/0264-9381/25/17/175010}{{\em
  Class. Quant. Grav.} {\bfseries 25} (2008) 175010},
\href{http://arxiv.org/abs/0711.1975}{{\ttfamily arXiv:0711.1975 [hep-th]}}.

\bibitem{Chow:2008ip}
D.~D.~K. Chow, ``{Charged rotating black holes in six-dimensional gauged
  supergravity},'' \href{http://dx.doi.org/10.1088/0264-9381/27/6/065004}{{\em
  Class. Quant. Grav.} {\bfseries 27} (2010) 065004},
\href{http://arxiv.org/abs/0808.2728}{{\ttfamily arXiv:0808.2728 [hep-th]}}.

\bibitem{Hristov:2019mqp}
K.~Hristov, S.~Katmadas, and C.~Toldo, ``{Matter-coupled supersymmetric
  Kerr-Newman-AdS$_4$ black holes},''
  \href{http://dx.doi.org/10.1103/PhysRevD.100.066016}{{\em Phys. Rev. D}
  {\bfseries 100} no.~6, (2019) 066016},
\href{http://arxiv.org/abs/1907.05192}{{\ttfamily arXiv:1907.05192 [hep-th]}}.

\bibitem{Witten:1995zh}
E.~Witten, ``{Some comments on string dynamics},'' in {\em {Future perspectives
  in string theory. Proceedings, Conference, Strings'95, Los Angeles, USA,
  March 13-18, 1995}}, pp.~501--523.
\newblock 1995.
\newblock
\href{http://arxiv.org/abs/hep-th/9507121}{{\ttfamily arXiv:hep-th/9507121
  [hep-th]}}.
\newblock

\bibitem{Klebanov:1998hh}
I.~R. Klebanov and E.~Witten, ``{Superconformal field theory on three-branes at
  a Calabi-Yau singularity},''
  \href{http://dx.doi.org/10.1016/S0550-3213(98)00654-3}{{\em Nucl. Phys. B}
  {\bfseries 536} (1998) 199--218},
\href{http://arxiv.org/abs/hep-th/9807080}{{\ttfamily arXiv:hep-th/9807080
  [hep-th]}}.

\bibitem{Cacciatori:2009iz}
S.~L. Cacciatori and D.~Klemm, ``{Supersymmetric AdS(4) black holes and
  attractors},'' \href{http://dx.doi.org/10.1007/JHEP01(2010)085}{{\em JHEP}
  {\bfseries 01} (2010) 085},
\href{http://arxiv.org/abs/0911.4926}{{\ttfamily arXiv:0911.4926 [hep-th]}}.

\bibitem{DallAgata:2010ejj}
G.~Dall'Agata and A.~Gnecchi, ``{Flow equations and attractors for black holes
  in N = 2 U(1) gauged supergravity},''
  \href{http://dx.doi.org/10.1007/JHEP03(2011)037}{{\em JHEP} {\bfseries 03}
  (2011) 037},
\href{http://arxiv.org/abs/1012.3756}{{\ttfamily arXiv:1012.3756 [hep-th]}}.

\bibitem{Hristov:2010ri}
K.~Hristov and S.~Vandoren, ``{Static supersymmetric black holes in AdS4 with
  spherical symmetry},'' \href{http://dx.doi.org/10.1007/JHEP04(2011)047}{{\em
  JHEP} {\bfseries 04} (2011) 047},
\href{http://arxiv.org/abs/1012.4314}{{\ttfamily arXiv:1012.4314 [hep-th]}}.

\bibitem{Katmadas:2014faa}
S.~Katmadas, ``{Static BPS black holes in U(1) gauged supergravity},''
  \href{http://dx.doi.org/10.1007/JHEP09(2014)027}{{\em JHEP} {\bfseries 09}
  (2014) 027},
\href{http://arxiv.org/abs/1405.4901}{{\ttfamily arXiv:1405.4901 [hep-th]}}.

\bibitem{Halmagyi:2014qza}
N.~Halmagyi, ``{Static BPS black holes in AdS$_{4}$ with general dyonic
  charges},'' \href{http://dx.doi.org/10.1007/JHEP03(2015)032}{{\em JHEP}
  {\bfseries 03} (2015) 032},
\href{http://arxiv.org/abs/1408.2831}{{\ttfamily arXiv:1408.2831 [hep-th]}}.

\bibitem{Hristov:2018spe}
K.~Hristov, S.~Katmadas, and C.~Toldo, ``{Rotating attractors and BPS black
  holes in $AdS_4$},'' \href{http://dx.doi.org/10.1007/JHEP01(2019)199}{{\em
  JHEP} {\bfseries 01} (2019) 199},
\href{http://arxiv.org/abs/1811.00292}{{\ttfamily arXiv:1811.00292 [hep-th]}}.

\bibitem{Sabra:1999ux}
W.~A. Sabra, ``{Anti-de Sitter BPS black holes in N=2 gauged supergravity},''
  \href{http://dx.doi.org/10.1016/S0370-2693(99)00564-X}{{\em Phys. Lett. B}
  {\bfseries 458} (1999) 36--42},
\href{http://arxiv.org/abs/hep-th/9903143}{{\ttfamily arXiv:hep-th/9903143
  [hep-th]}}.

\bibitem{Daniele:2019rpr}
N.~Daniele, F.~Faedo, D.~Klemm, and P.~F. Ramirez, ``{Rotating black holes in
  the FI-gauged $N=2$, $D=4$ $\overline{\mathbb{C}\text{P}}^n$ model},''
  \href{http://dx.doi.org/10.1007/JHEP03(2019)151}{{\em JHEP} {\bfseries 03}
  (2019) 151},
\href{http://arxiv.org/abs/1902.03113}{{\ttfamily arXiv:1902.03113 [hep-th]}}.

\bibitem{Romans:1991nq}
L.~J. Romans, ``{Supersymmetric, cold and lukewarm black holes in cosmological
  Einstein-Maxwell theory},''
  \href{http://dx.doi.org/10.1016/0550-3213(92)90684-4}{{\em Nucl. Phys. B}
  {\bfseries 383} (1992) 395--415},
\href{http://arxiv.org/abs/hep-th/9203018}{{\ttfamily arXiv:hep-th/9203018
  [hep-th]}}.

\bibitem{Hristov:2011ye}
K.~Hristov, C.~Toldo, and S.~Vandoren, ``{On BPS bounds in D=4 N=2 gauged
  supergravity},'' \href{http://dx.doi.org/10.1007/JHEP12(2011)014}{{\em JHEP}
  {\bfseries 12} (2011) 014},
\href{http://arxiv.org/abs/1110.2688}{{\ttfamily arXiv:1110.2688 [hep-th]}}.

\bibitem{Hristov:2011qr}
K.~Hristov, ``{On BPS Bounds in D=4 N=2 Gauged Supergravity II: General Matter
  couplings and Black Hole Masses},''
  \href{http://dx.doi.org/10.1007/JHEP03(2012)095}{{\em JHEP} {\bfseries 03}
  (2012) 095},
\href{http://arxiv.org/abs/1112.4289}{{\ttfamily arXiv:1112.4289 [hep-th]}}.

\bibitem{Klebanov:1999tb}
I.~R. Klebanov and E.~Witten, ``{AdS / CFT correspondence and symmetry
  breaking},'' \href{http://dx.doi.org/10.1016/S0550-3213(99)00387-9}{{\em
  Nucl. Phys. B} {\bfseries 556} (1999) 89--114},
\href{http://arxiv.org/abs/hep-th/9905104}{{\ttfamily arXiv:hep-th/9905104
  [hep-th]}}.

\bibitem{Witten:1988ze}
E.~Witten, ``{Topological Quantum Field Theory},''
\href{http://dx.doi.org/10.1007/BF01223371}{{\em Commun. Math. Phys.}
  {\bfseries 117} (1988) 353}.

\bibitem{Nekrasov:2003rj}
N.~Nekrasov and A.~Okounkov, ``{Seiberg-Witten theory and random partitions},''
  \href{http://dx.doi.org/10.1007/0-8176-4467-9_15}{{\em Prog. Math.}
  {\bfseries 244} (2006) 525--596},
\href{http://arxiv.org/abs/hep-th/0306238}{{\ttfamily arXiv:hep-th/0306238
  [hep-th]}}.

\bibitem{Silva:2006xv}
P.~J. Silva, ``{Thermodynamics at the BPS bound for Black Holes in AdS},''
  \href{http://dx.doi.org/10.1088/1126-6708/2006/10/022}{{\em JHEP} {\bfseries
  10} (2006) 022},
\href{http://arxiv.org/abs/hep-th/0607056}{{\ttfamily arXiv:hep-th/0607056
  [hep-th]}}.

\bibitem{Ooguri:2004zv}
H.~Ooguri, A.~Strominger, and C.~Vafa, ``{Black hole attractors and the
  topological string},''
  \href{http://dx.doi.org/10.1103/PhysRevD.70.106007}{{\em Phys. Rev. D}
  {\bfseries 70} (2004) 106007},
\href{http://arxiv.org/abs/hep-th/0405146}{{\ttfamily arXiv:hep-th/0405146
  [hep-th]}}.

\bibitem{Sen:2005wa}
A.~Sen, ``{Black hole entropy function and the attractor mechanism in higher
  derivative gravity},''
  \href{http://dx.doi.org/10.1088/1126-6708/2005/09/038}{{\em JHEP} {\bfseries
  09} (2005) 038},
\href{http://arxiv.org/abs/hep-th/0506177}{{\ttfamily arXiv:hep-th/0506177
  [hep-th]}}.

\bibitem{Sen:2008vm}
A.~Sen, ``{Quantum Entropy Function from AdS(2)/CFT(1) Correspondence},''
  \href{http://dx.doi.org/10.1142/S0217751X09045893}{{\em Int. J. Mod. Phys. A}
  {\bfseries 24} (2009) 4225--4244},
\href{http://arxiv.org/abs/0809.3304}{{\ttfamily arXiv:0809.3304 [hep-th]}}.

\bibitem{Sen:2009vz}
A.~Sen, ``{Arithmetic of Quantum Entropy Function},''
  \href{http://dx.doi.org/10.1088/1126-6708/2009/08/068}{{\em JHEP} {\bfseries
  08} (2009) 068},
\href{http://arxiv.org/abs/0903.1477}{{\ttfamily arXiv:0903.1477 [hep-th]}}.

\bibitem{Kinney:2005ej}
J.~Kinney, J.~M. Maldacena, S.~Minwalla, and S.~Raju, ``{An Index for 4
  dimensional super conformal theories},''
  \href{http://dx.doi.org/10.1007/s00220-007-0258-7}{{\em Commun. Math. Phys.}
  {\bfseries 275} (2007) 209--254},
\href{http://arxiv.org/abs/hep-th/0510251}{{\ttfamily arXiv:hep-th/0510251
  [hep-th]}}.

\bibitem{Benvenuti:2006qr}
S.~Benvenuti, B.~Feng, A.~Hanany, and Y.-H. He, ``{Counting BPS Operators in
  Gauge Theories: Quivers, Syzygies and Plethystics},''
  \href{http://dx.doi.org/10.1088/1126-6708/2007/11/050}{{\em JHEP} {\bfseries
  11} (2007) 050},
\href{http://arxiv.org/abs/hep-th/0608050}{{\ttfamily arXiv:hep-th/0608050
  [hep-th]}}.

\bibitem{Witten:1982df}
E.~Witten, ``{Constraints on Supersymmetry Breaking},''
\href{http://dx.doi.org/10.1016/0550-3213(82)90071-2}{{\em Nucl. Phys. B}
  {\bfseries 202} (1982) 253}.

\bibitem{Dabholkar:2010rm}
A.~Dabholkar, J.~Gomes, S.~Murthy, and A.~Sen, ``{Supersymmetric Index from
  Black Hole Entropy},'' \href{http://dx.doi.org/10.1007/JHEP04(2011)034}{{\em
  JHEP} {\bfseries 04} (2011) 034},
\href{http://arxiv.org/abs/1009.3226}{{\ttfamily arXiv:1009.3226 [hep-th]}}.

\bibitem{Ferrara:1996dd}
S.~Ferrara and R.~Kallosh, ``{Supersymmetry and attractors},''
  \href{http://dx.doi.org/10.1103/PhysRevD.54.1514}{{\em Phys. Rev. D}
  {\bfseries 54} (1996) 1514--1524},
\href{http://arxiv.org/abs/hep-th/9602136}{{\ttfamily arXiv:hep-th/9602136
  [hep-th]}}.

\bibitem{Ferrara:1995ih}
S.~Ferrara, R.~Kallosh, and A.~Strominger, ``{N=2 extremal black holes},''
  \href{http://dx.doi.org/10.1103/PhysRevD.52.R5412}{{\em Phys. Rev. D}
  {\bfseries 52} (1995) R5412--R5416},
\href{http://arxiv.org/abs/hep-th/9508072}{{\ttfamily arXiv:hep-th/9508072
  [hep-th]}}.

\bibitem{Gibbons:1976ue}
G.~W. Gibbons and S.~W. Hawking, ``{Action Integrals and Partition Functions in
  Quantum Gravity},''
\href{http://dx.doi.org/10.1103/PhysRevD.15.2752}{{\em Phys. Rev. D} {\bfseries
  15} (1977) 2752--2756}.

\bibitem{Hosseini:2017mds}
S.~M. Hosseini, K.~Hristov, and A.~Zaffaroni, ``{An extremization principle for
  the entropy of rotating BPS black holes in AdS$_{5}$},''
  \href{http://dx.doi.org/10.1007/JHEP07(2017)106}{{\em JHEP} {\bfseries 07}
  (2017) 106},
\href{http://arxiv.org/abs/1705.05383}{{\ttfamily arXiv:1705.05383 [hep-th]}}.

\bibitem{Hosseini:2018dob}
S.~M. Hosseini, K.~Hristov, and A.~Zaffaroni, ``{A note on the entropy of
  rotating BPS AdS$_7\times S^4$ black holes},''
  \href{http://dx.doi.org/10.1007/JHEP05(2018)121}{{\em JHEP} {\bfseries 05}
  (2018) 121},
\href{http://arxiv.org/abs/1803.07568}{{\ttfamily arXiv:1803.07568 [hep-th]}}.

\bibitem{Choi:2018fdc}
S.~Choi, C.~Hwang, S.~Kim, and J.~Nahmgoong, ``{Entropy Functions of BPS Black
  Holes in AdS$_{4}$ and AdS$_{6}$},''
  \href{http://dx.doi.org/10.3938/jkps.76.101}{{\em J. Korean Phys. Soc.}
  {\bfseries 76} no.~2, (2020) 101--108},
  \href{http://arxiv.org/abs/1811.02158}{{\ttfamily arXiv:1811.02158
  [hep-th]}}.

\bibitem{Cassani:2019mms}
D.~Cassani and L.~Papini, ``{The BPS limit of rotating AdS black hole
  thermodynamics},'' \href{http://dx.doi.org/10.1007/JHEP09(2019)079}{{\em
  JHEP} {\bfseries 09} (2019) 079},
\href{http://arxiv.org/abs/1906.10148}{{\ttfamily arXiv:1906.10148 [hep-th]}}.

\bibitem{Hosseini:2019iad}
S.~M. Hosseini, K.~Hristov, and A.~Zaffaroni, ``{Gluing gravitational blocks
  for AdS black holes},'' \href{http://dx.doi.org/10.1007/JHEP12(2019)168}{{\em
  JHEP} {\bfseries 12} (2019) 168},
\href{http://arxiv.org/abs/1909.10550}{{\ttfamily arXiv:1909.10550 [hep-th]}}.

\bibitem{Benini:2012cz}
F.~Benini and N.~Bobev, ``{Exact two-dimensional superconformal R-symmetry and
  c-extremization},''
  \href{http://dx.doi.org/10.1103/PhysRevLett.110.061601}{{\em Phys. Rev.
  Lett.} {\bfseries 110} no.~6, (2013) 061601},
\href{http://arxiv.org/abs/1211.4030}{{\ttfamily arXiv:1211.4030 [hep-th]}}.

\bibitem{Jafferis:2010un}
D.~L. Jafferis, ``{The Exact Superconformal R-Symmetry Extremizes Z},''
  \href{http://dx.doi.org/10.1007/JHEP05(2012)159}{{\em JHEP} {\bfseries 05}
  (2012) 159},
\href{http://arxiv.org/abs/1012.3210}{{\ttfamily arXiv:1012.3210 [hep-th]}}.

\bibitem{Intriligator:2003jj}
K.~A. Intriligator and B.~Wecht, ``{The Exact superconformal R symmetry
  maximizes a},'' \href{http://dx.doi.org/10.1016/S0550-3213(03)00459-0}{{\em
  Nucl. Phys. B} {\bfseries 667} (2003) 183--200},
\href{http://arxiv.org/abs/hep-th/0304128}{{\ttfamily arXiv:hep-th/0304128
  [hep-th]}}.

\bibitem{Benini:2016rke}
F.~Benini, K.~Hristov, and A.~Zaffaroni, ``{Exact microstate counting for
  dyonic black holes in AdS4},''
  \href{http://dx.doi.org/10.1016/j.physletb.2017.05.076}{{\em Phys. Lett. B}
  {\bfseries 771} (2017) 462--466},
\href{http://arxiv.org/abs/1608.07294}{{\ttfamily arXiv:1608.07294 [hep-th]}}.

\bibitem{Martelli:2005tp}
D.~Martelli, J.~Sparks, and S.-T. Yau, ``{The Geometric dual of a-maximisation
  for Toric Sasaki-Einstein manifolds},''
  \href{http://dx.doi.org/10.1007/s00220-006-0087-0}{{\em Commun. Math. Phys.}
  {\bfseries 268} (2006) 39--65},
\href{http://arxiv.org/abs/hep-th/0503183}{{\ttfamily arXiv:hep-th/0503183
  [hep-th]}}.

\bibitem{Couzens:2018wnk}
C.~Couzens, J.~P. Gauntlett, D.~Martelli, and J.~Sparks, ``{A geometric dual of
  $c$-extremization},'' \href{http://dx.doi.org/10.1007/JHEP01(2019)212}{{\em
  JHEP} {\bfseries 01} (2019) 212},
\href{http://arxiv.org/abs/1810.11026}{{\ttfamily arXiv:1810.11026 [hep-th]}}.

\bibitem{Gauntlett:2018dpc}
J.~P. Gauntlett, D.~Martelli, and J.~Sparks, ``{Toric geometry and the dual of
  $c$-extremization},'' \href{http://dx.doi.org/10.1007/JHEP01(2019)204}{{\em
  JHEP} {\bfseries 01} (2019) 204},
  \href{http://arxiv.org/abs/1812.05597}{{\ttfamily arXiv:1812.05597
  [hep-th]}}.

\bibitem{Gauntlett:2019roi}
J.~P. Gauntlett, D.~Martelli, and J.~Sparks, ``{Toric geometry and the dual of
  ${\cal I}$-extremization},''
  \href{http://dx.doi.org/10.1007/JHEP06(2019)140}{{\em JHEP} {\bfseries 06}
  (2019) 140},
\href{http://arxiv.org/abs/1904.04282}{{\ttfamily arXiv:1904.04282 [hep-th]}}.

\bibitem{Hosseini:2019ddy}
S.~M. Hosseini and A.~Zaffaroni, ``{Geometry of $\mathcal{I}$-extremization and
  black holes microstates},''
  \href{http://dx.doi.org/10.1007/JHEP07(2019)174}{{\em JHEP} {\bfseries 07}
  (2019) 174},
\href{http://arxiv.org/abs/1904.04269}{{\ttfamily arXiv:1904.04269 [hep-th]}}.

\bibitem{Kim:2019umc}
H.~Kim and N.~Kim, ``{Black holes with baryonic charge and
  $\mathcal{I}$-extremization},''
  \href{http://dx.doi.org/10.1007/JHEP11(2019)050}{{\em JHEP} {\bfseries 11}
  (2019) 050},
\href{http://arxiv.org/abs/1904.05344}{{\ttfamily arXiv:1904.05344 [hep-th]}}.

\bibitem{Benini:2015noa}
F.~Benini and A.~Zaffaroni, ``{A topologically twisted index for
  three-dimensional supersymmetric theories},''
  \href{http://dx.doi.org/10.1007/JHEP07(2015)127}{{\em JHEP} {\bfseries 07}
  (2015) 127},
\href{http://arxiv.org/abs/1504.03698}{{\ttfamily arXiv:1504.03698 [hep-th]}}.

\bibitem{Benini:2016hjo}
F.~Benini and A.~Zaffaroni, ``{Supersymmetric partition functions on Riemann
  surfaces},'' \href{http://dx.doi.org/10.1090/pspum/096}{{\em Proc. Symp. Pure
  Math.} {\bfseries 96} (2017) 13--46},
\href{http://arxiv.org/abs/1605.06120}{{\ttfamily arXiv:1605.06120 [hep-th]}}.

\bibitem{Okuda:2012nx}
S.~Okuda and Y.~Yoshida, ``{G/G gauged WZW model and Bethe Ansatz for the phase
  model},'' \href{http://dx.doi.org/10.1007/JHEP11(2012)146}{{\em JHEP}
  {\bfseries 11} (2012) 146},
\href{http://arxiv.org/abs/1209.3800}{{\ttfamily arXiv:1209.3800 [hep-th]}}.

\bibitem{Okuda:2013fea}
S.~Okuda and Y.~Yoshida, ``{G/G gauged WZW-matter model, Bethe Ansatz for
  q-boson model and Commutative Frobenius algebra},''
  \href{http://dx.doi.org/10.1007/JHEP03(2014)003}{{\em JHEP} {\bfseries 03}
  (2014) 003},
\href{http://arxiv.org/abs/1308.4608}{{\ttfamily arXiv:1308.4608 [hep-th]}}.

\bibitem{Nekrasov:2014xaa}
N.~A. Nekrasov and S.~L. Shatashvili, ``{Bethe/Gauge correspondence on curved
  spaces},'' \href{http://dx.doi.org/10.1007/JHEP01(2015)100}{{\em JHEP}
  {\bfseries 01} (2015) 100},
\href{http://arxiv.org/abs/1405.6046}{{\ttfamily arXiv:1405.6046 [hep-th]}}.

\bibitem{Gukov:2015sna}
S.~Gukov and D.~Pei, ``{Equivariant {Verlinde} formula from fivebranes and
  vortices},'' \href{http://dx.doi.org/10.1007/s00220-017-2931-9}{{\em Commun.
  Math. Phys.} {\bfseries 355} no.~1, (2017) 1--50},
\href{http://arxiv.org/abs/1501.01310}{{\ttfamily arXiv:1501.01310 [hep-th]}}.

\bibitem{Okuda:2015yea}
S.~Okuda and Y.~Yoshida, ``{Gauge/Bethe correspondence on $S^1 \times \Sigma_h$
  and index over moduli space},'' {\em arXiv e-prints} (2015) ,
\href{http://arxiv.org/abs/1501.03469}{{\ttfamily arXiv:1501.03469 [hep-th]}}.

\bibitem{Closset:2016arn}
C.~Closset and H.~Kim, ``{Comments on twisted indices in 3d supersymmetric
  gauge theories},'' \href{http://dx.doi.org/10.1007/JHEP08(2016)059}{{\em
  JHEP} {\bfseries 08} (2016) 059},
\href{http://arxiv.org/abs/1605.06531}{{\ttfamily arXiv:1605.06531 [hep-th]}}.

\bibitem{Gukov:2016gkn}
S.~Gukov, P.~Putrov, and C.~Vafa, ``{Fivebranes and 3-manifold homology},''
  \href{http://dx.doi.org/10.1007/JHEP07(2017)071}{{\em JHEP} {\bfseries 07}
  (2017) 071},
\href{http://arxiv.org/abs/1602.05302}{{\ttfamily arXiv:1602.05302 [hep-th]}}.

\bibitem{Closset:2017zgf}
C.~Closset, H.~Kim, and B.~Willett, ``{Supersymmetric partition functions and
  the three-dimensional A-twist},''
  \href{http://dx.doi.org/10.1007/JHEP03(2017)074}{{\em JHEP} {\bfseries 03}
  (2017) 074},
\href{http://arxiv.org/abs/1701.03171}{{\ttfamily arXiv:1701.03171 [hep-th]}}.

\bibitem{Closset:2017bse}
C.~Closset, H.~Kim, and B.~Willett, ``{$ \mathcal{N} $ = 1 supersymmetric
  indices and the four-dimensional A-model},''
  \href{http://dx.doi.org/10.1007/JHEP08(2017)090}{{\em JHEP} {\bfseries 08}
  (2017) 090},
\href{http://arxiv.org/abs/1707.05774}{{\ttfamily arXiv:1707.05774 [hep-th]}}.

\bibitem{Closset:2018ghr}
C.~Closset, H.~Kim, and B.~Willett, ``{Seifert fibering operators in 3d
  $\mathcal{N}=2$ theories},''
  \href{http://dx.doi.org/10.1007/JHEP11(2018)004}{{\em JHEP} {\bfseries 11}
  (2018) 004},
\href{http://arxiv.org/abs/1807.02328}{{\ttfamily arXiv:1807.02328 [hep-th]}}.

\bibitem{Nekrasov:2009rc}
N.~A. Nekrasov and S.~L. Shatashvili,
  \href{http://dx.doi.org/10.1142/9789814304634_0015}{``{Quantization of
  Integrable Systems and Four Dimensional Gauge Theories},''} in {\em
  {Proceedings, 16th International Congress on Mathematical Physics (ICMP09):
  Prague, Czech Republic, August 3-8, 2009}}, pp.~265--289.
\newblock 2009.
\newblock \href{http://arxiv.org/abs/0908.4052}{{\ttfamily arXiv:0908.4052
  [hep-th]}}.
\newblock
\url{https://inspirehep.net/record/829640/files/arXiv:0908.4052.pdf}.
\newblock

\bibitem{Festuccia:2011ws}
G.~Festuccia and N.~Seiberg, ``{Rigid Supersymmetric Theories in Curved
  Superspace},'' \href{http://dx.doi.org/10.1007/JHEP06(2011)114}{{\em JHEP}
  {\bfseries 06} (2011) 114},
\href{http://arxiv.org/abs/1105.0689}{{\ttfamily arXiv:1105.0689 [hep-th]}}.

\bibitem{Klare:2012gn}
C.~Klare, A.~Tomasiello, and A.~Zaffaroni, ``{Supersymmetry on Curved Spaces
  and Holography},'' \href{http://dx.doi.org/10.1007/JHEP08(2012)061}{{\em
  JHEP} {\bfseries 08} (2012) 061},
\href{http://arxiv.org/abs/1205.1062}{{\ttfamily arXiv:1205.1062 [hep-th]}}.

\bibitem{Dumitrescu:2012ha}
T.~T. Dumitrescu, G.~Festuccia, and N.~Seiberg, ``{Exploring Curved
  Superspace},'' \href{http://dx.doi.org/10.1007/JHEP08(2012)141}{{\em JHEP}
  {\bfseries 08} (2012) 141},
\href{http://arxiv.org/abs/1205.1115}{{\ttfamily arXiv:1205.1115 [hep-th]}}.

\bibitem{Closset:2012ru}
C.~Closset, T.~T. Dumitrescu, G.~Festuccia, and Z.~Komargodski,
  ``{Supersymmetric Field Theories on Three-Manifolds},''
  \href{http://dx.doi.org/10.1007/JHEP05(2013)017}{{\em JHEP} {\bfseries 05}
  (2013) 017},
\href{http://arxiv.org/abs/1212.3388}{{\ttfamily arXiv:1212.3388 [hep-th]}}.

\bibitem{Hristov:2013spa}
K.~Hristov, A.~Tomasiello, and A.~Zaffaroni, ``{Supersymmetry on
  Three-dimensional Lorentzian Curved Spaces and Black Hole Holography},''
  \href{http://dx.doi.org/10.1007/JHEP05(2013)057}{{\em JHEP} {\bfseries 05}
  (2013) 057},
\href{http://arxiv.org/abs/1302.5228}{{\ttfamily arXiv:1302.5228 [hep-th]}}.

\bibitem{Witten:1991zz}
E.~Witten, ``{Mirror manifolds and topological field theory},'' in {\em {Mirror
  Symmetry I}}, S.-T. Yau, ed., vol.~9 of {\em AMS/IP Studies in Advanced
  Mathematics}, pp.~121--160.
\newblock AMS/IP, 1991.
\newblock
\href{http://arxiv.org/abs/hep-th/9112056}{{\ttfamily arXiv:hep-th/9112056
  [hep-th]}}.
\newblock

\bibitem{Witten:1982im}
E.~Witten, ``{Supersymmetry and Morse theory},''
{\em J. Diff. Geom.} {\bfseries 17} no.~4, (1982) 661--692.

\bibitem{JeffreyKirwan}
L.~C. Jeffrey and F.~C. Kirwan, ``Localization for nonabelian group actions,''
  \href{http://dx.doi.org/10.1016/0040-9383(94)00028-J}{{\em Topology}
  {\bfseries 34} (1995) 291--327},
  \href{http://arxiv.org/abs/alg-geom/9307001}{{\ttfamily
  arXiv:alg-geom/9307001}}.

\bibitem{Benini:2013nda}
F.~Benini, R.~Eager, K.~Hori, and Y.~Tachikawa, ``{Elliptic genera of
  two-dimensional N=2 gauge theories with rank-one gauge groups},''
  \href{http://dx.doi.org/10.1007/s11005-013-0673-y}{{\em Lett. Math. Phys.}
  {\bfseries 104} (2014) 465--493},
\href{http://arxiv.org/abs/1305.0533}{{\ttfamily arXiv:1305.0533 [hep-th]}}.

\bibitem{Hori:2014tda}
K.~Hori, H.~Kim, and P.~Yi, ``{Witten Index and Wall Crossing},''
  \href{http://dx.doi.org/10.1007/JHEP01(2015)124}{{\em JHEP} {\bfseries 01}
  (2015) 124},
\href{http://arxiv.org/abs/1407.2567}{{\ttfamily arXiv:1407.2567 [hep-th]}}.

\bibitem{Closset:2015rna}
C.~Closset, S.~Cremonesi, and D.~S. Park, ``{The equivariant A-twist and gauged
  linear sigma models on the two-sphere},''
  \href{http://dx.doi.org/10.1007/JHEP06(2015)076}{{\em JHEP} {\bfseries 06}
  (2015) 076},
\href{http://arxiv.org/abs/1504.06308}{{\ttfamily arXiv:1504.06308 [hep-th]}}.

\bibitem{Aharony:1997bx}
O.~Aharony, A.~Hanany, K.~A. Intriligator, N.~Seiberg, and M.~J. Strassler,
  ``{Aspects of $N=2$ supersymmetric gauge theories in three-dimensions},''
  \href{http://dx.doi.org/10.1016/S0550-3213(97)00323-4}{{\em Nucl. Phys. B}
  {\bfseries 499} (1997) 67--99},
\href{http://arxiv.org/abs/hep-th/9703110}{{\ttfamily arXiv:hep-th/9703110
  [hep-th]}}.

\bibitem{Witten:1993yc}
E.~Witten, ``{Phases of N=2 theories in two-dimensions},''
  \href{http://dx.doi.org/10.1016/0550-3213(93)90033-L}{{\em Nucl. Phys. B}
  {\bfseries 403} (1993) 159--222},
  \href{http://arxiv.org/abs/hep-th/9301042}{{\ttfamily arXiv:hep-th/9301042
  [hep-th]}}.
[AMS/IP Stud. Adv. Math.1,143(1996)].

\bibitem{Benini:2018mlo}
F.~Benini and P.~Milan, ``{A Bethe Ansatz type formula for the superconformal
  index},'' \href{http://dx.doi.org/10.1007/s00220-019-03679-y}{{\em Commun.
  Math. Phys.} {\bfseries 376} no.~2, (2020) 1413--1440},
  \href{http://arxiv.org/abs/1811.04107}{{\ttfamily arXiv:1811.04107
  [hep-th]}}.

\bibitem{Hosseini:2017fjo}
S.~M. Hosseini, K.~Hristov, and A.~Passias, ``{Holographic microstate counting
  for AdS$_{4}$ black holes in massive IIA supergravity},''
  \href{http://dx.doi.org/10.1007/JHEP10(2017)190}{{\em JHEP} {\bfseries 10}
  (2017) 190},
\href{http://arxiv.org/abs/1707.06884}{{\ttfamily arXiv:1707.06884 [hep-th]}}.

\bibitem{Benini:2017oxt}
F.~Benini, H.~Khachatryan, and P.~Milan, ``{Black hole entropy in massive Type
  IIA},'' \href{http://dx.doi.org/10.1088/1361-6382/aa9f5b}{{\em Class. Quant.
  Grav.} {\bfseries 35} no.~3, (2018) 035004},
\href{http://arxiv.org/abs/1707.06886}{{\ttfamily arXiv:1707.06886 [hep-th]}}.

\bibitem{Bobev:2018uxk}
N.~Bobev, V.~S. Min, and K.~Pilch, ``{Mass-deformed ABJM and black holes in
  AdS$_{4}$},'' \href{http://dx.doi.org/10.1007/JHEP03(2018)050}{{\em JHEP}
  {\bfseries 03} (2018) 050},
\href{http://arxiv.org/abs/1801.03135}{{\ttfamily arXiv:1801.03135 [hep-th]}}.

\bibitem{Andrianopoli:1996cm}
L.~Andrianopoli, M.~Bertolini, A.~Ceresole, R.~D'Auria, S.~Ferrara, P.~Fre, and
  T.~Magri, ``{$N=2$ supergravity and $N=2$ super Yang-Mills theory on general
  scalar manifolds: Symplectic covariance, gaugings and the momentum map},''
  \href{http://dx.doi.org/10.1016/S0393-0440(97)00002-8}{{\em J. Geom. Phys.}
  {\bfseries 23} (1997) 111--189},
\href{http://arxiv.org/abs/hep-th/9605032}{{\ttfamily arXiv:hep-th/9605032
  [hep-th]}}.

\bibitem{Chimento:2015rra}
S.~Chimento, D.~Klemm, and N.~Petri, ``{Supersymmetric black holes and
  attractors in gauged supergravity with hypermultiplets},''
  \href{http://dx.doi.org/10.1007/JHEP06(2015)150}{{\em JHEP} {\bfseries 06}
  (2015) 150},
\href{http://arxiv.org/abs/1503.09055}{{\ttfamily arXiv:1503.09055 [hep-th]}}.

\bibitem{Hristov:2012bk}
K.~Hristov, {\em {Lessons from the Vacuum Structure of 4d N=2 Supergravity}}.
\newblock PhD thesis, Utrecht U., 2012.
\newblock
\href{http://arxiv.org/abs/1207.3830}{{\ttfamily arXiv:1207.3830 [hep-th]}}.
\newblock

\bibitem{Drukker:2010nc}
N.~Drukker, M.~Marino, and P.~Putrov, ``{From weak to strong coupling in ABJM
  theory},'' \href{http://dx.doi.org/10.1007/s00220-011-1253-6}{{\em Commun.
  Math. Phys.} {\bfseries 306} (2011) 511--563},
\href{http://arxiv.org/abs/1007.3837}{{\ttfamily arXiv:1007.3837 [hep-th]}}.

\bibitem{Klebanov:2009sg}
I.~R. Klebanov and G.~Torri, ``{M2-branes and AdS/CFT},''
  \href{http://dx.doi.org/10.1142/S0217751X10048652}{{\em Int. J. Mod. Phys. A}
  {\bfseries 25} (2010) 332--350},
\href{http://arxiv.org/abs/0909.1580}{{\ttfamily arXiv:0909.1580 [hep-th]}}.

\bibitem{Hosseini:2016tor}
S.~M. Hosseini and A.~Zaffaroni, ``{Large $N$ matrix models for 3d ${\cal N}=2$
  theories: twisted index, free energy and black holes},''
  \href{http://dx.doi.org/10.1007/JHEP08(2016)064}{{\em JHEP} {\bfseries 08}
  (2016) 064},
\href{http://arxiv.org/abs/1604.03122}{{\ttfamily arXiv:1604.03122 [hep-th]}}.

\bibitem{Herzog:2010hf}
C.~P. Herzog, I.~R. Klebanov, S.~S. Pufu, and T.~Tesileanu, ``{Multi-Matrix
  Models and Tri-Sasaki Einstein Spaces},''
  \href{http://dx.doi.org/10.1103/PhysRevD.83.046001}{{\em Phys. Rev. D}
  {\bfseries 83} (2011) 046001},
\href{http://arxiv.org/abs/1011.5487}{{\ttfamily arXiv:1011.5487 [hep-th]}}.

\bibitem{Jafferis:2011zi}
D.~L. Jafferis, I.~R. Klebanov, S.~S. Pufu, and B.~R. Safdi, ``{Towards the
  F-Theorem: N=2 Field Theories on the Three-Sphere},''
  \href{http://dx.doi.org/10.1007/JHEP06(2011)102}{{\em JHEP} {\bfseries 06}
  (2011) 102},
\href{http://arxiv.org/abs/1103.1181}{{\ttfamily arXiv:1103.1181 [hep-th]}}.

\bibitem{Jafferis:2012iv}
D.~L. Jafferis and S.~S. Pufu, ``{Exact results for five-dimensional
  superconformal field theories with gravity duals},''
  \href{http://dx.doi.org/10.1007/JHEP05(2014)032}{{\em JHEP} {\bfseries 05}
  (2014) 032},
\href{http://arxiv.org/abs/1207.4359}{{\ttfamily arXiv:1207.4359 [hep-th]}}.

\bibitem{Minahan:2013jwa}
J.~A. Minahan, A.~Nedelin, and M.~Zabzine, ``{5D super Yang-Mills theory and
  the correspondence to AdS$_7$/CFT$_6$},''
  \href{http://dx.doi.org/10.1088/1751-8113/46/35/355401}{{\em J. Phys. A}
  {\bfseries 46} (2013) 355401},
\href{http://arxiv.org/abs/1304.1016}{{\ttfamily arXiv:1304.1016 [hep-th]}}.

\bibitem{Hama:2011ea}
N.~Hama, K.~Hosomichi, and S.~Lee, ``{SUSY Gauge Theories on Squashed
  Three-Spheres},'' \href{http://dx.doi.org/10.1007/JHEP05(2011)014}{{\em JHEP}
  {\bfseries 05} (2011) 014},
\href{http://arxiv.org/abs/1102.4716}{{\ttfamily arXiv:1102.4716 [hep-th]}}.

\bibitem{Azzurli:2017kxo}
F.~Azzurli, N.~Bobev, P.~M. Crichigno, V.~S. Min, and A.~Zaffaroni, ``{A
  universal counting of black hole microstates in AdS$_{4}$},''
  \href{http://dx.doi.org/10.1007/JHEP02(2018)054}{{\em JHEP} {\bfseries 02}
  (2018) 054},
\href{http://arxiv.org/abs/1707.04257}{{\ttfamily arXiv:1707.04257 [hep-th]}}.

\bibitem{Halmagyi:2017hmw}
N.~Halmagyi and S.~Lal, ``{On the on-shell: the action of AdS$_{4}$ black
  holes},'' \href{http://dx.doi.org/10.1007/JHEP03(2018)146}{{\em JHEP}
  {\bfseries 03} (2018) 146},
\href{http://arxiv.org/abs/1710.09580}{{\ttfamily arXiv:1710.09580 [hep-th]}}.

\bibitem{Cabo-Bizet:2017xdr}
A.~Cabo-Bizet, U.~Kol, L.~A. Pando~Zayas, I.~Papadimitriou, and V.~Rathee,
  ``{Entropy functional and the holographic attractor mechanism},''
  \href{http://dx.doi.org/10.1007/JHEP05(2018)155}{{\em JHEP} {\bfseries 05}
  (2018) 155},
\href{http://arxiv.org/abs/1712.01849}{{\ttfamily arXiv:1712.01849 [hep-th]}}.

\bibitem{Bobev:2020pjk}
N.~Bobev, A.~M. Charles, and V.~S. Min, ``{Euclidean Black Saddles and AdS4
  Black Holes},'' \href{http://arxiv.org/abs/2006.01148}{{\ttfamily
  arXiv:2006.01148 [hep-th]}}.

\bibitem{Hanany:2008fj}
A.~Hanany, D.~Vegh, and A.~Zaffaroni, ``{Brane Tilings and M2 Branes},''
  \href{http://dx.doi.org/10.1088/1126-6708/2009/03/012}{{\em JHEP} {\bfseries
  03} (2009) 012},
\href{http://arxiv.org/abs/0809.1440}{{\ttfamily arXiv:0809.1440 [hep-th]}}.

\bibitem{Hanany:2008cd}
A.~Hanany and A.~Zaffaroni, ``{Tilings, {Chern-Simons} Theories and M2
  Branes},'' \href{http://dx.doi.org/10.1088/1126-6708/2008/10/111}{{\em JHEP}
  {\bfseries 10} (2008) 111},
\href{http://arxiv.org/abs/0808.1244}{{\ttfamily arXiv:0808.1244 [hep-th]}}.

\bibitem{Martelli:2008si}
D.~Martelli and J.~Sparks, ``{Moduli spaces of {Chern-Simons} quiver gauge
  theories and AdS(4)/CFT(3)},''
  \href{http://dx.doi.org/10.1103/PhysRevD.78.126005}{{\em Phys. Rev. D}
  {\bfseries 78} (2008) 126005},
\href{http://arxiv.org/abs/0808.0912}{{\ttfamily arXiv:0808.0912 [hep-th]}}.

\bibitem{Gaiotto:2009tk}
D.~Gaiotto and D.~L. Jafferis, ``{Notes on adding D6 branes wrapping RP3 in
  AdS(4) x CP3},'' \href{http://dx.doi.org/10.1007/JHEP11(2012)015}{{\em JHEP}
  {\bfseries 11} (2012) 015},
\href{http://arxiv.org/abs/0903.2175}{{\ttfamily arXiv:0903.2175 [hep-th]}}.

\bibitem{Benini:2009qs}
F.~Benini, C.~Closset, and S.~Cremonesi, ``{Chiral flavors and M2-branes at
  toric CY4 singularities},''
  \href{http://dx.doi.org/10.1007/JHEP02(2010)036}{{\em JHEP} {\bfseries 02}
  (2010) 036},
\href{http://arxiv.org/abs/0911.4127}{{\ttfamily arXiv:0911.4127 [hep-th]}}.

\bibitem{Jafferis:2008qz}
D.~L. Jafferis and A.~Tomasiello, ``{A Simple class of N=3 gauge/gravity
  duals},'' \href{http://dx.doi.org/10.1088/1126-6708/2008/10/101}{{\em JHEP}
  {\bfseries 10} (2008) 101},
\href{http://arxiv.org/abs/0808.0864}{{\ttfamily arXiv:0808.0864 [hep-th]}}.

\bibitem{Gulotta:2011vp}
D.~R. Gulotta, J.~P. Ang, and C.~P. Herzog, ``{Matrix Models for Supersymmetric
  {Chern-Simons} Theories with an ADE Classification},''
  \href{http://dx.doi.org/10.1007/JHEP01(2012)132}{{\em JHEP} {\bfseries 01}
  (2012) 132},
\href{http://arxiv.org/abs/1111.1744}{{\ttfamily arXiv:1111.1744 [hep-th]}}.

\bibitem{Crichigno:2012sk}
P.~M. Crichigno, C.~P. Herzog, and D.~Jain, ``{Free Energy of
  ${{\widehat{D}}_n}$ Quiver {Chern-Simons} Theories},''
  \href{http://dx.doi.org/10.1007/JHEP03(2013)039}{{\em JHEP} {\bfseries 03}
  (2013) 039},
\href{http://arxiv.org/abs/1211.1388}{{\ttfamily arXiv:1211.1388 [hep-th]}}.

\bibitem{Martelli:2009ga}
D.~Martelli and J.~Sparks, ``{AdS(4) / CFT(3) duals from M2-branes at
  hypersurface singularities and their deformations},''
  \href{http://dx.doi.org/10.1088/1126-6708/2009/12/017}{{\em JHEP} {\bfseries
  12} (2009) 017},
\href{http://arxiv.org/abs/0909.2036}{{\ttfamily arXiv:0909.2036 [hep-th]}}.

\bibitem{Guarino:2015jca}
A.~Guarino, D.~L. Jafferis, and O.~Varela, ``{String Theory Origin of Dyonic
  N=8 Supergravity and Its {Chern-Simons} Duals},''
  \href{http://dx.doi.org/10.1103/PhysRevLett.115.091601}{{\em Phys. Rev.
  Lett.} {\bfseries 115} no.~9, (2015) 091601},
\href{http://arxiv.org/abs/1504.08009}{{\ttfamily arXiv:1504.08009 [hep-th]}}.

\bibitem{Fluder:2015eoa}
M.~Fluder and J.~Sparks, ``{D2-brane {Chern-Simons} theories: F-maximization =
  a-maximization},'' \href{http://dx.doi.org/10.1007/JHEP01(2016)048}{{\em
  JHEP} {\bfseries 01} (2016) 048},
\href{http://arxiv.org/abs/1507.05817}{{\ttfamily arXiv:1507.05817 [hep-th]}}.

\bibitem{Hosseini:2016ume}
S.~M. Hosseini and N.~Mekareeya, ``{Large $N$ topologically twisted index:
  necklace quivers, dualities, and Sasaki-Einstein spaces},''
  \href{http://dx.doi.org/10.1007/JHEP08(2016)089}{{\em JHEP} {\bfseries 08}
  (2016) 089},
\href{http://arxiv.org/abs/1604.03397}{{\ttfamily arXiv:1604.03397 [hep-th]}}.

\bibitem{Jain:2019lqb}
D.~Jain and A.~Ray, ``{3d $\mathcal{N}=2$ $\widehat{ADE}$ {Chern-Simons}
  quivers},'' \href{http://dx.doi.org/10.1103/PhysRevD.100.046007}{{\em Phys.
  Rev. D} {\bfseries 100} no.~4, (2019) 046007},
\href{http://arxiv.org/abs/1902.10498}{{\ttfamily arXiv:1902.10498 [hep-th]}}.

\bibitem{Jain:2019euv}
D.~Jain, ``{Twisted Indices of more 3d Quivers},'' {\em arXiv e-prints} (2019)
  ,
\href{http://arxiv.org/abs/1908.03035}{{\ttfamily arXiv:1908.03035 [hep-th]}}.

\bibitem{Hosseini:2018qsx}
S.~M. Hosseini, {\em {Black hole microstates and supersymmetric localization}}.
\newblock PhD thesis, Milan Bicocca U., 2018-02.
\newblock
\href{http://arxiv.org/abs/1803.01863}{{\ttfamily arXiv:1803.01863 [hep-th]}}.
\newblock

\bibitem{Guarino:2017pkw}
A.~Guarino, ``{BPS black hole horizons from massive IIA},''
  \href{http://dx.doi.org/10.1007/JHEP08(2017)100}{{\em JHEP} {\bfseries 08}
  (2017) 100},
\href{http://arxiv.org/abs/1706.01823}{{\ttfamily arXiv:1706.01823 [hep-th]}}.

\bibitem{Benini:2015bwz}
F.~Benini, N.~Bobev, and P.~M. Crichigno, ``{Two-dimensional SCFTs from
  D3-branes},'' \href{http://dx.doi.org/10.1007/JHEP07(2016)020}{{\em JHEP}
  {\bfseries 07} (2016) 020},
\href{http://arxiv.org/abs/1511.09462}{{\ttfamily arXiv:1511.09462 [hep-th]}}.

\bibitem{Bobev:2017uzs}
N.~Bobev and P.~M. Crichigno, ``{Universal RG Flows Across Dimensions and
  Holography},'' \href{http://dx.doi.org/10.1007/JHEP12(2017)065}{{\em JHEP}
  {\bfseries 12} (2017) 065},
\href{http://arxiv.org/abs/1708.05052}{{\ttfamily arXiv:1708.05052 [hep-th]}}.

\bibitem{Liu:2017vbl}
J.~T. Liu, L.~A. Pando~Zayas, V.~Rathee, and W.~Zhao, ``{One-Loop Test of
  Quantum Black Holes in anti de Sitter Space},''
  \href{http://dx.doi.org/10.1103/PhysRevLett.120.221602}{{\em Phys. Rev.
  Lett.} {\bfseries 120} no.~22, (2018) 221602},
\href{http://arxiv.org/abs/1711.01076}{{\ttfamily arXiv:1711.01076 [hep-th]}}.

\bibitem{Liu:2017vll}
J.~T. Liu, L.~A. Pando~Zayas, V.~Rathee, and W.~Zhao, ``{Toward Microstate
  Counting Beyond Large N in Localization and the Dual One-loop Quantum
  Supergravity},'' \href{http://dx.doi.org/10.1007/JHEP01(2018)026}{{\em JHEP}
  {\bfseries 01} (2018) 026},
\href{http://arxiv.org/abs/1707.04197}{{\ttfamily arXiv:1707.04197 [hep-th]}}.

\bibitem{Jeon:2017aif}
I.~Jeon and S.~Lal, ``{Logarithmic Corrections to Entropy of Magnetically
  Charged AdS4 Black Holes},''
  \href{http://dx.doi.org/10.1016/j.physletb.2017.09.026}{{\em Phys. Lett. B}
  {\bfseries 774} (2017) 41--45},
\href{http://arxiv.org/abs/1707.04208}{{\ttfamily arXiv:1707.04208 [hep-th]}}.

\bibitem{Liu:2018bac}
J.~T. Liu, L.~A. Pando~Zayas, and S.~Zhou, ``{Subleading Microstate Counting in
  the Dual to Massive Type IIA},'' {\em arXiv e-prints} (2018) ,
\href{http://arxiv.org/abs/1808.10445}{{\ttfamily arXiv:1808.10445 [hep-th]}}.

\bibitem{PandoZayas:2019hdb}
L.~A. Pando~Zayas and Y.~Xin, ``{The Topologically Twisted Index in the {'t
  Hooft} Limit and the Dual AdS$_4$ Black Hole Entropy},''
  \href{http://dx.doi.org/10.1103/PhysRevD.100.126019}{{\em Phys. Rev. D}
  {\bfseries 100} (2019) 126019},
\href{http://arxiv.org/abs/1908.01194}{{\ttfamily arXiv:1908.01194 [hep-th]}}.

\bibitem{Dabholkar:2010uh}
A.~Dabholkar, J.~Gomes, and S.~Murthy, ``{Quantum black holes, localization and
  the topological string},''
  \href{http://dx.doi.org/10.1007/JHEP06(2011)019}{{\em JHEP} {\bfseries 06}
  (2011) 019},
\href{http://arxiv.org/abs/1012.0265}{{\ttfamily arXiv:1012.0265 [hep-th]}}.

\bibitem{Dabholkar:2011ec}
A.~Dabholkar, J.~Gomes, and S.~Murthy, ``{Localization \& Exact Holography},''
  \href{http://dx.doi.org/10.1007/JHEP04(2013)062}{{\em JHEP} {\bfseries 04}
  (2013) 062},
\href{http://arxiv.org/abs/1111.1161}{{\ttfamily arXiv:1111.1161 [hep-th]}}.

\bibitem{Iqbal:2003ds}
A.~Iqbal, N.~Nekrasov, A.~Okounkov, and C.~Vafa, ``{Quantum foam and
  topological strings},''
  \href{http://dx.doi.org/10.1088/1126-6708/2008/04/011}{{\em JHEP} {\bfseries
  04} (2008) 011}, \href{http://arxiv.org/abs/hep-th/0312022}{{\ttfamily
  arXiv:hep-th/0312022}}.

\bibitem{Hristov:2018lod}
K.~Hristov, I.~Lodato, and V.~Reys, ``{On the quantum entropy function in 4d
  gauged supergravity},'' \href{http://dx.doi.org/10.1007/JHEP07(2018)072}{{\em
  JHEP} {\bfseries 07} (2018) 072},
\href{http://arxiv.org/abs/1803.05920}{{\ttfamily arXiv:1803.05920 [hep-th]}}.

\bibitem{Hristov:2019xku}
K.~Hristov, I.~Lodato, and V.~Reys, ``{One-loop determinants for black holes in
  4d gauged supergravity},''
  \href{http://dx.doi.org/10.1007/JHEP11(2019)105}{{\em JHEP} {\bfseries 11}
  (2019) 105},
\href{http://arxiv.org/abs/1908.05696}{{\ttfamily arXiv:1908.05696 [hep-th]}}.

\bibitem{Hosseini:2018uzp}
S.~M. Hosseini, I.~Yaakov, and A.~Zaffaroni, ``{Topologically twisted indices
  in five dimensions and holography},''
  \href{http://dx.doi.org/10.1007/JHEP11(2018)119}{{\em JHEP} {\bfseries 11}
  (2018) 119},
\href{http://arxiv.org/abs/1808.06626}{{\ttfamily arXiv:1808.06626 [hep-th]}}.

\bibitem{Crichigno:2018adf}
P.~M. Crichigno, D.~Jain, and B.~Willett, ``{5d Partition Functions with A
  Twist},'' \href{http://dx.doi.org/10.1007/JHEP11(2018)058}{{\em JHEP}
  {\bfseries 11} (2018) 058},
\href{http://arxiv.org/abs/1808.06744}{{\ttfamily arXiv:1808.06744 [hep-th]}}.

\bibitem{Suh:2018tul}
M.~Suh, ``{Supersymmetric AdS$_{6}$ black holes from F(4) gauged
  supergravity},'' \href{http://dx.doi.org/10.1007/JHEP01(2019)035}{{\em JHEP}
  {\bfseries 01} (2019) 035},
\href{http://arxiv.org/abs/1809.03517}{{\ttfamily arXiv:1809.03517 [hep-th]}}.

\bibitem{Hosseini:2018usu}
S.~M. Hosseini, K.~Hristov, A.~Passias, and A.~Zaffaroni, ``{6D attractors and
  black hole microstates},''
  \href{http://dx.doi.org/10.1007/JHEP12(2018)001}{{\em JHEP} {\bfseries 2018}
  no.~12, (2018) },
\href{http://arxiv.org/abs/1809.10685}{{\ttfamily arXiv:1809.10685 [hep-th]}}.

\bibitem{Suh:2018szn}
M.~Suh, ``{Supersymmetric $AdS_6$ black holes from matter coupled $F(4)$ gauged
  supergravity},'' \href{http://dx.doi.org/10.1007/JHEP02(2019)108}{{\em JHEP}
  {\bfseries 02} (2019) 108},
\href{http://arxiv.org/abs/1810.00675}{{\ttfamily arXiv:1810.00675 [hep-th]}}.

\bibitem{Suh:2018qyv}
M.~Suh, ``{On-shell action and the Bekenstein-Hawking entropy of supersymmetric
  black holes in $AdS_6$},'' {\em arXiv e-prints} (2018) ,
\href{http://arxiv.org/abs/1812.10491}{{\ttfamily arXiv:1812.10491 [hep-th]}}.

\bibitem{Fluder:2019szh}
M.~Fluder, S.~M. Hosseini, and C.~F. Uhlemann, ``{Black hole microstate
  counting in Type IIB from 5d SCFTs},''
  \href{http://dx.doi.org/10.1007/JHEP05(2019)134}{{\em JHEP} {\bfseries 05}
  (2019) 134},
\href{http://arxiv.org/abs/1902.05074}{{\ttfamily arXiv:1902.05074 [hep-th]}}.

\bibitem{Cabo-Bizet:2017jsl}
A.~Cabo-Bizet, V.~I. Giraldo-Rivera, and L.~A. Pando~Zayas, ``{Microstate
  counting of AdS$_{4}$ hyperbolic black hole entropy via the topologically
  twisted index},'' \href{http://dx.doi.org/10.1007/JHEP08(2017)023}{{\em JHEP}
  {\bfseries 08} (2017) 023},
\href{http://arxiv.org/abs/1701.07893}{{\ttfamily arXiv:1701.07893 [hep-th]}}.

\bibitem{Toldo:2017qsh}
C.~Toldo and B.~Willett, ``{Partition functions on 3d circle bundles and their
  gravity duals},'' \href{http://dx.doi.org/10.1007/JHEP05(2018)116}{{\em JHEP}
  {\bfseries 05} (2018) 116},
\href{http://arxiv.org/abs/1712.08861}{{\ttfamily arXiv:1712.08861 [hep-th]}}.

\bibitem{Gang:2018hjd}
D.~Gang and N.~Kim, ``{Large $N$ twisted partition functions in 3d-3d
  correspondence and Holography},''
  \href{http://dx.doi.org/10.1103/PhysRevD.99.021901}{{\em Phys. Rev. D}
  {\bfseries 99} no.~2, (2019) 021901},
\href{http://arxiv.org/abs/1808.02797}{{\ttfamily arXiv:1808.02797 [hep-th]}}.

\bibitem{Gang:2019uay}
D.~Gang, N.~Kim, and L.~A. Pando~Zayas, ``{Precision Microstate Counting for
  the Entropy of Wrapped M5-branes},''
  \href{http://dx.doi.org/10.1007/JHEP03(2020)164}{{\em JHEP} {\bfseries 03}
  (2020) 164}, \href{http://arxiv.org/abs/1905.01559}{{\ttfamily
  arXiv:1905.01559 [hep-th]}}.

\bibitem{Hosseini:2019and}
S.~M. Hosseini, C.~Toldo, and I.~Yaakov, ``{Supersymmetric R\'enyi entropy and
  charged hyperbolic black holes},'' {\em arXiv e-prints} (2019) ,
\href{http://arxiv.org/abs/1912.04868}{{\ttfamily arXiv:1912.04868 [hep-th]}}.

\bibitem{Bae:2019poj}
J.-B. Bae, D.~Gang, and K.~Lee, ``{Magnetically charged AdS$_{5}$ black holes
  from class $ \mathcal{S} $ theories on hyperbolic 3-manifolds},''
  \href{http://dx.doi.org/10.1007/JHEP02(2020)158}{{\em JHEP} {\bfseries 02}
  (2020) 158}, \href{http://arxiv.org/abs/1907.03430}{{\ttfamily
  arXiv:1907.03430 [hep-th]}}.

\bibitem{Benini:2013cda}
F.~Benini and N.~Bobev, ``{Two-dimensional SCFTs from wrapped branes and
  c-extremization},'' \href{http://dx.doi.org/10.1007/JHEP06(2013)005}{{\em
  JHEP} {\bfseries 06} (2013) 005},
\href{http://arxiv.org/abs/1302.4451}{{\ttfamily arXiv:1302.4451 [hep-th]}}.

\bibitem{Brown:1986nw}
J.~D. Brown and M.~Henneaux, ``{Central charges in the canonical realization of
  asymptotic symmetries: An example from three dimensional gravity},''
\href{http://dx.doi.org/10.1007/BF01211590}{{\em Commun. Math. Phys.}
  {\bfseries 104} no.~2, (1986) 207--226}.

\bibitem{Henningson:1998gx}
M.~Henningson and K.~Skenderis, ``{The Holographic Weyl anomaly},''
  \href{http://dx.doi.org/10.1088/1126-6708/1998/07/023}{{\em JHEP} {\bfseries
  07} (1998) 023},
\href{http://arxiv.org/abs/hep-th/9806087}{{\ttfamily arXiv:hep-th/9806087
  [hep-th]}}.

\bibitem{Amariti:2017iuz}
A.~Amariti, L.~Cassia, and S.~Penati, ``{c-extremization from toric
  geometry},'' \href{http://dx.doi.org/10.1016/j.nuclphysb.2018.01.025}{{\em
  Nucl. Phys. B} {\bfseries 929} (2018) 137--170},
\href{http://arxiv.org/abs/1706.07752}{{\ttfamily arXiv:1706.07752 [hep-th]}}.

\bibitem{Hosseini:2019use}
S.~M. Hosseini and A.~Zaffaroni, ``{Proving the equivalence of
  $c$-extremization and its gravitational dual for all toric quivers},''
  \href{http://dx.doi.org/10.1007/JHEP03(2019)108}{{\em JHEP} {\bfseries 03}
  (2019) 108},
\href{http://arxiv.org/abs/1901.05977}{{\ttfamily arXiv:1901.05977 [hep-th]}}.

\bibitem{Closset:2013sxa}
C.~Closset and I.~Shamir, ``{The $\mathcal{N}=1$ Chiral Multiplet on $T^2\times
  S^2$ and Supersymmetric Localization},''
  \href{http://dx.doi.org/10.1007/JHEP03(2014)040}{{\em JHEP} {\bfseries 03}
  (2014) 040},
\href{http://arxiv.org/abs/1311.2430}{{\ttfamily arXiv:1311.2430 [hep-th]}}.

\bibitem{Hong:2018viz}
J.~Hong and J.~T. Liu, ``{The topologically twisted index of $ \mathcal{N} $ =
  4 super-Yang-Mills on T$^{2} \times S^{2}$ and the elliptic genus},''
  \href{http://dx.doi.org/10.1007/JHEP07(2018)018}{{\em JHEP} {\bfseries 07}
  (2018) 018},
\href{http://arxiv.org/abs/1804.04592}{{\ttfamily arXiv:1804.04592 [hep-th]}}.

\bibitem{ArabiArdehali:2019orz}
A.~Arabi~Ardehali, J.~Hong, and J.~T. Liu, ``{Asymptotic growth of the 4d
  $\mathcal N=4$ index and partially deconfined phases},''
  \href{http://arxiv.org/abs/1912.04169}{{\ttfamily arXiv:1912.04169
  [hep-th]}}.

\bibitem{Lezcano:2019pae}
A.~Gonz\'alez~Lezcano and L.~A. Pando~Zayas, ``{Microstate counting via Bethe
  Ansatze in the 4d $ \mathcal{N} $ = 1 superconformal index},''
  \href{http://dx.doi.org/10.1007/JHEP03(2020)088}{{\em JHEP} {\bfseries 03}
  (2020) 088}, \href{http://arxiv.org/abs/1907.12841}{{\ttfamily
  arXiv:1907.12841 [hep-th]}}.

\bibitem{Lanir:2019abx}
A.~Lanir, A.~Nedelin, and O.~Sela, ``{Black hole entropy function for toric
  theories via Bethe Ansatz},''
  \href{http://dx.doi.org/10.1007/JHEP04(2020)091}{{\em JHEP} {\bfseries 04}
  (2020) 091}, \href{http://arxiv.org/abs/1908.01737}{{\ttfamily
  arXiv:1908.01737 [hep-th]}}.

\bibitem{Cardy:1986ie}
J.~L. Cardy, ``{Operator Content of Two-Dimensional Conformally Invariant
  Theories},''
\href{http://dx.doi.org/10.1016/0550-3213(86)90552-3}{{\em Nucl. Phys. B}
  {\bfseries 270} (1986) 186--204}.

\bibitem{Hosseini:2019lkt}
S.~M. Hosseini, K.~Hristov, and A.~Zaffaroni, ``{Microstates of rotating
  AdS$_{5}$ strings},'' \href{http://dx.doi.org/10.1007/JHEP11(2019)090}{{\em
  JHEP} {\bfseries 11} (2019) 090},
\href{http://arxiv.org/abs/1909.08000}{{\ttfamily arXiv:1909.08000 [hep-th]}}.

\bibitem{Hosseini:2020vgl}
S.~M. Hosseini, K.~Hristov, Y.~Tachikawa, and A.~Zaffaroni, ``{Anomalies, Black
  strings and the charged Cardy formula},''
  \href{http://arxiv.org/abs/2006.08629}{{\ttfamily arXiv:2006.08629
  [hep-th]}}.

\bibitem{Dijkgraaf:2000fq}
R.~Dijkgraaf, J.~M. Maldacena, G.~W. Moore, and E.~P. Verlinde, ``{A Black hole
  Farey tail},'' {\em arXiv e-prints} (2000) ,
\href{http://arxiv.org/abs/hep-th/0005003}{{\ttfamily arXiv:hep-th/0005003
  [hep-th]}}.

\bibitem{Kraus:2006nb}
P.~Kraus and F.~Larsen, ``{Partition functions and elliptic genera from
  supergravity},'' \href{http://dx.doi.org/10.1088/1126-6708/2007/01/002}{{\em
  JHEP} {\bfseries 01} (2007) 002},
  \href{http://arxiv.org/abs/hep-th/0607138}{{\ttfamily arXiv:hep-th/0607138}}.

\bibitem{Manschot:2007ha}
J.~Manschot and G.~W. Moore, ``{A Modern Farey Tail},''
  \href{http://dx.doi.org/10.4310/CNTP.2010.v4.n1.a3}{{\em Commun. Num. Theor.
  Phys.} {\bfseries 4} (2010) 103--159},
\href{http://arxiv.org/abs/0712.0573}{{\ttfamily arXiv:0712.0573 [hep-th]}}.

\bibitem{Dabholkar:2012nd}
A.~Dabholkar, S.~Murthy, and D.~Zagier, ``{Quantum Black Holes, Wall Crossing,
  and Mock Modular Forms},'' {\em arXiv e-prints} (2012) ,
\href{http://arxiv.org/abs/1208.4074}{{\ttfamily arXiv:1208.4074 [hep-th]}}.

\bibitem{Banados:1992wn}
M.~Ba{\~n}ados, C.~Teitelboim, and J.~Zanelli, ``{The Black hole in
  three-dimensional space-time},''
  \href{http://dx.doi.org/10.1103/PhysRevLett.69.1849}{{\em Phys. Rev. Lett.}
  {\bfseries 69} (1992) 1849--1851},
\href{http://arxiv.org/abs/hep-th/9204099}{{\ttfamily arXiv:hep-th/9204099
  [hep-th]}}.

\bibitem{Hristov:2014eza}
K.~Hristov, ``{Dimensional reduction of BPS attractors in AdS gauged
  supergravities},'' \href{http://dx.doi.org/10.1007/JHEP12(2014)066}{{\em
  JHEP} {\bfseries 12} (2014) 066},
\href{http://arxiv.org/abs/1409.8504}{{\ttfamily arXiv:1409.8504 [hep-th]}}.

\bibitem{Romelsberger:2005eg}
C.~Romelsberger, ``{Counting chiral primaries in N = 1, d=4 superconformal
  field theories},''
  \href{http://dx.doi.org/10.1016/j.nuclphysb.2006.03.037}{{\em Nucl. Phys. B}
  {\bfseries 747} (2006) 329--353},
\href{http://arxiv.org/abs/hep-th/0510060}{{\ttfamily arXiv:hep-th/0510060
  [hep-th]}}.

\bibitem{Assel:2014paa}
B.~Assel, D.~Cassani, and D.~Martelli, ``{Localization on Hopf surfaces},''
  \href{http://dx.doi.org/10.1007/JHEP08(2014)123}{{\em JHEP} {\bfseries 08}
  (2014) 123},
\href{http://arxiv.org/abs/1405.5144}{{\ttfamily arXiv:1405.5144 [hep-th]}}.

\bibitem{Assel:2015nca}
B.~Assel, D.~Cassani, L.~Di~Pietro, Z.~Komargodski, J.~Lorenzen, and
  D.~Martelli, ``{The Casimir Energy in Curved Space and its Supersymmetric
  Counterpart},'' \href{http://dx.doi.org/10.1007/JHEP07(2015)043}{{\em JHEP}
  {\bfseries 07} (2015) 043},
\href{http://arxiv.org/abs/1503.05537}{{\ttfamily arXiv:1503.05537 [hep-th]}}.

\bibitem{Lorenzen:2014pna}
J.~Lorenzen and D.~Martelli, ``{Comments on the Casimir energy in
  supersymmetric field theories},''
  \href{http://dx.doi.org/10.1007/JHEP07(2015)001}{{\em JHEP} {\bfseries 07}
  (2015) 001},
\href{http://arxiv.org/abs/1412.7463}{{\ttfamily arXiv:1412.7463 [hep-th]}}.

\bibitem{Genolini:2016sxe}
P.~Benetti~Genolini, D.~Cassani, D.~Martelli, and J.~Sparks, ``{The holographic
  supersymmetric Casimir energy},''
  \href{http://dx.doi.org/10.1103/PhysRevD.95.021902}{{\em Phys. Rev. D}
  {\bfseries 95} no.~2, (2017) 021902},
\href{http://arxiv.org/abs/1606.02724}{{\ttfamily arXiv:1606.02724 [hep-th]}}.

\bibitem{Martelli:2015kuk}
D.~Martelli and J.~Sparks, ``{The character of the supersymmetric Casimir
  energy},'' \href{http://dx.doi.org/10.1007/JHEP08(2016)117}{{\em JHEP}
  {\bfseries 08} (2016) 117},
\href{http://arxiv.org/abs/1512.02521}{{\ttfamily arXiv:1512.02521 [hep-th]}}.

\bibitem{Closset:2019ucb}
C.~Closset, L.~Di~Pietro, and H.~Kim, ``{{'t Hooft} anomalies and the
  holomorphy of supersymmetric partition functions},''
  \href{http://dx.doi.org/10.1007/JHEP08(2019)035}{{\em JHEP} {\bfseries 08}
  (2019) 035},
\href{http://arxiv.org/abs/1905.05722}{{\ttfamily arXiv:1905.05722 [hep-th]}}.

\bibitem{Dolan:2008qi}
F.~A. Dolan and H.~Osborn, ``{Applications of the Superconformal Index for
  Protected Operators and q-Hypergeometric Identities to N=1 Dual Theories},''
  \href{http://dx.doi.org/10.1016/j.nuclphysb.2009.01.028}{{\em Nucl. Phys. B}
  {\bfseries 818} (2009) 137--178},
\href{http://arxiv.org/abs/0801.4947}{{\ttfamily arXiv:0801.4947 [hep-th]}}.

\bibitem{Felder_2000}
G.~Felder and A.~Varchenko, ``The elliptic gamma function and $sl(3,\mathbb{Z})
  \rtimes \mathbb{Z}_3$,'' \href{http://dx.doi.org/10.1006/aima.2000.1951}{{\em
  Advances in Mathematics} {\bfseries 156} no.~1, (Dec, 2000) 44 -- 76}.

\bibitem{Bhattacharya:2008zy}
J.~Bhattacharya, S.~Bhattacharyya, S.~Minwalla, and S.~Raju, ``{Indices for
  Superconformal Field Theories in 3,5 and 6 Dimensions},''
  \href{http://dx.doi.org/10.1088/1126-6708/2008/02/064}{{\em JHEP} {\bfseries
  02} (2008) 064},
\href{http://arxiv.org/abs/0801.1435}{{\ttfamily arXiv:0801.1435 [hep-th]}}.

\bibitem{Kim:2009wb}
S.~Kim, ``{The Complete superconformal index for N=6 {Chern-Simons} theory},''
  \href{http://dx.doi.org/10.1016/j.nuclphysb.2012.07.015,
  10.1016/j.nuclphysb.2009.06.025}{{\em Nucl. Phys. B} {\bfseries 821} (2009)
  241--284}, \href{http://arxiv.org/abs/0903.4172}{{\ttfamily arXiv:0903.4172
  [hep-th]}}.
[Erratum: Nucl. Phys.B864,884(2012)].

\bibitem{Imamura:2011su}
Y.~Imamura and S.~Yokoyama, ``{Index for three dimensional superconformal field
  theories with general R-charge assignments},''
  \href{http://dx.doi.org/10.1007/JHEP04(2011)007}{{\em JHEP} {\bfseries 04}
  (2011) 007},
\href{http://arxiv.org/abs/1101.0557}{{\ttfamily arXiv:1101.0557 [hep-th]}}.

\bibitem{Kapustin:2011jm}
A.~Kapustin and B.~Willett, ``{Generalized Superconformal Index for Three
  Dimensional Field Theories},'' {\em arXiv e-prints} (2011) ,
\href{http://arxiv.org/abs/1106.2484}{{\ttfamily arXiv:1106.2484 [hep-th]}}.

\bibitem{Dimofte:2011py}
T.~Dimofte, D.~Gaiotto, and S.~Gukov, ``{3-Manifolds and 3d Indices},''
  \href{http://dx.doi.org/10.4310/ATMP.2013.v17.n5.a3}{{\em Adv. Theor. Math.
  Phys.} {\bfseries 17} no.~5, (2013) 975--1076},
\href{http://arxiv.org/abs/1112.5179}{{\ttfamily arXiv:1112.5179 [hep-th]}}.

\bibitem{Grant:2008sk}
L.~Grant, P.~A. Grassi, S.~Kim, and S.~Minwalla, ``{Comments on $1/16$ BPS
  Quantum States and Classical Configurations},''
  \href{http://dx.doi.org/10.1088/1126-6708/2008/05/049}{{\em JHEP} {\bfseries
  05} (2008) 049},
\href{http://arxiv.org/abs/0803.4183}{{\ttfamily arXiv:0803.4183 [hep-th]}}.

\bibitem{Chang:2013fba}
C.-M. Chang and X.~Yin, ``{1/16 BPS states in $\mathcal N=$ 4 super-Yang-Mills
  theory},'' \href{http://dx.doi.org/10.1103/PhysRevD.88.106005}{{\em Phys.
  Rev. D} {\bfseries 88} no.~10, (2013) 106005},
\href{http://arxiv.org/abs/1305.6314}{{\ttfamily arXiv:1305.6314 [hep-th]}}.

\bibitem{Yokoyama:2014qwa}
S.~Yokoyama, ``{More on BPS States in $ \mathcal{N}=4 $ Supersymmetric
  Yang-Mills Theory on R $\times$ S$^{3}$},''
  \href{http://dx.doi.org/10.1007/JHEP12(2014)163}{{\em JHEP} {\bfseries 12}
  (2014) 163},
\href{http://arxiv.org/abs/1406.6694}{{\ttfamily arXiv:1406.6694 [hep-th]}}.

\bibitem{Larsen:2019oll}
F.~Larsen, J.~Nian, and Y.~Zeng, ``{AdS$_{5}$ black hole entropy near the BPS
  limit},'' \href{http://dx.doi.org/10.1007/JHEP06(2020)001}{{\em JHEP}
  {\bfseries 06} (2020) 001}, \href{http://arxiv.org/abs/1907.02505}{{\ttfamily
  arXiv:1907.02505 [hep-th]}}.

\bibitem{Kim:2006he}
S.~Kim and K.-M. Lee, ``{1/16-BPS Black Holes and Giant Gravitons in the AdS(5)
  X S5 Space},'' \href{http://dx.doi.org/10.1088/1126-6708/2006/12/077}{{\em
  JHEP} {\bfseries 12} (2006) 077},
\href{http://arxiv.org/abs/hep-th/0607085}{{\ttfamily arXiv:hep-th/0607085
  [hep-th]}}.

\bibitem{Brandhuber:1999np}
A.~Brandhuber and Y.~Oz, ``{The D-4 - D-8 brane system and five-dimensional
  fixed points},'' \href{http://dx.doi.org/10.1016/S0370-2693(99)00763-7}{{\em
  Phys. Lett. B} {\bfseries 460} (1999) 307--312},
\href{http://arxiv.org/abs/hep-th/9905148}{{\ttfamily arXiv:hep-th/9905148
  [hep-th]}}.

\bibitem{Seiberg:1996bd}
N.~Seiberg, ``{Five-dimensional SUSY field theories, nontrivial fixed points
  and string dynamics},''
  \href{http://dx.doi.org/10.1016/S0370-2693(96)01215-4}{{\em Phys. Lett. B}
  {\bfseries 388} (1996) 753--760},
\href{http://arxiv.org/abs/hep-th/9608111}{{\ttfamily arXiv:hep-th/9608111
  [hep-th]}}.

\bibitem{Bobev:2015kza}
N.~Bobev, M.~Bullimore, and H.-C. Kim, ``{Supersymmetric Casimir Energy and the
  Anomaly Polynomial},'' \href{http://dx.doi.org/10.1007/JHEP09(2015)142}{{\em
  JHEP} {\bfseries 09} (2015) 142},
\href{http://arxiv.org/abs/1507.08553}{{\ttfamily arXiv:1507.08553 [hep-th]}}.

\bibitem{Pasquetti:2011fj}
S.~Pasquetti, ``{Factorisation of N = 2 Theories on the Squashed 3-Sphere},''
  \href{http://dx.doi.org/10.1007/JHEP04(2012)120}{{\em JHEP} {\bfseries 04}
  (2012) 120},
\href{http://arxiv.org/abs/1111.6905}{{\ttfamily arXiv:1111.6905 [hep-th]}}.

\bibitem{Beem:2012mb}
C.~Beem, T.~Dimofte, and S.~Pasquetti, ``{Holomorphic Blocks in Three
  Dimensions},'' \href{http://dx.doi.org/10.1007/JHEP12(2014)177}{{\em JHEP}
  {\bfseries 12} (2014) 177},
\href{http://arxiv.org/abs/1211.1986}{{\ttfamily arXiv:1211.1986 [hep-th]}}.

\bibitem{Nieri:2013yra}
F.~Nieri, S.~Pasquetti, and F.~Passerini, ``{3d and 5d Gauge Theory Partition
  Functions as $q$-deformed CFT Correlators},''
  \href{http://dx.doi.org/10.1007/s11005-014-0727-9}{{\em Lett. Math. Phys.}
  {\bfseries 105} no.~1, (2015) 109--148},
\href{http://arxiv.org/abs/1303.2626}{{\ttfamily arXiv:1303.2626 [hep-th]}}.

\bibitem{Pasquetti:2016dyl}
S.~Pasquetti, ``{Holomorphic blocks and the 5d AGT correspondence},''
  \href{http://dx.doi.org/10.1088/1751-8121/aa60fe}{{\em J. Phys.} {\bfseries
  A50} no.~44, (2017) 443016},
\href{http://arxiv.org/abs/1608.02968}{{\ttfamily arXiv:1608.02968 [hep-th]}}.

\bibitem{Nekrasov:2003vi}
N.~A. Nekrasov, ``{Localizing gauge theories},'' in {\em {Mathematical physics.
  Proceedings, 14th International Congress, ICMP 2003, Lisbon, Portugal, July
  28-August 2, 2003}}, pp.~645--654.
\newblock
2003.
\newblock

\bibitem{Bershtein:2016mxz}
M.~Bershtein, G.~Bonelli, M.~Ronzani, and A.~Tanzini, ``{Gauge theories on
  compact toric surfaces, conformal field theories and equivariant Donaldson
  invariants},'' \href{http://dx.doi.org/10.1016/j.geomphys.2017.01.012}{{\em
  J. Geom. Phys.} {\bfseries 118} (2017) 40--50},
\href{http://arxiv.org/abs/1606.07148}{{\ttfamily arXiv:1606.07148 [hep-th]}}.

\bibitem{Kim:2013nva}
H.-C. Kim, S.~Kim, S.-S. Kim, and K.~Lee, ``{The general M5-brane
  superconformal index},'' {\em arXiv e-prints} (2013) ,
\href{http://arxiv.org/abs/1307.7660}{{\ttfamily arXiv:1307.7660 [hep-th]}}.

\bibitem{Qiu:2016dyj}
J.~Qiu and M.~Zabzine, ``{Review of localization for 5d supersymmetric gauge
  theories},'' \href{http://dx.doi.org/10.1088/1751-8121/aa5ef0}{{\em J. Phys.
  A} {\bfseries 50} no.~44, (2017) 443014},
\href{http://arxiv.org/abs/1608.02966}{{\ttfamily arXiv:1608.02966 [hep-th]}}.

\bibitem{Festuccia:2018rew}
G.~Festuccia, J.~Qiu, J.~Winding, and M.~Zabzine, ``{Twisting with a Flip (the
  Art of Pestunization)},''
  \href{http://dx.doi.org/10.1007/s00220-020-03681-9}{{\em Commun. Math. Phys.}
  {\bfseries 377} no.~1, (2020) 341--385},
  \href{http://arxiv.org/abs/1812.06473}{{\ttfamily arXiv:1812.06473
  [hep-th]}}.

\bibitem{Choi:2019zpz}
S.~Choi, C.~Hwang, and S.~Kim, ``{Quantum vortices, M2-branes and black
  holes},'' \href{http://arxiv.org/abs/1908.02470}{{\ttfamily arXiv:1908.02470
  [hep-th]}}.

\bibitem{Choi:2019dfu}
S.~Choi and C.~Hwang, ``{Universal 3d Cardy Block and Black Hole Entropy},''
  \href{http://dx.doi.org/10.1007/JHEP03(2020)068}{{\em JHEP} {\bfseries 03}
  (2020) 068}, \href{http://arxiv.org/abs/1911.01448}{{\ttfamily
  arXiv:1911.01448 [hep-th]}}.

\bibitem{Cabo-Bizet:2019eaf}
A.~Cabo-Bizet and S.~Murthy, ``{Supersymmetric phases of 4d $N=4$ SYM at large
  $N$},'' {\em arXiv e-prints} (2019) ,
\href{http://arxiv.org/abs/1909.09597}{{\ttfamily arXiv:1909.09597 [hep-th]}}.

\bibitem{Benini:2020gjh}
F.~Benini, E.~Colombo, S.~Soltani, A.~Zaffaroni, and Z.~Zhang,
  ``{Superconformal indices at large $N$ and the entropy of AdS$_5$ $\times$
  SE$_5$ black holes},'' \href{http://arxiv.org/abs/2005.12308}{{\ttfamily
  arXiv:2005.12308 [hep-th]}}.

\bibitem{Choi:2018vbz}
S.~Choi, J.~Kim, S.~Kim, and J.~Nahmgoong, ``{Comments on deconfinement in
  AdS/CFT},'' {\em arXiv e-prints} (2018) ,
\href{http://arxiv.org/abs/1811.08646}{{\ttfamily arXiv:1811.08646 [hep-th]}}.

\bibitem{Honda:2019cio}
M.~Honda, ``{Quantum Black Hole Entropy from 4d Supersymmetric Cardy
  formula},'' \href{http://dx.doi.org/10.1103/PhysRevD.100.026008}{{\em Phys.
  Rev. D} {\bfseries 100} no.~2, (2019) 026008},
\href{http://arxiv.org/abs/1901.08091}{{\ttfamily arXiv:1901.08091 [hep-th]}}.

\bibitem{ArabiArdehali:2019tdm}
A.~Arabi~Ardehali, ``{Cardy-like asymptotics of the 4d $ \mathcal{N}=4 $ index
  and AdS$_{5}$ blackholes},''
  \href{http://dx.doi.org/10.1007/JHEP06(2019)134}{{\em JHEP} {\bfseries 06}
  (2019) 134},
\href{http://arxiv.org/abs/1902.06619}{{\ttfamily arXiv:1902.06619 [hep-th]}}.

\bibitem{Murthy:2020rbd}
S.~Murthy, ``{The growth of the $\frac{1}{16}$-BPS index in 4d $\mathcal{N}=4$
  SYM},'' \href{http://arxiv.org/abs/2005.10843}{{\ttfamily arXiv:2005.10843
  [hep-th]}}.

\bibitem{Agarwal:2020zwm}
P.~Agarwal, S.~Choi, J.~Kim, S.~Kim, and J.~Nahmgoong, ``{AdS black holes and
  finite N indices},'' \href{http://arxiv.org/abs/2005.11240}{{\ttfamily
  arXiv:2005.11240 [hep-th]}}.

\bibitem{Benvenuti:2006xg}
S.~Benvenuti, L.~A. Pando~Zayas, and Y.~Tachikawa, ``{Triangle anomalies from
  Einstein manifolds},''
  \href{http://dx.doi.org/10.4310/ATMP.2006.v10.n3.a4}{{\em Adv. Theor. Math.
  Phys.} {\bfseries 10} no.~3, (2006) 395--432},
\href{http://arxiv.org/abs/hep-th/0601054}{{\ttfamily arXiv:hep-th/0601054
  [hep-th]}}.

\bibitem{Kim:2019yrz}
J.~Kim, S.~Kim, and J.~Song, ``{A 4d $N=1$ Cardy Formula},'' {\em arXiv
  e-prints} (2019) ,
\href{http://arxiv.org/abs/1904.03455}{{\ttfamily arXiv:1904.03455 [hep-th]}}.

\bibitem{Cabo-Bizet:2019osg}
A.~Cabo-Bizet, D.~Cassani, D.~Martelli, and S.~Murthy, ``{The asymptotic growth
  of states of the 4d $ \mathcal{N}=1 $ superconformal index},''
  \href{http://dx.doi.org/10.1007/JHEP08(2019)120}{{\em JHEP} {\bfseries 08}
  (2019) 120},
\href{http://arxiv.org/abs/1904.05865}{{\ttfamily arXiv:1904.05865 [hep-th]}}.

\bibitem{DiPietro:2014bca}
L.~Di~Pietro and Z.~Komargodski, ``{Cardy formulae for SUSY theories in $d =$ 4
  and $d =$ 6},'' \href{http://dx.doi.org/10.1007/JHEP12(2014)031}{{\em JHEP}
  {\bfseries 12} (2014) 031},
\href{http://arxiv.org/abs/1407.6061}{{\ttfamily arXiv:1407.6061 [hep-th]}}.

\bibitem{Amariti:2019mgp}
A.~Amariti, I.~Garozzo, and G.~Lo~Monaco, ``{Entropy function from toric
  geometry},'' {\em arXiv e-prints} (2019) ,
\href{http://arxiv.org/abs/1904.10009}{{\ttfamily arXiv:1904.10009 [hep-th]}}.

\bibitem{Cabo-Bizet:2020nkr}
A.~Cabo-Bizet, D.~Cassani, D.~Martelli, and S.~Murthy, ``{The large-$N$ limit
  of the 4d $\mathcal{N}=1$ superconformal index},''
  \href{http://arxiv.org/abs/2005.10654}{{\ttfamily arXiv:2005.10654
  [hep-th]}}.

\bibitem{Nian:2019pxj}
J.~Nian and L.~A. Pando~Zayas, ``{Microscopic entropy of rotating electrically
  charged AdS$_{4}$ black holes from field theory localization},''
  \href{http://dx.doi.org/10.1007/JHEP03(2020)081}{{\em JHEP} {\bfseries 03}
  (2020) 081}, \href{http://arxiv.org/abs/1909.07943}{{\ttfamily
  arXiv:1909.07943 [hep-th]}}.

\bibitem{Choi:2019miv}
S.~Choi and S.~Kim, ``{Large AdS$_6$ black holes from CFT$_5$},''
  \href{http://arxiv.org/abs/1904.01164}{{\ttfamily arXiv:1904.01164
  [hep-th]}}.

\bibitem{Kantor:2019lfo}
G.~Kantor, C.~Papageorgakis, and P.~Richmond, ``{AdS$_7$ Black-Hole Entropy and
  5D $\mathcal{N}=2$ Yang-Mills},''
  \href{http://dx.doi.org/10.1007/JHEP01(2020)017}{{\em JHEP} {\bfseries 01}
  (2020) 017},
\href{http://arxiv.org/abs/1907.02923}{{\ttfamily arXiv:1907.02923 [hep-th]}}.

\bibitem{Nahmgoong:2019hko}
J.~Nahmgoong, ``{6d superconformal {Cardy} formulas},'' {\em arXiv e-prints}
  (2019) ,
\href{http://arxiv.org/abs/1907.12582}{{\ttfamily arXiv:1907.12582 [hep-th]}}.

\bibitem{Benini:2019dyp}
F.~Benini, D.~Gang, and L.~A. Pando~Zayas, ``{Rotating Black Hole Entropy from
  M5 Branes},'' \href{http://dx.doi.org/10.1007/JHEP03(2020)057}{{\em JHEP}
  {\bfseries 03} (2020) 057}, \href{http://arxiv.org/abs/1909.11612}{{\ttfamily
  arXiv:1909.11612 [hep-th]}}.

\bibitem{Bobev:2019zmz}
N.~Bobev and P.~M. Crichigno, ``{Universal spinning black holes and theories of
  class R},'' \href{http://dx.doi.org/10.1007/JHEP12(2019)054}{{\em JHEP}
  {\bfseries 12} (2019) 054},
\href{http://arxiv.org/abs/1909.05873}{{\ttfamily arXiv:1909.05873 [hep-th]}}.

\bibitem{Goldstein:2019gpz}
K.~Goldstein, V.~Jejjala, Y.~Lei, S.~van Leuven, and W.~Li, ``{Probing the EVH
  limit of supersymmetric AdS black holes},''
  \href{http://dx.doi.org/10.1007/JHEP02(2020)154}{{\em JHEP} {\bfseries 02}
  (2020) 154}, \href{http://arxiv.org/abs/1910.14293}{{\ttfamily
  arXiv:1910.14293 [hep-th]}}.

\end{thebibliography}\endgroup
\end{document}